\documentclass[12pt,preprint]{aastex}

\newcommand{\che} {\log\ ({\rm C/He})}

\newcommand{\msun} {$M_{\odot}$}

\newcommand{\Te} {T_{\rm eff}}
\newcommand{\logg} {\log g}
\newcommand{\nh} {\log\ ({\rm H/He})}

\begin{document}

\title{On the Spectral Evolution of Cool, Helium-Atmosphere\\ White Dwarfs:
Detailed Spectroscopic and Photometric\\ Analysis of DZ Stars}

\author{P. Dufour\altaffilmark{1}, P. Bergeron\altaffilmark{1},
James Liebert\altaffilmark{2},
H. C. Harris\altaffilmark{3},
G.R. Knapp\altaffilmark{4},
S.F. Anderson\altaffilmark{5},
Patrick B. Hall\altaffilmark{6},
Michael A. Strauss\altaffilmark{4},
Matthew J. Collinge\altaffilmark{4}, and
Matt C. Edwards\altaffilmark{4}}

\altaffiltext{1}{D\'{e}partement de Physique, Universit\'{e} 
de Montr\'{e}al, C.P. 6128, Succ. Centre-Ville, Montr\'{e}al, Qu\'{e}bec, 
Canada H3C 3J7; dufourpa@astro.umontreal.ca,
bergeron@astro.umontreal.ca}
\altaffiltext{2}{Steward Observatory, University of Arizona, 933 North Cherry Avenue, Tucson, AZ 85721; liebert@as.arizona.edu}
\altaffiltext{3}{US Naval Observatory, P.O. Box 1149, Flagstaff, AZ 86002-1149; hch@nofs.navy.mil}
\altaffiltext{4}{Princeton Univ. Obs., Peyton Hall,Princeton, NJ 08544; gk@astro.princeton.edu}
\altaffiltext{5}{Department of Astronomy, University of Washington, Box 351580, Seattle, WA 98195; anderson@astro.washington.edu}
\altaffiltext{6}{Department of Physics and Astronomy, York University, 128 Petrie Science and Engineering Building, 4700 Keele Street, Toronto, ON M3J 1P3, Canada}

\begin{abstract}

We present a detailed analysis of a large spectroscopic and
photometric sample of DZ white dwarfs based on our latest model
atmosphere calculations. We revise the atmospheric parameters of the
trigonometric parallax sample of Bergeron, Leggett, \& Ruiz (12 stars)
and analyze 147 new DZ white dwarfs discovered in the Sloan Digital
Sky Survey. The inclusion of metals and hydrogen in our model
atmosphere calculations leads to different atmospheric parameters than
those derived from pure helium models. Calcium
abundances are found in the range from 
$\log\ ({\rm Ca/He}) = -12$ to $-8$. We also find that fits of
the coolest objects show peculiarities, suggesting that our
physical models may not correctly describe the
conditions of high atmospheric pressure encountered in the coolest DZ
stars. We find that the mean mass of the 11 DZ stars with
trigonometric parallaxes, $\langle M \rangle= 0.63$ \msun, is
significantly lower than that obtained from pure helium models,
$\langle M \rangle=0.78$ \msun, and in much better agreement with the
mean mass of other types of white dwarfs. We determine hydrogen
abundances for $27 \%$ of the DZ stars in our sample, while only upper
limits are obtained for objects with low signal-to-noise ratio spectroscopic
data. We confirm with a high level of confidence that the accretion
rate of hydrogen is at least two orders of magnitude smaller than
that of metals (and up to five in some cases) to be
compatible with the observations. We find a correlation between
the hydrogen abundance and the effective temperature, suggesting for
the first time empirical evidence of a lower temperature boundary for
the hydrogen screening mechanism. Finally, we speculate on the
possibility that the DZA white dwarfs could be the result of the
convective mixing of thin hydrogen-rich atmospheres with the
underlying helium convection zone.

\end{abstract}


\section{INTRODUCTION}

Cool helium-rich white dwarfs showing traces of heavy elements (other
than carbon) in their optical spectra are collectively known as DZ
stars (stars with $\Te\geq12,000$~K show He~\textsc{i} lines and are
thus classified DBZ stars). They are easily recognized by the presence
of the strong resonance Ca~\textsc{ii} H \& K doublet in the optical
and sometimes Ca~\textsc{i} $\lambda$4226, Mg~\textsc{i} $\lambda3835$
or Fe~\textsc{i} $\lambda3730$ (see \citealt{SionAtlas},
\citealt{WAtlas}, and \citealt{harris03} for representative spectra).

Analyses of DZ white dwarfs were pioneered by \citet{weidemann1960}, 
\citet{wegner72}, and \citet{grenfell74}, who were the first to analyze the 
classical DZ stars vMa2, Ross~640, and L745-46A using model atmosphere
calculations to reproduce the optical spectrum. Other
important analyses of DZ stars relying on optical data only include those 
of \citet{liebert77}, \citet{WL1980}, 
\citet{kapranidis86}, \citet{LW87}, \citet{Sion90}, and
\citet{dufour06}. Further progress in our understanding of these stars
has also been achieved from ultraviolet observations with the {\it
International Ultraviolet Explorer} (IUE) by \citet{cottrell80a},
\citet{cottrell80b}, \citet{Zeidler86}, \citet{weidemann89}, and 
\citet{koester90}, and with the {\it Hubble Space Telescope} (HST) by 
\citet{KW00} and \citet{WKL02}. The UV region of the energy distribution 
is most useful for a detailed abundance analysis of elements with no
strong resonance lines in the optical, and thus most of our current
knowledge of DZ stars comes from UV observations.

Since the heavy elements present in the atmospheric regions sink below
the photosphere in a timescale much shorter than the white dwarf
cooling time \citep{paquette86}, the presence of metals in DZ stars
must be explained in terms of episodic accretion from the
interstellar medium, a model put on a more quantitative basis in a
series of papers by
\citet{dupuis1,dupuis2,dupuis3}. Following the discussion by
\citet{KW00}, the most notable problem with this scenario is the low
hydrogen abundances observed in these objects. Being the lightest
element, hydrogen should only accumulate in the atmospheric regions of
the star over time, but the observed abundances (or limits in some
cases) indicate that the hydrogen accretion rate must be at least two
orders of magnitude lower than that of metals \citep{WKL02,dupuis3}.
This conclusion is based on a relatively small sample of DBZ/DZ stars,
albeit this was the largest homogeneous data set available at that
time, and the analysis of a larger sample is required to
advance our understanding of the accretion problem in cool DZ
stars. Such a large sample has recently become available, thanks to
the discovery of hundreds of new DZ white dwarfs by the Sloan Digital
Sky Survey
\citep[SDSS, York et al. 2000; see][for typical SDSS DZ spectra]{harris03}. Since these
spectra cover only the optical range, their analysis will be
restricted to the determination of the calcium and hydrogen
abundances, since UV observations of these faint (17 $\le g \le$
21) SDSS stars will not be possible in the near future. Nevertheless,
the addition of these new objects will allow a clear picture to be
drawn of the chemical evolution of white dwarf stars and to improve
the statistical significance of earlier results since, to our
knowledge, only 17 DZ stars with both $\Te$ and Ca/He measurements are
found in the literature.

One of the goals of the theory of spectral evolution is to explain
quantitatively the chemical evolution of cool white dwarfs. For
instance, the ratio of hydrogen- to helium-dominated atmospheres is
observed to vary significantly over the white dwarf cooling sequence,
indicating the existence of a physical mechanism to convert
one type into another. Moreover, the spectroscopic
features observed in helium-rich white dwarfs vary quite significantly with
temperature from the DB spectral type to the cooler DQ,
DC, and DZ types. The exact reason for this evolution into either of
these types is still not fully understood. To complicate the picture
even further, in a fraction of cool DA stars, the upper hydrogen layers
is expected to convectively
mix with the underlying helium layers to form helium-rich atmosphere
white dwarf stars.

Recently, \citet{brl97} and \citet[][hereafter referred to as
BRL97 and BLR01, respectively]{blr01} analyzed the energy distributions
of a large sample of cool white dwarfs with the aim of improving our
understanding of their chemical evolution. Among
their sample, we find several DZ stars that were analyzed under the
assumption of pure helium compositions, the only models available at
that time. \citet{dufour05} showed that the analysis of the energy
distribution of DQ stars based on pure helium models
overestimates the effective temperature compared to that obtained from
models including carbon. This effect is due to
an increase of the He$^-$ free-free opacity resulting from
the additional free electrons from carbon. A similar effect is thus expected in
the case of DZ white dwarfs. In this paper, we first include explicitly
the presence of metals in our model atmosphere calculations and
reanalyze the available photometric and spectroscopic data of DZ stars
from BRL97 and BLR01. We then present a similar analysis of a much
larger sample of DZ stars discovered in the SDSS \citep{gunn98,gunn06,pier03,tucker06,stoughton02,abazajian03,adelman06}. In
\S~\ref{observation}, we describe the observations. Our theoretical
framework including our model atmosphere and synthetic spectrum
calculations are presented in \S~\ref{theoretical}. The detailed
analysis follows in \S~\ref{analysis}, and the results are
interpreted and discussed in \S~\ref{results}. Our conclusions are
summarized in \S~\ref{conclusion}.

\section{OBSERVATIONS}\label{observation}

The first sample used for this study is drawn from the BRL97 and BLR01
analyses, which include 12 DZ stars with optical $BVRI$ and infrared
$JHK$ photometry as well as trigonometric parallax measurements, with
the exception of ESO 445-271 (WD~1338$-$311), for which the parallax
and $JHK$ measurements are not available. The photometric data can be
found in Tables 1 of BRL97 and BLR01. Our analysis also relies on
spectroscopic observations taken from various sources. New high
signal-to-noise ratio (S/N) spectroscopic observations 
at a resolution of $\sim 6$ \AA\ FWHM, covering the
Ca~\textsc{ii} H \& K doublet region have been secured with the Mont
M\'egantic Observatory 1.6 m telescope in 2004 September (WD~0046+051,
WD~0802+386, WD~1626$-$368, and WD~1705+030) and with the Steward
Observatory 2.3 m telescope in 2004 May (WD~2251$-$070 and
WD~2312$-$024). Optical spectra for the remaining stars were secured
by BRL97 and BLR01 (details of the observations are provided in these
references).

Our second sample consists of DZ white dwarfs spectroscopically
identified in the SDSS. We selected all DZ stars from the First and
Fourth Data Release white dwarf catalogs \citep{kleinman04,
Eisenstein06}. These catalogs are not complete in any way (see below)
and several more DZ
white dwarfs are certainly present in the SDSS archive. Additional DZ
stars have also been added to our list as they were discovered
serendipitously by examination of the SDSS spectroscopic archive. Our
final sample consists of 147 SDSS DZ stars with spectra covering the
3800$-$9200 \AA\ region at a resolution of $\sim$ 3~\AA\ FWHM. Also
available are SDSS photometric observations on the $ugriz$ system
\citep{fukugita96, hogg01,smith02,ivezic04}. Combining the two data sets,
we have a large sample of 159 DZ white dwarfs that can
be analyzed in a {\it homogeneous} fashion. This is a
considerably larger sample than the 17 DZ stars with both $\Te$ and
Ca/He determinations analyzed by various groups in the last $\sim$50
years or so.

\section{MODEL ATMOSPHERE AND SYNTHETIC SPECTRUM CALCULATIONS}\label{theoretical}

Our LTE model atmosphere code is similar to that described in
\citet{dufour06} for the study of the DZ white dwarf G165-7. It 
is based on a modified version of the code described at length in
\citet{BSW95}, which is appropriate for pure hydrogen and pure
helium atmospheric compositions, as well as mixed hydrogen and helium
compositions, while energy transport by convection is treated within
the mixing-length theory. One important modification is that metals
and molecules are now included in the equation-of-state and opacity
calculations \citep[see][for details]{dufour05,dufour06}. As was the
case for DQ stars, He$^-$ free-free absorption is found to be the dominant
source of opacity in DZ stars. It is thus important to account for all
possible sources of electrons in the equation-of-state, and we have
included here all elements with $Z\le26$. The chemical abundances
cannot be determined individually, however, since most of these
elements are not observed spectroscopically. We thus initially assume that
the relative abundances are consistent with solar ratios; this is a
reasonable assumption, at least for the observed elements, according
to our analysis of G165-7 \citep{dufour06} and that of
\citet[][]{WKL02} for several DZ stars (see their Fig.~7). We show below that this assumption 
is not very critical for our atmospheric parameter
determinations.

More critical is the assumed hydrogen abundance for stars not showing
H$\alpha$ absorption. To explore this issue, we calculated 4 grids with
respectively $\nh=-3$, $-4$, $-5$, and $-30$ respectively. Our model 
grids cover a
range of atmospheric parameters from $\Te=4000$ to 12,000~K in steps of
500 K at a fixed value of $\logg=8$, and $\log\ ({\rm Ca/He})= -12$ to
$-7$ in steps of 0.5 dex, while the relative abundances of all
elements heavier than helium are set with respect to the calcium
abundance in solar ratios. These assumptions will need to be verified
a posteriori. We also calculated a smaller grid with $\nh=-6$,
$\Te=9000$ to 12,000~K, and $\log\ ({\rm Ca/He})=-11$ to $-8$ (with
identical steps as above). In order to assess the influence of the
Ly$\alpha$ line wing opacity, we calculated an additional grid with
$\nh=-3$ including the Ly$\alpha$ profile calculations described in
\citet{KW00}, kindly made available to us by D.~Koester (2005, private
communication). To take into account, in an approximate fashion, the
non-ideal effects on heavy elements, we also calculated models at
$\nh=-5$ with the occupation probability formalism of
\citet{HM88}. This formalism may not be adequate for stars with extremely 
high atmospheric pressure, however \citep[see, e.g.,][]{saumon95}, and it is used
mainly to allow a better comparison with other models in the
literature
\citep{KW00,WKL02}. Finally, we computed additional models with
different relative abundances of metals and values of $\logg$, to
explore the sensitivity of our results to these assumptions.

In addition to the increased He$^-$
free-free continuum opacity, important metal absorption features in the
ultraviolet may potentially affect the energy distributions and thus
the atmospheric structures compared to those obtained from pure helium
models. Thus, over 4000 of the strongest metal lines --- $\sim 2600$
lines from Fe~~\textsc{i} alone --- are included explicitly in both
the model and synthetic spectrum calculations. These lines are
selected by taking all lines contributing more than one tenth of the
He$^-$ free-free opacity in the range $\tau_R = 0.1 - 1.0$ from
several models at $\log{\rm Ca/He}=-7$ and $\Te$ between 5000 and
12,000 K. We are confident that this line list includes all the
important contributors to the atomic opacity, since spectra calculated
by increasing the number of lines by an order of magnitude did not
have any detectable effect on the emergent spectrum. The line absorption
coefficient is calculated using a Voigt profile for every line at
every depth point. The line broadening is treated within the impact
approximation with van der Waals broadening by neutral helium. Central
wavelengths of the transitions, $gf$ values, energy levels, and
damping constants are extracted from the GFALL line list of
R.~L.~Kurucz\footnote{see http://kurucz.harvard.edu/LINELISTS.html}.

Illustrative spectra from our model grid at $\logg=8$ are displayed in
Figure \ref{fg:f1} for various values of the effective temperature,
metal, and hydrogen abundances. We first notice that large hydrogen
abundances of $\nh \sim -3$ lead to a reduction of the width and depth
of the Ca~\textsc{ii} H \& K doublet compared to hydrogen-free
models. This can be explained in terms of the increased He$^-$
free-free opacity produced by the free electrons coming from hydrogen
when its abundance is sufficiently large, which actually outnumber
the contribution from metals. Since the pressure gradient is inversely
proportional to the Rosseland mean opacity, an increase in the opacity
results in a drop of the atmospheric pressure and corresponding
pressure broadening of the atomic lines. The competition between
metals and hydrogen as electron donors is illustrated in Figure
\ref{fg:f1} by comparing spectra at $\Te=9000$~K with $\log\ ({\rm
Ca/He})=-8.0$ and $-10.0$. The effect of the presence of hydrogen in
the latter case can already be observed at $\nh=-5$, while the
contribution of hydrogen can only be seen at much larger abundances in
the $\log\ ({\rm Ca/He})=-8.0$ model. At 12,000 K, the effect is 
less pronounced since the contribution from helium to the free
electrons becomes significant, and models with $\nh=-5$ are
practically identical to those without hydrogen. At $\nh=-3$, hydrogen
becomes the principal electron donor at 12,000~K (note also the
presence of H$\delta$ at 4101 \AA). Of course, if one is interested in
the determination of the hydrogen abundance in this temperature range,
direct observations at H$\alpha$ are certainly more useful since this
line can be detected for hydrogen abundances as low as $\nh\sim-5$ at
$\Te=10,000$~K (not shown here).

At lower effective temperatures, the hydrogen abundance can be
directly determined through H$\alpha$ only when the line is
spectroscopically visible, which occurs at relatively large abundances
--- approximately $\nh\sim-3$ at 6500 K and $\nh\sim-6$ near 10,000
K. However, in the intermediate temperature range between $\Te = 6000$
and 8500~K, the hydrogen abundance can be determined {\it indirectly}
from observations of the core depth and wing profiles of the
Ca~\textsc{ii} H \& K doublet, as demonstrated with the 7500 K spectra
shown in Figure \ref{fg:f1}. At lower temperatures, the effect of
hydrogen on the synthetic spectrum becomes less important since the
low ionization potential of metals relative to hydrogen favors the
contribution of free electrons from metals rather than
hydrogen. Finally, metallic absorption features become very strong at
low effective temperatures ($\Te\sim 6000$~K) for abundances typical
of those found in hotter DZ stars, $\log ({\rm Ca/He}) \sim-8.0$,
producing objects similar to the unique DZ star G165-7. Also at low
effective temperatures, collision induced absorption (CIA) by
molecular hydrogen quickly dominates the infrared opacity, providing a
severe constraint on the presence of hydrogen (see below).

\section{DETAILED ANALYSIS}\label{analysis}

\subsection{Atmospheric Parameter Determination}

The method used to fit the photometric and spectroscopic data is
similar to that described at length in \citet{dufour05} and
\citet{dufour06}. Briefly, we first estimate the effective
temperature of the star by fitting the global energy distribution as
provided by the $BVRI$ and $JHK$ (or $ugriz$ for SDSS stars)
photometric observations. The fitting procedure relies on the
nonlinear least-squares method of Levenberg-Marquardt
\citep{pressetal92}. Here, both $\Te$ and the solid angle
$\pi(R/D)^2$, which relates the flux at the surface of the star to
that received at Earth, are considered free parameters ($R$ is the
radius of the star and $D$ its distance from Earth). Since the
temperature obtained from the energy distribution depends on the
assumed chemical composition, we use the spectroscopic observations to
constrain the metal abundances. We thus assume the $\Te$ value
obtained from the photometry and determine the chemical composition by
fitting the Ca~\textsc{ii} H \& K doublet region with our grid of
synthetic spectra. We also measure the hydrogen abundance by fitting
the H$\alpha$ absorption line if present; otherwise, we proceed at a
fixed value of $\nh$ (see below). A new estimate of the effective
temperature is then obtained by fitting the photometric observations
with models interpolated at the metal and hydrogen abundances
determined from the spectroscopic fit (all metal abundances are
assumed solar with respect to calcium). The procedure is then iterated
until the atmospheric parameters have converged to a consistent
photometric and spectroscopic solution.

For 11 DZ stars with known distances obtained from trigonometric
parallax measurements, we can obtain the radius of the star from the
solid angle from our modelling, which in turn can be converted into
$\logg$ (or mass) using evolutionary models similar to those described
in \citet{fon01} but with C/O cores, $q({\rm He})\equiv \log M_{\rm
He}/M_{\star}=10^{-2}$, and $q({\rm H})=10^{-10}$, which are
representative of helium-atmosphere white dwarfs\footnote{see
http://www.astro.umontreal.ca/~bergeron/CoolingModels/}.

\subsection{Reappraisal of the BRL97 and BLR01 analyses}

We begin by re-evaluating the atmospheric parameters of the 12 DZ stars
in the BRL97 and BLR01 samples using our new models that
incorporate metals. In addition to the contribution of metals as
electron donors in the atmospheres of DZ stars, the contribution from
hydrogen also needs to be properly evaluated. Hydrogen abundances can
be determined directly from spectroscopic observations for only three
stars in our sample, namely Ross~640, L745-46A, and G165-7, which
explicitly show H$\alpha$. For the remaining DZ stars, upper limits 
on the hydrogen abundance can be obtained from the absence
of a detectable H$\alpha$ feature, or for the coolest stars, from the
infrared energy distribution, since the CIA opacity becomes
particularly important in a mixed H/He atmosphere due to the
collisions of H$_2$ with neutral helium.  The assumed hydrogen
abundance for the analysis of DZ stars is more important than it was
for DQ stars (see \citealt{dufour05}) because the high carbon
abundances (relative to hydrogen) in DQ stars make hydrogen a minor
contributor to the total electron population. The uncertainties of our
atmospheric parameter determination introduced by the use of models
with various hydrogen abundances is discussed at length below.

For DZ stars showing H$\alpha$ (the DZA stars), we use the observed
line profile to determine $\nh$, $\log\ ({\rm Ca/He})$, and $\Te$ in a
consistent manner as described above. When H$\alpha$ is not detected
spectroscopically, we fit each star with $\nh$ fixed at $-30$, $-5$,
$-4$, and $-3$, and carefully examine each solution and reject those
that predict a detectable H$\alpha$ absorption feature 
given the S/N of each spectrum.  We also
reject solutions that predict an infrared flux deficiency due to the
CIA opacity that is incompatible with the observed photometry at
$JHK$.  As discussed above (see Fig.~\ref{fg:f1}), we find some cases
where the Ca~\textsc{ii} H \& K lines (core and/or wings) are better
reproduced with models including a trace of hydrogen. This is a direct
consequence of the increased opacity (more free electrons from
hydrogen), which reduces the atmospheric pressure, producing narrower
line profiles. Note that the surface gravity may also affect the
atmospheric pressure, but models calculated at $\logg=8.0
\pm 0.25$ reveal that the Ca~\textsc{ii} H \& K line wings are not as 
strongly affected.

Figure \ref{fg:f2} and \ref{fg:f3} illustrate examples of solutions
with various hydrogen abundances for the DZ stars WD~1705+030 and
WD~2312$-$024, where we can clearly see that the
solutions with $\nh= -3$ and $-4$, respectively, provide a better
match to the observed calcium lines 
than do models without hydrogen. Such {\it indirect} hydrogen abundance
determinations are inherently more uncertain than a direct
determination through the H$\alpha$ line profile, however, since it is
possible for elements spectroscopically invisible to contribute to the
number of free electrons in a significant way. Thus for those
particular stars with no detectable H$\alpha$, we obtain new
atmospheric parameters by fitting explicitly the Ca~\textsc{ii} H \& K
profiles, but this time by considering the hydrogen abundance as a
free parameter in our fitting procedure.

We present in Table 1 our atmospheric parameter determinations
(effective temperature, surface gravity, stellar mass, and calcium
abundance) for the BRL97 and BLR01 samples. Quantities in parentheses
represent the formal 1$\sigma$ uncertainties of each parameter
obtained from our fitting procedure, with the exception of $\logg$ and
the mass, for which the uncertainties are obtained by propagating the
error of the trigonometric parallax measurement.  Note that these
formal errors are relatively small for some $\log\ ({\rm Ca/He})$ and
$\nh$ values, and this certainly does not reflect the true
uncertainties. Indeed, the values given here represent only the formal
internal uncertainties of the fitted atmospheric parameters obtained
from the covariance matrix (see Press et al.~1992 for details).  We
find, from visual inspection, that changing the abundance by about 0.2
dex still gives adequate fits. We thus estimate instead that an
uncertainty of $\pm 0.2$ dex is probably more realistic (except for the
coolest stars, see below). A similar conclusion was reached
by \citet[][see their Fig.~6]{provencal02} in the case of Procyon B.
To bracket the possible range of solutions for stars whose hydrogen abundance
could not be determined directly or indirectly, we provide two
solutions in Table 1, one with the maximum amount of hydrogen allowed
by the photometric and spectroscopic observations, and the other
solution without any hydrogen (``no H'' in Table 1). However,
\citet{WKL02} found, on the basis of their analysis of UV observations, that 
hydrogen is present in virtually all white dwarfs in their sample. We
thus believe that one should not artificially separate the DZ stars
showing traces of hydrogen (the DZA stars) from those showing no
hydrogen features, since this distinction is mostly due to the level of
visibility of hydrogen and the S/N of the spectra. Hence, our 
solutions that include hydrogen in
Table 1 are probably more realistic.

The corresponding fits to the $BVRI$ and $JHK$ photometric
observations and calcium lines are displayed in Figures \ref{fg:f4}
and \ref{fg:f5}. To our knowledge, the calcium abundances for the DZ
stars WD~0552$-$041 (G99-44), WD~1313$-$198 (LHS 2710), WD~1338$-$311
(ESO 445-271), WD~1705+030 (G139-13), and WD~2345$-$447 (ESO 292-43)
are determined here for the first time. Some stars deserve additional
comments:

\noindent {\it WD~0046+051 (vMa 2):}  Our solution, $\Te=6220$~K, 
$\log\ ({\rm Ca/He})=-10$, and $\nh=-3.19$, differs slightly from that
obtained by \citet{WKL02}, $\Te=5700$~K, $\log\ ({\rm Ca/He})=-10.65$,
and $\nh=-5$. Their solution requires a small trace of hydrogen in
order to reproduce the IUE flux level, which is reduced significantly
by their improved Ly$\alpha$ line profile calculations. Our hydrogen
abundance determination, on the other hand, is obtained indirectly
from the Ca~\textsc{ii} H \& K profiles. If we assume instead a value
of $\nh=-5$, we find $\Te=6010$~K, $\log\ ({\rm Ca/He})= -10.65$,
closer to the Wolff et al.~solution, although our fits to the calcium
line cores and wings are not as satisfactory with this smaller
hydrogen abundance.  Note that Wolff et al.~also include non-ideal
effects using the Hummer-Mihalas occupation probability formalism. To
evaluate the effects of the different input physics between
the two sets of models, we calculated a single model with the Wolff et al.
atmospheric parameters by also including the Hummer-Mihalas formalism
for metals and the same Ly$\alpha$ theoretical profiles.  Our best fit
to this single model with our original grid at $\nh=-3$ yields
$\Te=6147$~K and $\log\ ({\rm Ca/He})=-10.48$, close to our original
estimate of the effective temperature. In other words, had we calculated
a full grid with the Hummer-Mihalas occupation probability formalism, we would 
probably have found atmospheric parameters close to that obtained by 
\citet{WKL02}. Hence the differences
between both solutions can be partially explained in terms of the small
differences in the input physics. 

\noindent {\it WD~0552$-$041 (LP 658-2):} \citet{WKL02} obtain a better 
match to {\it Faint Object Spectrograph} (FOS) observations by
including a small trace of hydrogen of H/He$= 5\times10^{-4}$, but
they do this at $\Te=5050$ K, the effective temperature obtained by
BLR01 under the assumption of a pure helium composition. Metals must
have an influence on both the temperature structure and UV absorptions
that are not taken into account in the BLR01 solution. Our iterative
procedure does indeed suggest a much lower effective temperature for
this object, $\Te\sim4300$~K. Note that at this temperature, $JHK$
photometry would be affected by the collision induced opacity (not
included in \citealt{WKL02}) if $\nh$ is higher than $\sim -5$.  Our
metal abundance is obtained from the extremely weak Ca~\textsc{ii} H
\& K lines. However, our solution also predicts a strong Ca~\textsc{i}
$\lambda$4226 line that is not observed spectroscopically. We can only
speculate at this point on the reason for this discrepancy. One
possibility is that neutral calcium could be pressure ionized under
the extreme conditions found in this star, although it is strange to
see a strong Ca~\textsc{i} $\lambda$4226 feature in WD~2251$-$070, a
DZ star with presumably even higher photospheric pressures. We were
also not able to achieve an acceptable fit using Koester's Ly$\alpha$
profiles or the occupation probability formalism in our model
calculations. A good fit to both sets of lines is possible only by
changing the effective temperature in such a way that the photometric
fit is no longer acceptable. Clearly, more effort in the modeling
of high atmospheric pressure white dwarfs is required.

\noindent {\it WD~0738$-$172 (L745-46A), WD~1626+368 (Ross 640):} 
Our solutions for these stars, $\Te=7590$~K, $\log\ ({\rm
Ca/He})=-10.91$ and $\Te=8440$~K, $\log\ ({\rm Ca/He})=-8.83$,
respectively, are very close to those obtained by \citet{KW00},
$\Te=7500$~K, $\log\ ({\rm Ca/He})=-10.6$ and $\Te=8500$~K, $\log\
({\rm Ca/He})=-8.65$, respectively. The latter were obtained from fits
to the Ca~\textsc{ii} lines and from FOS spectra. In order to
reproduce the observed fluxes in the UV, Wolff et al.~had to include
their improved Ly$\alpha$ theoretical calculations. Although our
models do not include this opacity and fail to reproduce the UV
fluxes, our atmospheric parameters do not differ significantly from
those of Wolff et al. The reason for this result is that the UV flux
absorbed by the Ly$\alpha$ wings represents only a small fraction of
the total flux, and its redistribution to other wavelengths affects
the thermodynamic structure only slightly. To test this hypothesis, we
calculated several models including the Ly$\alpha$ profile and fitted
the resulting synthetic spectra and energy distributions with our
original grid. The differences in $\Te$ and $\log\ ({\rm Ca/He})$ are
always smaller than $\sim$ 150 K and 0.20 dex, respectively, well
within the measurement uncertainties for these objects. Given that
these differences are small, we refrain from including systematically
this Ly$\alpha$ profile in our model calculations since the approach
used by Wolff et al.~has been successfully tested for only a few stars
that are relatively hot (7500 and 8500 K), and it is not clear whether
this formalism remains valid at lower temperatures.

\noindent {\it WD~0802+386 (LP 257-28):} This object is the hottest DZ 
star ($\Te = 10,980$~K) in the BRL97/BLR01 sample. At this temperature,
helium-rich white dwarfs are expected to show H$\alpha$ even for
hydrogen abundances as low as $\nh=-5.0$. Our featureless spectrum
near H$\alpha$ is thus extremely surprising considering that accretion
of hydrogen with a rate as low as $10^{-20}\ M_{\odot}$ yr$^{-1}$
should provide enough hydrogen to be easily detected
spectroscopically. We postpone the discussion of this object after the
analysis of the DZ stars from the SDSS, which include several white
dwarfs similar to LP 257-28.

\noindent {\it WD~1328+307 (G165-7):} This star has a magnetic field 
and has been analyzed separately by \citet{dufour06}. The atmospheric
parameters given in Table 1 and the fit with magnetic models displayed
in Figure \ref{fg:f4} are taken from that analysis.

\noindent {\it WD~2251$-$070 (LP 701-29):} The best fit
obtained for this star by \citet{kapranidis86} reproduces the
Ca~\textsc{i} $\lambda$4226 line but fails to provide an acceptable
fit to the Ca~\textsc{ii} lines. The authors argue that pressure
ionization probably affects the metal ionization equilibrium, but an
exact treatment was not available at that time and could not be tested
further. We find that the photometric and spectroscopic observations
are fairly well reproduced by our hydrogen-free models; note that our
temperature estimate for this star, $\Te=4000$~K, is at the limit of
our grid. In this temperature range, even though the H$\alpha$ line is
spectroscopically invisible, the hydrogen abundance can be easily
constrained from the infrared $JHK$ photometry since the H$_2$-He CIA
opacity becomes important even for hydrogen abundances as low as
$\nh=-5$.  We note also that \citet{kapranidis86} used pure helium
stratifications for their synthetic spectrum calculations. Our models
reveal significant differences between the pressure and temperature
structures of pure helium models and those including metals.  It is
unclear, however, how this would translate in the models of Kapranidis
\& Liebert calculated using a different theoretical framework based on
a Thomas-Fermi equation-of-state, which also includes electron thermal
conduction as an energy transfer mechanism. Although not perfect, our
best fit reproduces the neutral calcium line well, and shows a hint of
ionized calcium. Our models fail, however, to reproduce the metallic
blend near 4500 \AA. It is clearly inconsistent to invoke pressure
ionization to explain the absence of Ca~\textsc{i} $\lambda$4226 in
WD~0552$-$041 when the same line is very strong in a star
characterized with even higher atmospheric pressures. This result
suggests that the physics of cool helium-rich models at high densities
is not fully understood yet, and that significant improvements are
still required.

The mean mass for the 11 DZ stars with trigonometric parallax
measurements analyzed in this paper is $\langle M \rangle = 0.63$
\msun. As discussed above, however, the atmospheric parameters of the 
five coolest stars are certainly more uncertain than those of the rest of the 
sample. If we exclude these
objects, the mean mass rise slightly to $\langle M \rangle = 0.66$ \msun,
significantly lower than the average mass obtained by BRL97/BLR01,
$\langle M \rangle = 0.78$ (or $\langle M \rangle = 0.82$~\msun\ if we
exclude L745-46A and Ross~640 which were analyzed with mixed H/He
models by BRL97 and BLR01). This difference in average mass can be
readily explained in terms of the free electrons provided by the
metals as well as hydrogen, which both increase the
opacity in our model calculations.

\citet{Beauchamp95} found that the mass distribution of DBA stars 
significantly differs from that of DB stars, with a higher fraction of
massive white dwarfs above $M\sim0.65$ \msun. Moreover, the three most
massive stars in their sample were DBA stars, suggesting a possible
link between the presence of hydrogen and the mass of the star. We do
not find such a trend in our sample of DZ stars, the possible
descendents of DB white dwarfs. In fact, one of the most massive stars
in our sample, WD~0802+386, has an extremely small upper limit on its
hydrogen content. However, considering the limited size of our sample,
we remain cautious and will refrain from drawing any definitive
conclusion on the interpretation of the mass distribution of DZ white
dwarfs.

\subsection{Analysis of DZ Stars from the SDSS}

Prior to this work, abundance analyses of DZ white dwarfs ($\Te <
12,000$~K) had been carried out for only 17 stars. Our SDSS sample
alone contains 147 new DZ stars. These additional objects discovered
in the SDSS allow a significant improvement in the statistics of DZ
stars. The method we adopt to fit the SDSS data is similar to that
described above, with the exception that the SDSS $ugriz$ photometry
is used instead of the $BVRI$ measurements. These photometric bands
cover the entire optical range from the UV atmospheric cutoff (3200
\AA) to the red sensitivity cutoff of the detector ($\sim10,000$ \AA).
Furthermore, since trigonometric parallax measurements are not
available for the SDSS stars, we assume a value of $\logg = 8.0$ for
all objects. As was the case for the BRL97 and BLR01 samples, the
hydrogen abundance for most stars is unknown. We thus fit all stars
with our model grids with various hydrogen abundances, and then reject
the solutions that are incompatible with the observations at
H$\alpha$, or determine directly the hydrogen abundances when
H$\alpha$ is visible, or indirectly through the Ca~\textsc{ii} H \& K
line profiles whenever possible (i.e. for spectra with good enough 
S/N ratio).

Our adopted atmospheric parameters are presented in Table 2. We have
not applied any correction for Galactic extinction since the stars are
relatively close, therefore only a small fraction of the absorption
should be applied \citep[see][for more on this
issue]{dufour05}. Figures \ref{fg:f6} to \ref{fg:f24} show our best
fits to the energy distribution, to the Ca~\textsc{ii} H \& K lines,
and to the H$\alpha$ region for the maximum hydrogen abundance allowed
by the spectroscopic observations, or for the hydrogen abundance
obtained from fits to H$\alpha$ when it is visible spectroscopically.

There is another set of Ca~\textsc{ii} lines available in the red
portion of the SDSS spectra at 8498, 8542, and 8662 \AA\ (the
``infrared triplet'') that can be used as an internal consistency
check. However, these lines are not as strong as the H \& K lines, and
they are detected only at high calcium abundances above $\log\ ({\rm
Ca/He})>-9$ at the S/N typical of our spectra. Our models predict the
presence of these lines in several objects, but a direct comparison
could only be achieved for stars with good signal-to-noise ratio
spectra. Figure \ref{fg:f25} shows the DZ stars with recognizable
Ca~\textsc{ii} lines together with our synthetic spectra interpolated
at the atmospheric parameter solution obtained from the Ca~\textsc{ii}
H \& K lines. We find a good internal consistency between the two sets
of lines for three out of four stars with sufficiently high S/N to
allow a useful comparison (the bottom four objects). Higher S/N
spectroscopic observations would be required for the other DZ stars in
our sample. For SDSS J103809.19$-$003622.5, we observe a discrepancy
between both sets of lines, the Ca~\textsc{ii} lines in the red
favoring a higher effective temperature than that inferred from the
photometry (our photometric and spectroscopic fits shown in
Fig.~\ref{fg:f15} are not very good for this object). One possible solution is
that this star is an unresolved degenerate binary. We also find one
peculiar object (SDSS J155852.59+031242.9) showing strong absorption
features at the position of the Ca~\textsc{ii} lines that are at odds
with the predicted profiles.

The calcium lines displayed in Figure \ref{fg:f25} are also more
sensitive to the presence of a weak magnetic field since the
separation of the Zeeman components
is proportional to the square of the central wavelength of the
line, $\Delta\lambda = 9.34\times10^{-13} \lambda^2_cg_{eff}B_s$
\citep[see for instance the analysis of G165-7 by][]{dufour06}. The 
absence of any line splitting translates into an upper limit of
$\sim150$~kG for the four stars shown in Figure \ref{fg:f25}. Higher
S/N spectroscopic observations of the other stars in our sample could
potentially provide interesting limits on the presence of magnetic
fields in DZ white dwarfs.

\subsection{Comments on the Assumed Solar Composition}

Our synthetic spectrum calculations assume that the relative abundance
of heavy elements is solar; the metal content is thus fixed by our fit
to the calcium lines. Is this a reasonable assumption? Most heavy
elements are not observed in the optical spectrum and thus there is no
way to determine individual abundances. As for the elements that are
visible only occasionally (e.g., magnesium, iron), detailed abundance
analyses in the optical are extremely limited, and one has to rely on
UV observations to obtain a reasonable estimate of the metal
content. Unfortunately, UV observations are available only for a small
number of DZ stars, the most complete analysis being that of
\citet{WKL02}.  According to their Figure 7, most cool stars exhibit
metal-to-metal ratios that are compatible with solar ratios within 1
dex. Our calculations indicate that small variations of metal
abundances (other than calcium) with respect to solar ratios do not
have any significant effect on our atmospheric parameter
determinations. It thus seems reasonable to assume that the metals are
present in the atmospheres of DZ stars in solar ratios, or at least
not too far from solar. Also, models calculated with the abundances of
all elements never observed in DZ stars set to zero are
practically identical to those calculated with solar values with
respect to calcium. This is not surprising in view of the fact that
hydrogen, magnesium, and iron are always the principal electron
donors.

Our SDSS sample contains only a few objects showing elements other
than calcium in their optical spectrum (see, e.g., the Mg and Fe lines
near $\sim3850$ \AA\ in Figs.~\ref{fg:f6} to \ref{fg:f24}). Although
the S/N ratio is relatively low, the predicted and observed metallic
features for these few stars are at least compatible with solar
ratios. The DZ star SDSS J095645.15+591240.6 shown in Figures
\ref{fg:f13} (blue part) and \ref{fg:f26} (red part) represents an
extreme example of a DZ star with metal abundances consistent with
solar ratios based on an examination of the various iron and magnesium
lines.  Also displayed in Figure \ref{fg:f26} are white dwarf stars
that exhibit the Mg~\textsc{i} ``b'' blend at $\sim 5175$ \AA. The
feature is strongest in SDSS J095645.15+591240.6, where the famous
asymmetry of the blue wing can clearly be seen. \citet{dufour06}
argued that the asymmetry in the magnetic DZ star G165-7 was possibly
due to molecular absorption by MgH. However, at $\Te=8230$~K, SDSS
J095645.15+591240.6 is too hot for this molecule to be visible, and
another explanation must be sought. Note that an increase of the
magnesium abundance would produce lines in the 3800 \AA\ region that
would be incompatible with the observations. SDSS
J103809.19$-$003622.5, discussed in the previous section, is also
problematic here.

A metal-to-metal ratio analysis similar to those of \citet{dupuis3} and
\citet{WKL02} is not possible here since our sample contains only a few
objects with more than one heavy element observed. However, our study
can be used to identify the best candidates for future UV observations
(for abundance analyses) and IR observations \citep[to find possible
debris disks or dust clouds around white dwarfs; see similar studies
around DAZ stars in][]{Reach05,Kilic05,Becklin05,Kilic06b}. In
conclusion, models calculated with solar abundance ratios reproduce
all features observed in the optical (with the exception of the Mg
~\textsc{i} ``b'' blend), and they do not predict absorption features
that are not observed.

\section{Results}\label{results}

\subsection{Calcium Abundances}

Our results are summarized in Figure \ref{fg:f27} where the calcium
abundance for all DZ stars analyzed in this paper are shown as a
function of effective temperature. The various symbols are described
in the figure caption.  The two continuous curves correspond to the
predicted equilibrium abundances reached in the low and high phases of
the two-phase accretion scenario proposed by \citet{dupuis3}, which
correspond to accretion rates of $5\times10^{-20}$ and
$5\times10^{-15}\ M_{\odot}$ yr$^{-1}$, respectively. Whether the
origin of metals is related to encounters with interstellar clouds or
attributed to comets or asteroids, it is comforting to find that the
accretion rates required to explain the calcium abundances in DZ stars
are about the same as for the DAZ stars \citep{KW06}, a result that
suggests that the metals observed in DZ and DAZ stars have the same
origin.

We first notice that the atmospheric parameters of our DZ stars fill
most of the Ca/He -- $\Te$ plane between the two continuous curves,
with the exception of some regions discussed here. First and most
obvious is the empty triangular region at high effective temperatures
and low calcium abundances. No stars are found in this region simply
because the Ca~\textsc{ii} H \& K transitions are not sufficiently excited
to be detected spectroscopically. Stars in this particular region
would thus appear as DC white dwarfs, and only ultraviolet
observations would allow one to fill this part of the plane. The dashed
line represents the detection threshold for the Ca~\textsc{ii} H \& K
lines calculated for a limit set at 5 \AA\ of total equivalent width (a
typical value for calcium lines observed in low S/N spectra).

We also find a region near $\sim6500$~K and high calcium abundances
where the number of stars is significantly reduced (the only star in
this region is G165-7 discussed below). Figure
\ref{fg:f28} shows that cool DZ stars form a parabola in the ($g-r$,
$u-g$) color-color diagram at a given effective temperature, and that
models with high calcium abundances overlap the stellar locus
region (highest density point region). Objects in the SDSS are
selected for follow-up spectroscopic observations mainly on the basis
of their colors. High priority is given to potential QSOs that have
colors outside the stellar locus
\citep{Richards02}. The paucity of stars in the regions of the Ca/He
-- $\Te$ plane discussed above can thus be explained in part as a
selection effect in the SDSS targeting procedure. The absence of cool
stars can also be explained by the fact that most of our stars come
from the DR4 catalog of \citet{Eisenstein06}. As shown by their Figure
1, a color cut was made to eliminate objects close or in the stellar
locus before they search for white dwarfs candidates. Cool DZ stars
in these regions, if they exist, should be found by a more
careful search of the SDSS spectroscopic archive, or with the help
of reduced proper motion surveys similar to that undertaken by
\citet{kilic06a}.

The only star in the upper-right corner of Figure
\ref{fg:f27} is G165-7. However, the low effective temperature and
high metallic content of G165-7 produce strong absorption features and
unusual colors that put it on the other side of the stellar locus, far
to the right and outside the color range displayed in Figure
\ref{fg:f28}. Note that the parameters given in Table 1 suggest that 
G165-7 would be near the stellar locus. However, these parameters were
obtained partially with $BVRI$ and $JHK$ photometry, and the fit to $u$ and $g$
in Figure 6 of \citet{dufour06} is not very good.

Finally, we find no DZ stars between $\Te\sim 5000$ and 6000 K. Note
that vMa 2 should probably be in that temperature range according to
the analysis of \citet{WKL02} based on models calculated with a
non-ideal equation of state.  As explained above, such cool stars have
colors that overlap the stellar locus, and it is therefore not
surprising to not find any DZ star in this particular range of
temperature in the SDSS sample. There is also no DZ star in the BRL97
and BLR01 samples in the same temperature range. However, a
significant deficiency of helium atmosphere white dwarfs has been
observed, the so-called non DA gap, in the ($V-I$, $V-K$) color-color
diagram shown in Figure 9 of BLR01. But since our sample contains only
a few DZ stars in the vicinity of the gap, and since the physics
in this high pressure regime is not fully understood, the
absence of DZ stars between $\sim 5000$ and 6000 K is not
statistically significant, and no firm conclusion concerning the
nature of the gap can be reached from our results.

\subsection{Hydrogen Abundances}

Our SDSS sample contains 27 objects with a detectable H$\alpha$
feature, while we achieve a better fit to the Ca~\textsc{ii} H \& K
lines for 10 additional objects if hydrogen is included, for a total
of 37/147 stars (or $\sim 25 \%$) with hydrogen detected directly or
indirectly. The SDSS sample is nowhere complete in volume or
magnitude, but since SDSS white dwarfs are selected randomly and
independently of the spectral type, stars with hydrogen should not be
preferentially chosen, so we believe that this ratio is representative
of the entire DZ population.  For the BRL97 and BLR01 samples, there
are three stars showing H$\alpha$, and three more for which the hydrogen
abundance could be determined indirectly. We thus have a total
of 43/159 stars (or $27 \%$) with hydrogen abundances determined 
in our entire sample.

Figure \ref{fg:f29} shows our hydrogen abundance
determinations (or upper limits) as a function of effective
temperature for our full sample. Also indicated
are the hydrogen-to-helium ratios predicted from continuous accretion
from the ISM in solar proportions at various rates, which are needed to
reproduce the amount of hydrogen measured (or constrained) if the
accretion starts at $\Te$=20,000 K. The exact choice for the starting
temperature is not critical since the cooling time scales increase
rapidly with decreasing effective temperatures. We find that
most DZ stars in Figure \ref{fg:f29} with hydrogen abundance
determinations are consistent with accretion rates in the range from
$10^{-20}$ to $10^{-18}\ M_{\odot}$ yr$^{-1}$. This is about two to
four orders of magnitude smaller than the average metal accretion rate
of $\sim 10^{-16}\ M_{\odot}$ yr$^{-1}$ used by \citet[][]{dupuis3}.
Our results confirm the conclusions reached by
\citet{dupuis3} and \citet{WKL02} that the hydrogen accretion rate 
must be at least two orders of magnitude lower that that of metals.

The conclusion that the hydrogen accretion rate must be significantly
lower than that of metals is further demonstrated in Figure
\ref{fg:f30}, where we show the Ca/H abundance ratios as a function of
effective temperature for our sample of DZ stars. The apparent
correlation of Ca/H with $\Te$ is due in part to the fact that there
are no stars with low calcium abundances at high effective
temperatures, since the Ca~\textsc{ii} lines are not observable in
this range of temperature (see Fig.~\ref{fg:f27}). The dotted line
indicates the solar Ca/H abundance ratio. Because hydrogen can only
accumulate with time in the mixed H/He convection zone, the maximum
Ca/H abundance ratio can only become smaller with decreasing effective
temperature.  Thus, the observed Ca/H abundance ratios that are close to the
solar value in a few stars are only coincidental; this does not
necessarily imply accretion in solar proportions, but indicates
instead that the accretion of hydrogen was reduced relative to
metals. The maximum value of Ca/H that can be expected from accretion
of material with solar composition is indicated by the solid line
\citep[from equation 6 of][]{dupuis3}.

\subsection{Correlation with Effective Temperature}

More importantly, the results shown in Figure
\ref{fg:f29} reveal for the first time a correlation between the
hydrogen abundances and the effective temperature. Such a correlation
has never been observed before because the number of known DZ stars
was simply too small. Moreover, as noted above, the absence of
H$\alpha$ in WD~0802+386 near $\Te\sim 11,000$~K implies an
unexpectedly low hydrogen abundance of $\nh < -6.0$. We find seven
additional DZ stars in the SDSS sample with $\Te > 9500$~K and similar
low hydrogen abundances. Figure \ref{fg:f29} indicates that the
accretion rate for these stars must be as low as $10^{-21}\ M_{\odot}$
yr$^{-1}$. An accretion rate slightly above $10^{-20}\ M_{\odot}$
yr$^{-1}$ would be sufficient to produce a detectable
H$\alpha$ absorption feature at these temperatures. The presence of
such small amounts of hydrogen is very surprising, especially
considering that several DZ stars in our sample do show H$\alpha$ in
the same temperature range. It seems very unlikely that white dwarfs
experience such a diversity of average conditions while traveling
large distances through the interstellar medium over their lifetime.

The inverse problem occurs at the other end of the diagram, where cool
DZ stars with $\nh\sim-3$ can be accounted for with accretion rates
around $10^{-19}\ M_{\odot}$ yr$^{-1}$ and above.  However, DZ stars
accreting hydrogen at such a high rate should be easily recognized at
higher effective temperatures if the accretion rate has remained
constant during the cooling of the star. Yet, our sample contains no
hot ($\Te >10,000$~K) DZ star with high hydrogen abundances around
$\nh\sim- 4$ expected from the high accretion rate inferred from the
cool DZ stars. Actually, the fact that the accretion rate required to
explain the hydrogen abundances in Figure \ref{fg:f29} increases from
$\sim 10^{-20}\ M_{\odot}$ yr$^{-1}$ at $\Te=11,000$~K up to $\sim
10^{-19}\ M_{\odot}$ yr$^{-1}$ at $\Te=7000$~K could be interpreted as
evidence that the physical mechanism that prevents hydrogen from being
accreted onto the surface of these stars becomes less efficient with
decreasing effective temperature.  For instance, \citet{WT82} proposed
a model where protons are prevented from accreting onto the surface of
the white dwarf by a rotating magnetic field, while metals, most
probably in the form of grains, are unaffected by this mechanism and
thus reach the surface.  Our results suggest that this so-called
propeller mechanism may become less effective below
$\Te\sim9000$~K. If this interpretation is correct, this would
represent the first empirical evidence for a decrease in efficiency
with temperature of the hydrogen screening mechanism at work in cool
helium atmosphere white dwarfs. This mechanism is still operating
quite efficiently, however, since the inferred accretion rates at low
effective temperatures remain several orders of magnitude lower than
those required to explain the presence of metals in DZ stars.

\subsection{Connection with Hotter DB Stars}

As mentioned above, $27\%$ of the DZ stars in our sample contain
hydrogen, a higher ratio than the $\sim 16 \%$ found in DBA stars
(\citealt{shipman87} gives 6/32, but after removal of the subdwarf
PG~2224+034, we get 5/31, or $16 \%$). It is generally believed that
this is the ratio among DB3 ($\Te\sim 12,000-19,000$~K) white
dwarfs. However, these statistics for DB stars are based on the
detection of H$\beta$ and H$\gamma$ only.

As part of a campaign undertaken by the Montreal group \citep[see,
e.g.,][]{hunter01}, nearly 80 bright northern DB white dwarfs have been
spectroscopically observed at medium and high resolutions. This
sample also includes all known DB2 stars with spectra obtained at
H$\alpha$, a more sensitive line to study the ratio of DBA to DB
stars. H$\alpha$ is found in 7/18 stars (or $38 \%$), more than twice the
ratio obtained by \citet{shipman87} for the cooler DB3 stars. If this
fraction for DB2 stars extends to the cooler DB3 stars, a systematic
search for H$\alpha$ in these objects is expected to perhaps double
the number of DBA3 stars.  Indeed, the atmospheres of DB stars become
increasingly transparent with lower effective temperatures, which
makes H$\alpha$ easier to detect spectroscopically, even if the mass
of the helium convection zone has significantly increased and further
diluted the amount of hydrogen present in the photospheric regions.

We thus argue that the canonical $20 \%$ ratio of DBA to DB white dwarfs
is certainly underestimated. Some authors have speculated that {\it
all} DB stars must contain at least some traces of hydrogen, although
explicit spectroscopic determinations are difficult \citep[from
observations of Ly$\alpha$, see][]{KW89,provencal00}, if not
impossible, because the amounts of hydrogen expected are below the
visibility limit. At the cooler end of the white dwarf sequence,
\citet{WKL02} detected hydrogen in nearly all the DBZ and DZ stars in
their sample. Several studies now favor the accretion of hydrogen from
the ISM (as opposed to primordial origin) to explain the DBA
phenomenon, since the amount of hydrogen found in cool DBA stars is
incompatible with the scenario that involves the transformation of DA
stars into DB stars near the red edge of the DB gap at
$\Te\sim30,000$~K.  However, a measure of the hydrogen accretion rate
in DB stars is complicated by the fact that the mass of the helium
convection zone varies rapidly with decreasing effective temperature,
and also by the lack of accurate determinations of hydrogen abundances
for large samples of DB and DBA stars.

Depending on the evolutionary models used, the total mass of hydrogen
present in DBA stars ranges from about $10^{-14}$ to $10^{-11}\
M_{\odot}$. Such small amounts of hydrogen imply that cool helium-rich
white dwarfs below $\Te\sim12,000$~K {\it must have completely forgotten}
their past hydrogen history as they become DC, DZ or DQ stars, since
the increasing depth of the helium convection zone with decreasing
effective temperature \citep[see Fig.~10 of][]{tassoul90}
will dilute hydrogen to
extremely small abundances, typically to $\nh < -5$.  Thus the evidence
to date is that a large fraction --- probably larger than 35\%
--- of DB stars do accrete hydrogen at a rate we cannot yet estimate
precisely, but which probably lies in the range from $10^{-22}$ to
$10^{-19} \ M_{\odot}$ yr$^{-1}$, and that $\sim27 \%$ of DZ stars have
hydrogen abundances that can only be accounted for by invoking
unusually low hydrogen accretion rates (this ratio for DZ stars will
eventually go up when higher S/N ratio spectroscopy around H$\alpha$
becomes available for SDSS stars).

\subsection{Speculations on Spectral Evolution}

We discuss here an alternative, and very
speculative scenario aimed at explaining various problems
related to the abundance patterns observed in helium-rich white dwarfs. 
\citet{koester76} argued that the accretion rate from the ISM should be 
lower than the fluid rate obtained from the Bondi-Hoyle equation since
the physical conditions required for the hydrodynamic treatment are
perhaps not met in a low interacting medium.  If we assume that the accretion
rate is indeed much lower than the fluid rate ($\dot M < 10^{-21}\
M_{\odot}$ yr$^{-1}$), we could naturally explain the low hydrogen
abundances of $\nh<-$6 determined for the hot DZ stars in our
study. Such a low hydrogen accretion rate would also be compatible
with the hydrogen abundance pattern observed in the hotter DB white
dwarfs. The presence of metals in DZ stars would then have to be
explained by other means than accretion from the ISM, such as cometary
material or tidal disruption of planets or asteroids, models that are
growing in popularity since the recent discoveries of infrared
excesses in DAZ white dwarfs \citep[see, e.g.,][and references
therein]{Kilic06b}.

In Figure \ref{fg:f31}, we show the results already
displayed in Figure \ref{fg:f29} in a different way by plotting as a
function of effective temperature the {\it total mass of hydrogen}
present in each star, obtained by combining the mass of the helium
convection zone at a given effective temperature with our
determinations of the H/He abundance ratios.  Also reproduced in
Figure \ref{fg:f31} are the hydrogen layer masses predicted for
continuous accretion from the ISM in solar proportions at various
rates needed to account for the
amount of hydrogen measured (or constrained) if the accretion starts
at $\Te=20,000$~K.

We speculate here that the high hydrogen abundances, $\nh=-5$ to
$-3$, observed in a fraction of our DZA stars could be explained in
terms of convectively mixed DA stars at low effective
temperatures. This process is believed to occur when the bottom of the
thin hydrogen convection zone in a hydrogen-atmosphere white dwarf
penetrates the deeper and more massive underlying helium layer. Figure
\ref{fg:f32} illustrates the extent of the hydrogen convection zone as
a function of effective temperature in a 0.6 $M_{\odot}$ DA white
dwarf. A DA star with a very thin hydrogen layer of $q({\rm
H})\equiv\Delta M/M_\star = 10^{-11}$ would thus mix at an effective
temperature near 10,000~K, while a DA star with a much thicker
hydrogen layer of $q({\rm H})=10^{-8}$ would mix at a lower
temperature of $\sim 6500$~K. The mixing process, although not yet well
understood, would presumably turn a hydrogen-atmosphere white dwarf
into a helium-dominated atmosphere with only small traces of hydrogen,
since the mass of the helium convection zone is much larger than the
hydrogen layer mass.  Although there are no detailed quantitative
calculations available in the literature, the expected H/He abundance
ratios as a function of effective temperature would follow, at least
qualitatively, the abundance pattern observed in Figures \ref{fg:f29}
and \ref{fg:f31}.

If the convective mixing scenario proposed here to explain the
hydrogen abundances observed in cool DZ stars survives closer
examination, it could also naturally explain the 
absence of a correlation between the position and motion of DZ stars
with the distribution of local interstellar matter, as discussed by
\citet{aannestad93}. The mixing scenario could also provide an
explanation for another fact that has not received much attention in
the literature: the lack of DQ white dwarfs with hydrogen
abundances comparable to those observed in DZ stars. If we combine the
40 new DQ stars analyzed by \citet{koester06} with the 56 analyzed by
\citet{dufour05} (and other DQ stars from the literature), we have a
sample of over a hundred DQ stars, none of which show H$\alpha$. We
note that the increased opacity provided by carbon affects the
detection limit of hydrogen features compared to DZA stars, especially
for DQ stars with high carbon abundances ($\che > -4$), but according
to our models, H$\alpha$ should still be easily detected in DQ stars
with $\nh \sim -4$ to $-3$ at $\Te=7000$ -- 9000~K. In fact, only one
DQ star is known to contain hydrogen, G99-37, and it is only
indirectly detected from the presence of a CH molecular band. If
accretion from the interstellar medium is responsible for the presence
of hydrogen in DZ white dwarfs, we would then expect at least a few DQ
stars to show H$\alpha$. We propose instead that DQ white dwarfs do
not show large amounts of hydrogen in their atmospheres simply because
they are the direct descendants of the DB and DBA stars, while DZ and
DZA white dwarfs originate respectively from two separate populations
of white dwarfs, namely DB stars that have not accreted a detectable
quantity of hydrogen and convectively mixed DA stars. This scenario
also predicts that the proportion of DZA stars should increase with
decreasing effective temperature. However, it is premature to test
this idea with our sample since too many stars may have small traces
of hydrogen that will become detectable only through higher S/N
ratio spectroscopy.

\section{SUMMARY AND CONCLUSIONS}\label{conclusion}

We presented a detailed photometric and spectroscopic analysis of 159
DZ white dwarfs drawn from two samples (12 stars from BRL97 and BLR01,
and 147 from the SDSS). This is more than a ninefold increase in the
number of DZ stars ever analyzed and represents the largest set
analyzed in a {\it homogeneous} fashion.

Our reanalysis of the DZ stars from the BRL97/BLR01 sample reveals
that the effective temperatures and stellar masses derived with models
including hydrogen and metals are significantly different from the
values obtained from pure helium models. For instance, our mean mass
for the 11 DZ stars with trigonometric parallax measurements, $\langle
M \rangle = 0.63$ \msun, is significantly lower than the average mass
obtained by BRL97/BLR01, $\langle M \rangle = 0.78$ \msun, and closer
to the mean mass of other types of white dwarfs. The atmospheric
parameter determinations for the coolest DZ stars in our sample are
more uncertain due to possible pressure effects in white dwarf
atmospheres below $\Te\sim6000$~K. As such, the coolest DZ stars
represent a useful probe of non-ideal effects at
high gas densities.

Hydrogen is present in 43 of the 159 DZ stars (or $27\%$) in
our complete sample, a fraction that will most certainly rise when
higher signal-to-noise follow-up spectroscopy near H$\alpha$ becomes
available for SDSS objects. Our analysis also revealed for the first
time a correlation between the hydrogen abundance and the effective
temperature of DZ stars. The amount of hydrogen measured in the
photospheric regions of DZ stars can be explained from accretion from
the ISM only if the accretion rate is at least two (and
possibly up to five) orders of magnitude lower than the corresponding rate
for metals. Also, the hydrogen accretion rate inferred from the
results of our analysis suggests an increase in the rate of about one
order of magnitude when the effective temperature decreases, providing
perhaps the first empirical evidence of a lower temperature boundary
for the hydrogen screening mechanism. We finally speculated about an
alternative scenario where the hydrogen pattern in DZA white dwarfs
could be explained as the result of the convective mixing of the thin
hydrogen layer in DA stars with the more massive underlying helium
convective zone.

\acknowledgements
We wish to thank D.~Koester for providing us with his theoretical
Ly$\alpha$ line profile calculations, A.~Gianninas for a careful
reading of our manuscript, and an anonymous referee for many
constructive comments. We would also like to thank the director and
staff of Steward Observatory for the use of their facilities. This
work was supported in part by the NSERC Canada.  P. Bergeron is a
Cottrell Scholar of Research Corporation.

Funding for the SDSS and SDSS-II has been provided by the Alfred P.
Sloan Foundation, the Participating Institutions, the National Science
Foundation, the U.S. Department of Energy, the National Aeronautics
and Space Administration, the Japanese Monbukagakusho, the Max Planck
Society, and the Higher Education Funding Council for England.
 
The SDSS is managed by the Astrophysical Research Consortium for the
Participating Institutions. The Participating Institutions are the
American Museum of Natural History, Astrophysical Institute Potsdam,
University of Basel, Cambridge University, Case Western Reserve
University, University of Chicago, Drexel University, Fermilab, the
Institute for Advanced Study, the Japan Participation Group, Johns
Hopkins University, the Joint Institute for Nuclear Astrophysics, the
Kavli Institute for Particle Astrophysics and Cosmology, the Korean
Scientist Group, the Chinese Academy of Sciences (LAMOST), Los Alamos
National Laboratory, the Max-Planck-Institute for Astronomy (MPIA),
the Max-Planck-Institute for Astrophysics (MPA), New Mexico State
University, Ohio State University, University of Pittsburgh,
University of Portsmouth, Princeton University, the United States
Naval Observatory, and the University of Washington.

\clearpage

 \clearpage
 \begin{deluxetable}{llccccc}
 \tabletypesize{\footnotesize}
 \tablecolumns{6}
 \tablewidth{0pt}
 \tablecaption{Atmospheric Parameters of DZ Stars from the BRL97 and BLR01 Samples}
 \tablehead{
 \colhead{WD} &
 \colhead{Name} &
 \colhead{$T_{\rm eff}$(K)} &
 \colhead{log $g$} &
 \colhead{$M/M_{\odot}$}&
 \colhead{log (Ca/He)} &
 \colhead{log (H/He)}}
 \startdata
0046$+$051 &vMa 2         &  6220 (240)& 8.19 (0.04)& 0.69 (0.02)& $-$10.00 (0.05)&$-$3.19 (0.18)$^{\rm a}$     \\
0552$-$041 &LP 658-2      &  4270 ( 70)& 7.80 (0.02)& 0.45 (0.01)& $-$10.92 (0.04)&$<-$5.0             \\
&&  4350 ( 60)& 7.87 (0.02)& 0.49 (0.01)& $-$10.99 (0.03)&no H                \\
0738$-$172 &L745-46A      &  7590 (220)& 8.07 (0.03)& 0.62 (0.02)& $-$10.91 (0.03)&$-$3.41 (0.03)      \\
0802$+$386 &LP 257-28     & 10980 (490)& 8.31 (0.19)& 0.78 (0.12)& $-$ 9.76 (0.09)&$<-$6.0             \\
1313$-$198 &LHS 2710      &  4520 (160)& 7.86 (0.08)& 0.48 (0.04)& $-$11.18 (0.03)&$<-$5.0             \\
&&  4570 ( 90)& 7.89 (0.08)& 0.50 (0.04)& $-$11.23 (0.04)&no H                \\
1328$+$307 &G165-7        &  6440 (210)& 7.99 (0.29)& 0.57 (0.17)& $-$ 8.10 (0.15)&$-$3.0 (0.20)       \\
1338$-$311 &ESO 445-271   &  8210 (460)& 8.00& 0.58& $-$10.03 (0.02)&$<-$4.0             \\
&&  8560 (600)& 8.00& 0.58& $-$10.43 (0.02)&no H                \\
1626$+$368 &Ross 640      &  8440 (320)& 8.02 (0.05)& 0.59 (0.03)& $-$ 8.83 (0.04)&$-$3.63 (0.05)      \\
1705$+$030 &G139-13       &  6580 (200)& 8.20 (0.15)& 0.70 (0.09)& $-$10.05 (0.05)&$-$3.55 (0.14)$^{\rm a}$     \\
2251$-$070 &LP 701-29     &  4000 (200)& 8.01 (0.06)& 0.58 (0.04)& $-$10.45 (0.02)&$<-$6.0             \\
2312$-$024 &LHS 3917      &  6220 (190)& 8.19 (0.26)& 0.69 (0.16)& $-$10.55 (0.05)&$-$4.90 (0.30)$^{\rm a}$     \\
2345$-$447 &ESO 292-43    &  4620 (110)& 8.35 (0.09)& 0.80 (0.06)& $-$11.35 (0.04)&$<-$5.0             \\
&&  4650 ( 70)& 8.36 (0.08)& 0.81 (0.06)& $-$11.43 (0.04)&no H                \\
 \enddata
\tablenotetext{a}{Hydrogen abundance determined indirectly from the 
Ca~\textsc{ii} H $\&$ K profiles.} \end{deluxetable}
 \clearpage

 \clearpage
 \begin{deluxetable}{lccrcccc}
 \tabletypesize{\footnotesize}
 \tablecolumns{8}
 \tablewidth{0pt}
 \tablecaption{Atmospheric Parameters of DZ Stars from the SDSS}
 \tablehead{
 \colhead{Name} &
 \colhead{Plate} &
 \colhead{MJD} &
 \colhead{Fiber} &
 \colhead{$T_{\rm eff}$(K)} &
 \colhead{log (Ca/He)}&
 \colhead{$D$ (pc)}&
 \colhead{log (H/He)}}
 \startdata
SDSS J000557.20$+$001833.3 &  388&51793&  394&  7970 (190)&$-$9.62 (0.07)& 136& $<-$4.0             \\
SDSS J001849.43$+$001204.7 & 1491&52996&   27&  9600 (170)&$-$9.83 (0.07)& 157& $<-$5.0             \\
SDSS J003601.38$-$111214.0 &  655&52162&  300&  7280 ( 70)&$-$9.26 (0.04)&  55& $<-$4.0             \\
SDSS J004123.58$+$151109.0 &  419&51868&  395&  7770 (170)&$-$10.72 (0.16)& 177& $<-$4.0             \\
SDSS J004646.16$+$002430.9 &  691&52199&  503&  8770 (220)&$-$8.57 (0.09)& 225& $<-$4.0             \\
 \\
SDSS J005906.77$+$001725.2 &  395&51783&  508& 10400 (420)&$-$9.80 (0.19)& 248& $<-$5.0             \\
SDSS J010629.85$-$010344.2 &  396&51816&   55& 10240 (350)&$-$9.21 (0.09)& 192& $-$4.30 (0.13)      \\
SDSS J011338.35$+$000632.7 &  694&52209&   24&  7840 (160)&$-$9.79 (0.09)& 179& $<-$4.0             \\
SDSS J011358.98$-$095913.3 &  660&52177&  277& 10610 (290)&$-$8.87 (0.05)& 130& $-$5.06 (0.06)      \\
SDSS J012339.77$+$132433.4 &  424&51893&  132&  7300 (270)&$-$9.44 (0.22)& 223& $<-$3.0             \\
 \\
SDSS J013831.12$+$003101.6 &  698&52203&  500&  9060 (270)&$-$10.34 (0.11)& 225& $<-$5.0             \\
SDSS J020001.99$+$004018.4 &  701&52179&  430&  9860 (170)&$-$9.90 (0.09)& 173& $<-$5.0             \\
SDSS J020132.24$-$003932.0 &  404&51812&  303&  9650 (210)&$-$8.33 (0.05)& 110& $<-$5.0             \\
SDSS J021836.70$-$091945.0 &  668&52162&  138&  9560 (300)&$-$10.63 (0.07)&  97& $<-$5.0             \\
SDSS J022851.97$-$000938.8 &  406&51869&   31&  8310 (310)&$-$8.08 (0.22)& 279& $<-$3.0             \\
 \\
SDSS J030800.40$-$065659.9 &  459&51924&  345&  8520 (400)&$-$8.81 (0.23)& 285& $<-$3.0             \\
SDSS J031448.24$-$082755.2 &  459&51924&   46& 10290 (610)&$-$9.51 (0.31)& 358& $<-$5.0             \\
SDSS J041145.90$-$054848.5 &  465&51910&   52&  9320 (580)&$-$8.95 (0.32)& 359& $<-$3.0             \\
SDSS J073835.97$+$384438.2 &  431&51877&  601&  9530 (360)&$-$8.36 (0.13)& 271& $-$4.42 (0.35)      \\
SDSS J074743.56$+$400110.1 &  432&51884&   36&  8520 (200)&$-$10.87 (0.17)& 161& $<-$4.0             \\
 \\
SDSS J074751.39$+$373217.0 &  433&51873&   13&  9600 (390)&$-$10.00 (0.14)& 269& $<-$5.0             \\
SDSS J074821.86$+$350648.7 &  542&51993&  515& 10130 (430)&$-$10.43 (0.18)& 356& $<-$5.0             \\
SDSS J074942.87$+$312424.6 &  890&52583&  268&  6910 ( 80)&$-$9.86 (0.13)& 102& $-$3.14 (0.23)$^{\rm a}$     \\
SDSS J074958.26$+$434306.0 &  434&51885&  401&  8950 (470)&$-$9.62 (0.18)& 350& $<-$5.0             \\
SDSS J075846.90$+$322523.3 &  890&52583&  588& 10090 (460)&$-$10.24 (0.20)& 336& $<-$5.0             \\
 \\
SDSS J080211.42$+$301256.7 &  860&52319&  388&  9400 (470)&$-$9.17 (0.19)& 385& $<-$4.0             \\
SDSS J080331.46$+$450257.8 &  439&51877&  351&  9290 (320)&$-$9.53 (0.23)& 276& $-$3.85 (0.28)      \\
SDSS J080537.64$+$383212.4 &  758&52253&  383& 10660 ( 40)&$-$10.03 (0.03)&  49& $<-$6.0             \\
SDSS J080602.91$+$374720.6 &  758&52253&  122&  9430 (280)&$-$8.38 (0.14)& 244& $<-$5.0             \\
SDSS J080615.25$+$364018.1 &  758&52253&   80& 10130 (570)&$-$9.23 (0.24)& 389& $<-$5.0             \\
 \\
 \\
SDSS J082720.59$+$330437.9 &  932&52620&  372&  7870 (210)&$-$8.97 (0.12)& 221& $<-$5.0             \\
SDSS J082927.85$+$075911.4 & 1758&53084&  154&  9670 (180)&$-$8.68 (0.07)& 176& $<-$5.0             \\
SDSS J083434.68$+$464130.6 &  549&51981&  560&  6610 (100)&$-$9.85 (0.06)& 110& $<-$5.0             \\
SDSS J083556.31$+$090619.4 & 1759&53081&  338&  8390 (150)&$-$9.35 (0.09)& 165& $<-$4.0             \\
SDSS J084200.24$+$362540.0 &  864&52320&  101& 10320 (240)&$-$9.08 (0.08)& 218& $-$4.78 (0.13)      \\
 \\
SDSS J084502.70$+$411547.6 &  829&52296&  367&  8300 (220)&$-$8.55 (0.11)& 218& $<-$4.0             \\
SDSS J084525.00$+$535208.8 &  446&51899&  176&  7180 (160)&$-$10.71 (0.13)& 144& $<-$3.0             \\
SDSS J084709.11$+$450734.8 &  763&52235&  146&  8520 (350)&$-$9.47 (0.18)& 267& $<-$4.0             \\
SDSS J084828.00$+$521422.5 &  447&51877&  499&  8330 (190)&$-$10.83 (0.05)& 110& $<-$5.0             \\
SDSS J084849.42$+$354857.8 &  934&52672&  553&  8090 ( 80)&$-$10.46 (0.04)& 105& $<-$4.0             \\
 \\
SDSS J084857.88$+$002834.9 &  467&51901&  150& 11950 (320)&$-$8.25 (0.13)& 221& $<-$6.0             \\
SDSS J084906.69$+$071030.0 & 1298&52964&   29&  8320 (260)&$-$8.37 (0.14)& 237& $-$3.36 (0.25)      \\
SDSS J084911.86$+$403649.7 &  830&52293&  253&  9230 (140)&$-$10.35 (0.07)& 134& $<-$5.0             \\
SDSS J085141.72$+$053852.1 & 1189&52668&  139&  7480 (240)&$-$9.03 (0.14)& 212& $<-$4.0             \\
SDSS J090517.71$+$013307.6 &  471&51924&  221&  8180 (190)&$-$10.21 (0.11)& 193& $<-$4.0             \\
 \\
SDSS J090556.22$+$523533.1 &  553&51999&  380&  7510 (180)&$-$9.15 (0.09)& 151& $<-$4.0             \\
SDSS J091643.07$+$010531.5 &  472&51955&  577&  7500 (340)&$-$9.77 (0.24)& 240& $<-$3.0             \\
SDSS J092801.78$+$612434.1 &  485&51909&  527&  8570 (340)&$-$8.71 (0.21)& 288& $<-$4.0             \\
SDSS J093210.54$+$485601.7 &  901&52641&  259&  8210 (250)&$-$8.71 (0.19)& 262& $<-$4.0             \\
SDSS J093423.17$+$082225.3 & 1304&52993&  222&  9130 (110)&$-$9.82 (0.03)& 117& $<-$5.0             \\
 \\
SDSS J093545.45$+$003750.9 &  476&52314&  417&  9030 (470)&$-$8.67 (0.22)& 369& $<-$4.0             \\
SDSS J093704.99$+$364647.2 & 1275&52996&  114& 11190 (430)&$-$9.94 (0.26)& 263& $-$4.89 (0.18)      \\
SDSS J093942.30$+$555048.7 &  556&51991&  167&  8680 ( 80)&$-$8.51 (0.02)&  65& $-$4.20 (0.05)      \\
SDSS J094148.75$+$502214.5 &  901&52641&  590& 10940 (420)&$-$9.12 (0.17)& 294& $<-$5.0             \\
SDSS J094206.21$+$575556.0 &  452&51911&   32& 11430 (400)&$-$8.54 (0.23)& 312& $<-$6.0             \\
 \\
SDSS J094210.50$+$074354.7 & 1234&52724&  220&  9140 (190)&$-$9.32 (0.09)& 190& $<-$5.0             \\
SDSS J094415.33$+$393943.0 & 1215&52725&  576& 11190 (560)&$-$8.97 (0.14)& 314& $<-$6.0             \\
SDSS J094451.59$+$440856.7 &  941&52709&  326& 10460 (470)&$-$9.35 (0.22)& 288& $<-$5.0             \\
SDSS J094530.20$+$084624.8 & 1234&52724&  514& 10550 (220)&$-$10.03 (0.15)& 227& $<-$6.0             \\
SDSS J094743.09$+$423841.3 &  941&52709&  131&  9680 (320)&$-$8.50 (0.14)& 270& $<-$4.0             \\
 \\
 \\
SDSS J095119.85$+$403322.4 & 1216&52709&  141&  8370 ( 60)&$-$10.27 (0.04)&  84& $-$4.16 (0.13)      \\
SDSS J095435.86$+$563518.2 &  557&52253&  105&  8940 (260)&$-$9.79 (0.13)& 219& $<-$5.0             \\
SDSS J095645.15$+$591240.6 &  453&51915&  621&  8230 ( 80)&$-$7.54 (0.06)& 128& $-$3.20 (0.09)      \\
SDSS J100237.36$+$031325.5 &  573&52325&  291&  9800 (200)&$-$9.19 (0.12)& 212& $<-$5.0             \\
SDSS J100406.64$+$423151.3 & 1217&52672&  577&  8270 (340)&$-$9.98 (0.14)& 242& $<-$4.0             \\
 \\
SDSS J100421.25$+$045117.2 &  573&52325&  363&  7910 (180)&$-$9.21 (0.13)& 193& $<-$4.0             \\
SDSS J100823.75$+$024840.0 &  502&51957&  325&  7940 (220)&$-$11.26 (0.12)& 120& $<-$4.0             \\
SDSS J101007.84$+$394852.2 & 1356&53033&  640&  8470 ( 90)&$-$8.66 (0.05)& 128& $-$3.94 (0.16)      \\
SDSS J101558.21$+$414131.7 & 1357&53034&  523&  9390 (200)&$-$8.90 (0.10)& 206& $-$4.66 (0.25)      \\
SDSS J101805.17$+$034435.6 &  574&52355&  104&  8690 (370)&$-$8.62 (0.17)& 275& $<-$4.0             \\
 \\
SDSS J103126.19$+$120340.4 & 1599&53089&  455&  8350 ( 70)&$-$8.62 (0.04)& 102& $<-$4.0             \\
SDSS J103300.11$+$624747.8 &  772&52375&  542&  7590 (120)&$-$9.92 (0.45)& 137& $-$3.71 (0.89)$^{\rm a}$     \\
SDSS J103651.09$+$483754.0 &  875&52354&  122& 11460 (330)&$-$9.02 (0.20)& 292& $-$5.21 (0.26)      \\
SDSS J103809.19$-$003622.5 &  274&51913&  265&  6770 ( 30)&$-$9.44 (0.06)&  49& $<-$5.0             \\
SDSS J103941.86$+$461224.3 &  962&52620&  132&  6770 (120)&$-$10.05 (0.17)& 109& $-$5.62 (0.55)$^{\rm a}$     \\
 \\
SDSS J104511.21$+$625442.2 &  773&52376&  554&  9020 (470)&$-$9.77 (0.32)& 340& $<-$4.0             \\
SDSS J104911.53$+$515423.5 & 1010&52649&   84&  6680 ( 40)&$-$9.43 (0.05)&  72& $-$4.41 (0.24)$^{\rm a}$     \\
SDSS J104915.06$-$000706.2 &  275&51910&  111&  8680 (160)&$-$8.53 (0.05)& 171& $-$3.91 (0.14)      \\
SDSS J105221.56$+$065915.4 & 1001&52670&  557& 11000 (440)&$-$9.60 (0.32)& 317& $<-$5.0             \\
SDSS J105601.50$+$012825.0 &  507&52353&   14& 10200 (240)&$-$9.24 (0.06)& 122& $-$4.76 (0.08)      \\
 \\
SDSS J105616.90$-$000449.4 &  277&51908&  311& 10950 (460)&$-$9.17 (0.25)& 355& $-$4.63 (0.32)      \\
SDSS J105641.71$+$571448.9 &  949&52427&   99&  7420 ( 70)&$-$10.70 (0.08)&  99& $<-$4.0             \\
SDSS J105853.70$+$604136.8 &  774&52286&  262&  8870 (460)&$-$8.96 (0.32)& 316& $<-$4.0             \\
SDSS J110438.38$+$071129.8 & 1003&52641&  172& 10910 (160)&$-$10.45 (0.15)& 169& $<-$6.0             \\
SDSS J111215.05$+$070052.3 & 1004&52723&  111&  6890 (120)&$-$9.02 (0.08)& 142& $<-$3.0             \\
 \\
SDSS J112258.34$+$504146.7 &  879&52365&  289&  8740 (200)&$-$7.93 (0.12)& 217& $-$4.40 (0.34)      \\
SDSS J112617.16$+$524155.1 &  879&52365&  407&  9460 (150)&$-$9.50 (0.06)& 159& $-$4.86 (0.16)      \\
SDSS J112956.98$-$015229.5 &  326&52375&  590&  8480 (190)&$-$9.04 (0.10)& 192& $-$3.33 (0.17)      \\
SDSS J113001.61$+$103614.8 & 1223&52781&  570&  8620 (170)&$-$8.59 (0.08)& 168& $<-$4.0             \\
SDSS J113711.28$+$034324.7 &  838&52378&  292&  7120 ( 60)&$-$10.91 (0.08)&  73& $<-$3.0             \\
 \\
 \\
SDSS J114054.87$+$532827.4 & 1015&52734&   95&  9860 (250)&$-$8.54 (0.12)& 232& $-$4.51 (0.23)      \\
SDSS J114712.18$+$492801.0 &  967&52636&   28&  7810 ( 80)&$-$9.70 (0.05)& 111& $<-$4.0             \\
SDSS J115547.42$+$432751.4 & 1447&53120&  538&  8390 (120)&$-$10.46 (0.07)& 125& $-$4.00 (0.00)$^{\rm a}$     \\
SDSS J121218.69$+$540938.7 & 1019&52707&  293&  8710 (140)&$-$9.41 (0.08)& 142& $<-$5.0             \\
SDSS J121837.12$+$002304.0 &  288&52000&  423&  6090 (100)&$-$9.92 (0.10)& 122& $<-$4.0             \\
 \\
SDSS J122204.48$+$634354.5 &  780&52370&  367&  8790 (340)&$-$10.11 (0.17)& 247& $<-$4.0             \\
SDSS J122733.45$+$633029.5 &  780&52370&  443&  7270 ( 60)&$-$8.71 (0.04)&  95& $<-$4.0             \\
SDSS J122929.04$+$425414.4 & 1452&53112&  541& 11090 (690)&$-$9.48 (0.33)& 429& $<-$5.0             \\
SDSS J122953.17$+$512925.2 &  884&52374&   94&  9230 (250)&$-$9.60 (0.16)& 218& $<-$4.0             \\
SDSS J123455.96$-$033047.1 &  335&52000&  247&  8860 (180)&$-$10.83 (0.06)& 114& $<-$5.0             \\
 \\
SDSS J124006.36$-$003700.9 &  290&51941&   24&  8370 (170)&$-$9.23 (0.09)& 118& $<-$5.0             \\
SDSS J125413.79$-$023608.5 &  337&51997&  153&  6960 (120)&$-$10.45 (0.19)&  89& $-$3.34 (0.50)$^{\rm a}$     \\
SDSS J130746.34$+$030742.0 &  524&52027&  589&  8170 (180)&$-$9.64 (0.06)& 130& $-$3.55 (0.10)      \\
SDSS J130905.26$+$491359.7 & 1281&52753&  439&  8620 ( 60)&$-$10.16 (0.03)&  83& $<-$5.0             \\
SDSS J131336.96$+$573800.5 & 1319&52791&  409&  8900 ( 50)&$-$9.39 (0.02)&  68& $-$4.70 (0.09)      \\
 \\
SDSS J132506.96$+$652132.2 &  603&52056&  401& 11650 (370)&$-$8.97 (0.19)& 262& $-$4.41 (0.21)      \\
SDSS J133824.97$-$013017.1 &  912&52427&  457& 10730 (120)&$-$9.44 (0.06)& 147& $<-$6.0             \\
SDSS J134144.10$-$011238.4 &  299&51671&  137&  8090 (220)&$-$8.73 (0.11)& 222& $<-$4.0             \\
SDSS J134226.93$+$052248.6 &  854&52373&  371&  7960 (110)&$-$8.97 (0.10)& 146& $<-$4.0             \\
SDSS J134459.35$+$650512.5 &  497&51989&  124& 10010 (450)&$-$9.65 (0.20)& 320& $<-$5.0             \\
 \\
SDSS J135118.47$+$425316.0 & 1345&52814&   32&  6770 (120)&$-$11.35 (0.08)&  45& $<-$3.0             \\
SDSS J135137.07$+$613607.0 &  786&52319&  571&  7190 (150)&$-$10.26 (0.14)& 166& $<-$3.0             \\
SDSS J135637.78$+$404703.4 & 1378&53061&  444&  7100 ( 70)&$-$8.89 (0.05)& 125& $<-$3.0             \\
SDSS J140316.91$-$002450.0 &  301&51942&   30&  4660 (110)&$-$10.92 (0.08)&  67& $<-$4.0             \\
SDSS J140445.10$-$023237.2 &  915&52443&  114&  7620 (150)&$-$9.36 (0.11)& 143& $<-$4.0             \\
 \\
SDSS J141426.55$-$011354.7 &  303&51615&  124&  8940 (200)&$-$10.75 (0.06)& 109& $<-$5.0             \\
SDSS J142516.43$-$005048.7 &  305&51613&  197&  7220 ( 90)&$-$11.42 (0.11)&  81& $<-$4.0             \\
SDSS J142931.17$+$583927.9 &  789&52342&   93&  9220 (310)&$-$8.52 (0.15)& 235& $<-$4.0             \\
SDSS J143235.82$+$035423.3 &  585&52027&  143& 10760 (430)&$-$8.91 (0.13)& 233& $-$4.35 (0.18)      \\
SDSS J144022.52$-$023222.2 &  919&52409&   36&  6860 ( 80)&$-$10.44 (0.20)& 100& $-$3.25 (0.50)$^{\rm a}$     \\
 \\
 \\
SDSS J144453.67$+$574147.2 &  791&52435&  419&  9120 (400)&$-$7.44 (0.36)& 360& $<-$4.0             \\
SDSS J144516.24$-$020849.6 &  920&52411&  197&  6530 ( 80)&$-$10.71 (0.17)&  94& $-$3.53 (0.58)$^{\rm a}$     \\
SDSS J151054.84$+$381450.0 & 1399&53172&  198&  9130 (240)&$-$8.55 (0.13)& 229& $-$4.17 (0.26)      \\
SDSS J151441.87$+$501209.7 & 1330&52822&  533&  9930 (510)&$-$9.40 (0.36)& 377& $<-$4.0             \\
SDSS J153129.26$+$424015.7 & 1052&52466&  291&  7130 ( 80)&$-$10.36 (0.04)&  88& $<-$5.0             \\
 \\
SDSS J154025.18$+$514921.6 &  796&52401&  397&  8070 (250)&$-$9.45 (0.17)& 238& $<-$4.0             \\
SDSS J154540.22$+$492145.3 &  796&52401&   77&  9130 (510)&$-$8.79 (0.35)& 334& $<-$4.0             \\
SDSS J154623.07$+$392723.6 & 1680&53171&  588&  8680 (220)&$-$8.92 (0.10)& 192& $<-$5.0             \\
SDSS J154953.26$+$023929.7 &  594&52045&   65&  7990 (380)&$-$9.07 (0.31)& 286& $<-$3.0             \\
SDSS J155852.59$+$031242.9 &  595&52023&  484&  7170 (110)&$-$9.30 (0.08)& 136& $<-$4.0             \\
 \\
SDSS J161801.34$+$445220.8 &  814&52443&  152&  9560 (290)&$-$7.98 (0.11)& 226& $<-$5.0             \\
SDSS J164026.65$+$315453.9 & 1340&52781&  484&  6840 (100)&$-$9.89 (0.08)& 110& $<-$3.0             \\
SDSS J165741.59$+$373824.2 &  820&52433&  443&  7650 (160)&$-$9.79 (0.11)& 182& $<-$4.0             \\
SDSS J165741.59$+$373824.2 &  820&52438&  448&  7660 (160)&$-$9.87 (0.10)& 182& $<-$4.0             \\
SDSS J171139.82$+$220152.9 & 1689&53177&   33&  9160 (350)&$-$8.93 (0.16)& 286& $<-$4.0             \\
 \\
SDSS J171436.43$+$550112.4 &  367&51997&  304&  9190 (510)&$-$9.52 (0.23)& 377& $<-$4.0             \\
SDSS J171513.64$+$280403.8 &  980&52431&  351&  8240 (150)&$-$10.83 (0.11)& 139& $<-$4.0             \\
SDSS J171900.63$+$562350.5 &  367&51997&  402&  9300 (210)&$-$9.29 (0.05)& 192& $<-$5.0             \\
SDSS J205059.11$-$011021.9 & 1115&52914&  281&  9560 (290)&$-$7.59 (0.15)& 291& $<-$5.0             \\
SDSS J210303.47$-$010842.5 & 1114&53179&    4&  8140 (170)&$-$8.88 (0.16)& 182& $<-$4.0             \\
 \\
SDSS J210733.93$+$005557.7 &  985&52431&  100&  7510 (100)&$-$10.19 (0.10)& 124& $-$3.10 (0.18)      \\
SDSS J212424.69$-$011452.5 &  987&52523&   74&  7690 (110)&$-$9.18 (0.08)& 125& $<-$4.0             \\
SDSS J222802.05$+$120733.3 &  737&52518&    9&  6760 ( 20)&$-$9.96 (0.02)&  34& $<-$4.0             \\
SDSS J223222.32$+$010920.7 &  376&52143&  565&  6900 (140)&$-$10.46 (0.14)& 133& $<-$3.0             \\
SDSS J223841.05$+$010150.3 &  674&52201&  531&  7140 (140)&$-$10.30 (0.25)& 132& $-$4.97 (0.79)$^{\rm a}$     \\
 \\
SDSS J234245.90$+$000632.4 &  682&52525&   27&  8570 (410)&$-$9.15 (0.15)& 316& $-$3.20 (0.18)      \\
SDSS J235516.61$+$143136.1 &  749&52226&  119&  6940 (110)&$-$10.80 (0.25)& 129& $-$3.67 (0.89)$^{\rm a}$     \\
 \enddata
\tablenotetext{a}{Hydrogen abundance determined indirectly from the 
Ca~\textsc{ii} H $\&$ K profiles.}
 \end{deluxetable}
 \clearpage

\clearpage

\figcaption[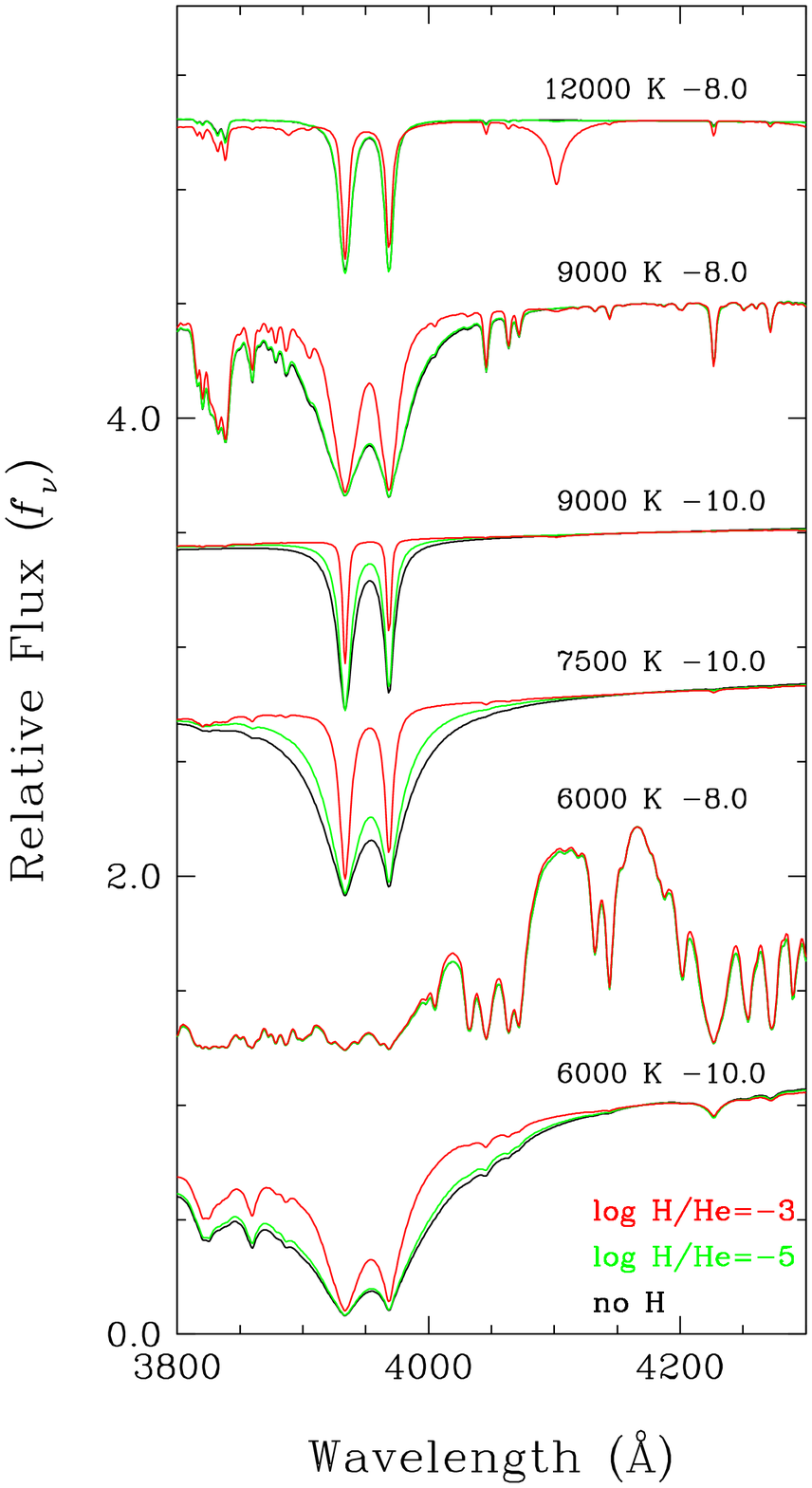] {Representative synthetic spectra of DZ white dwarfs 
taken from our model grid at $\logg = 8.0$ for various metal and
hydrogen abundances. The spectra are normalized to unity at 4170 \AA\
and offset by an arbitrary factor for clarity. The labels indicate the
effective temperature and the calcium abundance, $\log({\rm Ca/He})$,
while the color of each spectrum corresponds to a different hydrogen
abundance, as indicated in the lower right corner.\label{fg:f1}}

\figcaption[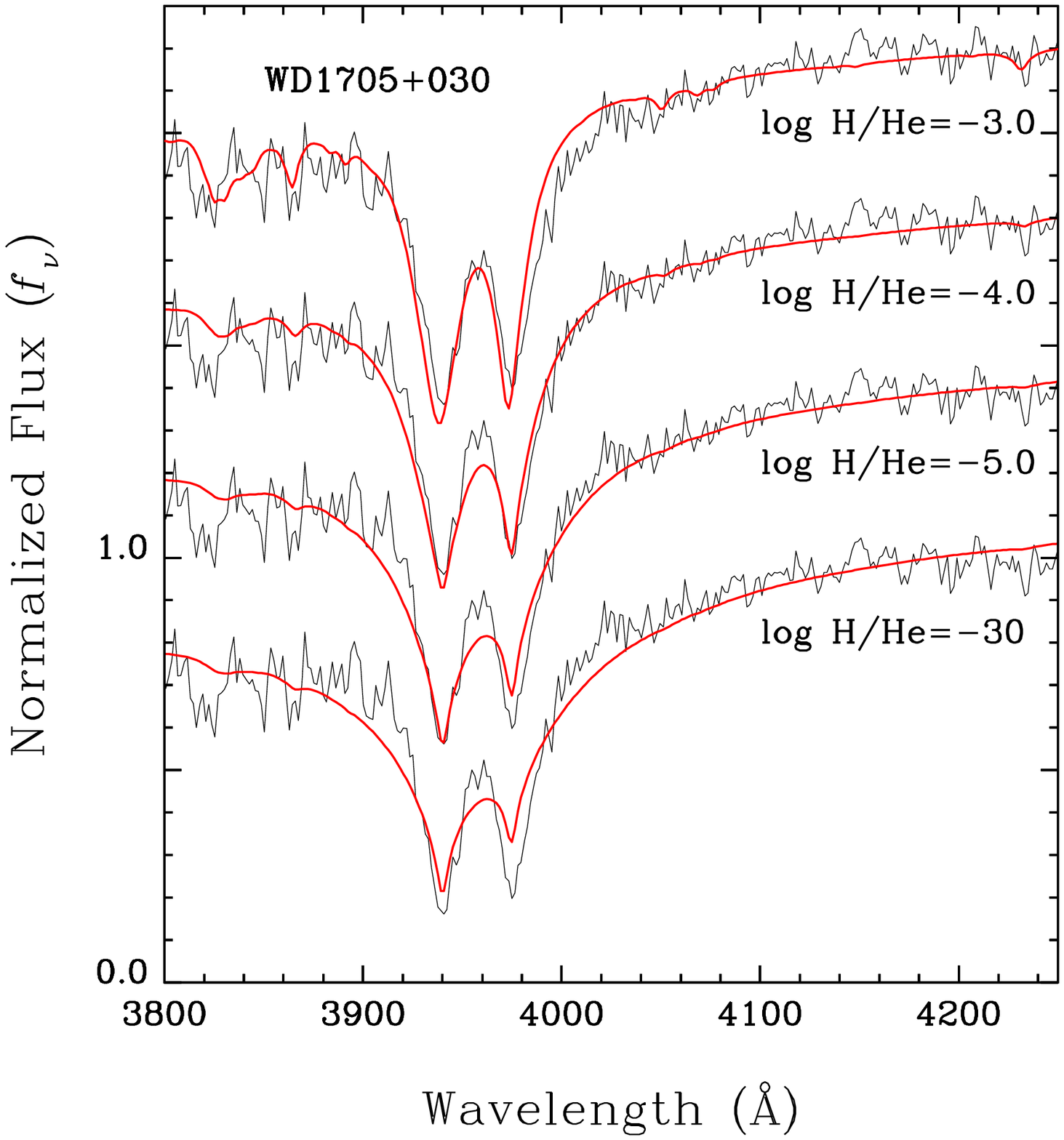] {Fits to the Ca~\textsc{ii} H \& K lines with 
various hydrogen abundances for the DZ star G139-13 (WD~1705+030).
[{\it See the electronic version of the Journal for a color version
of this figure.}]\label{fg:f2}}

\figcaption[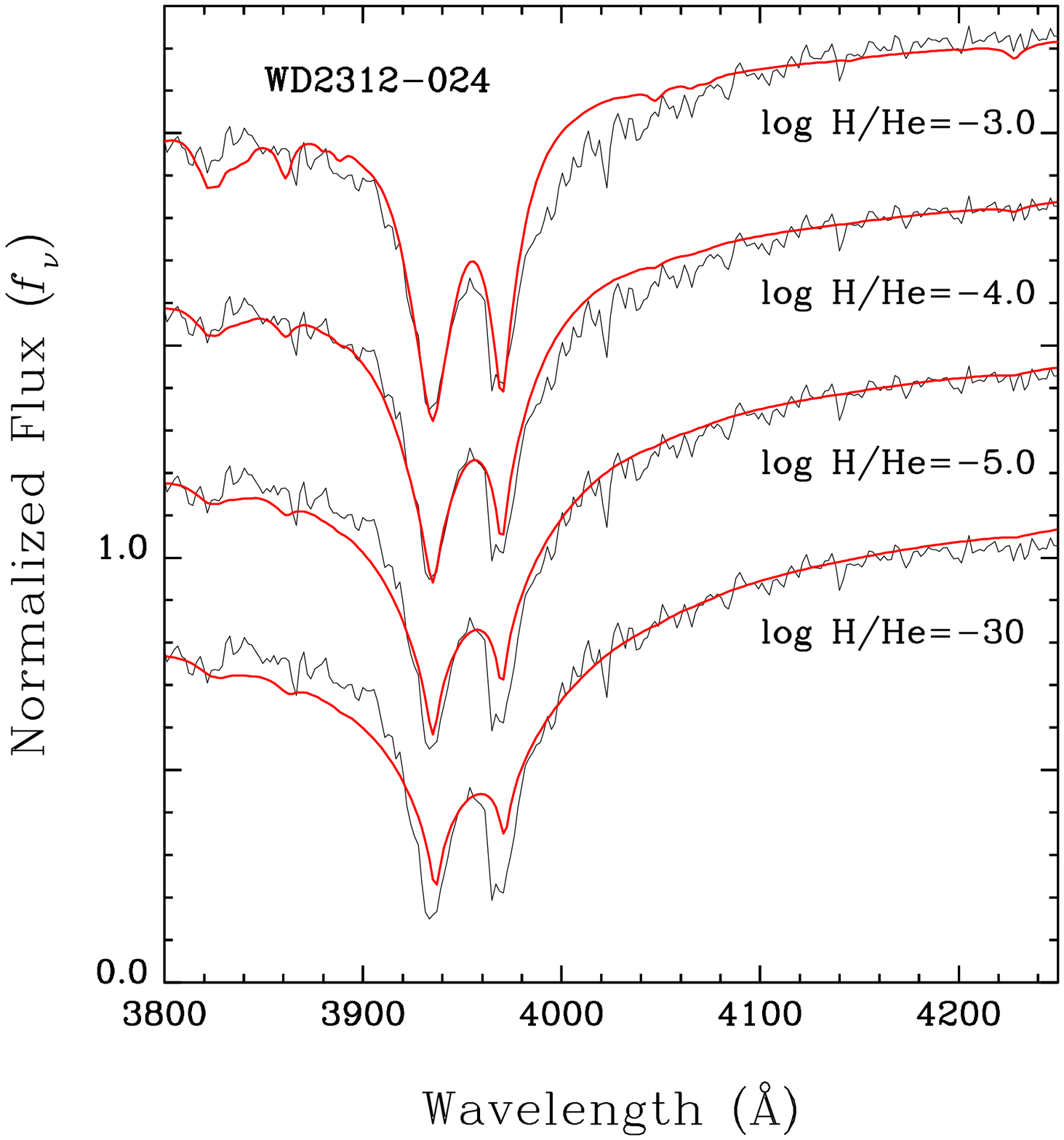] {Fits to the Ca~\textsc{ii} H \& K lines with 
various hydrogen abundances for the DZ star G157-35
(WD~2312$-$024). [{\it See the electronic version of the Journal for a color version
of this figure.}]\label{fg:f3}}

\figcaption[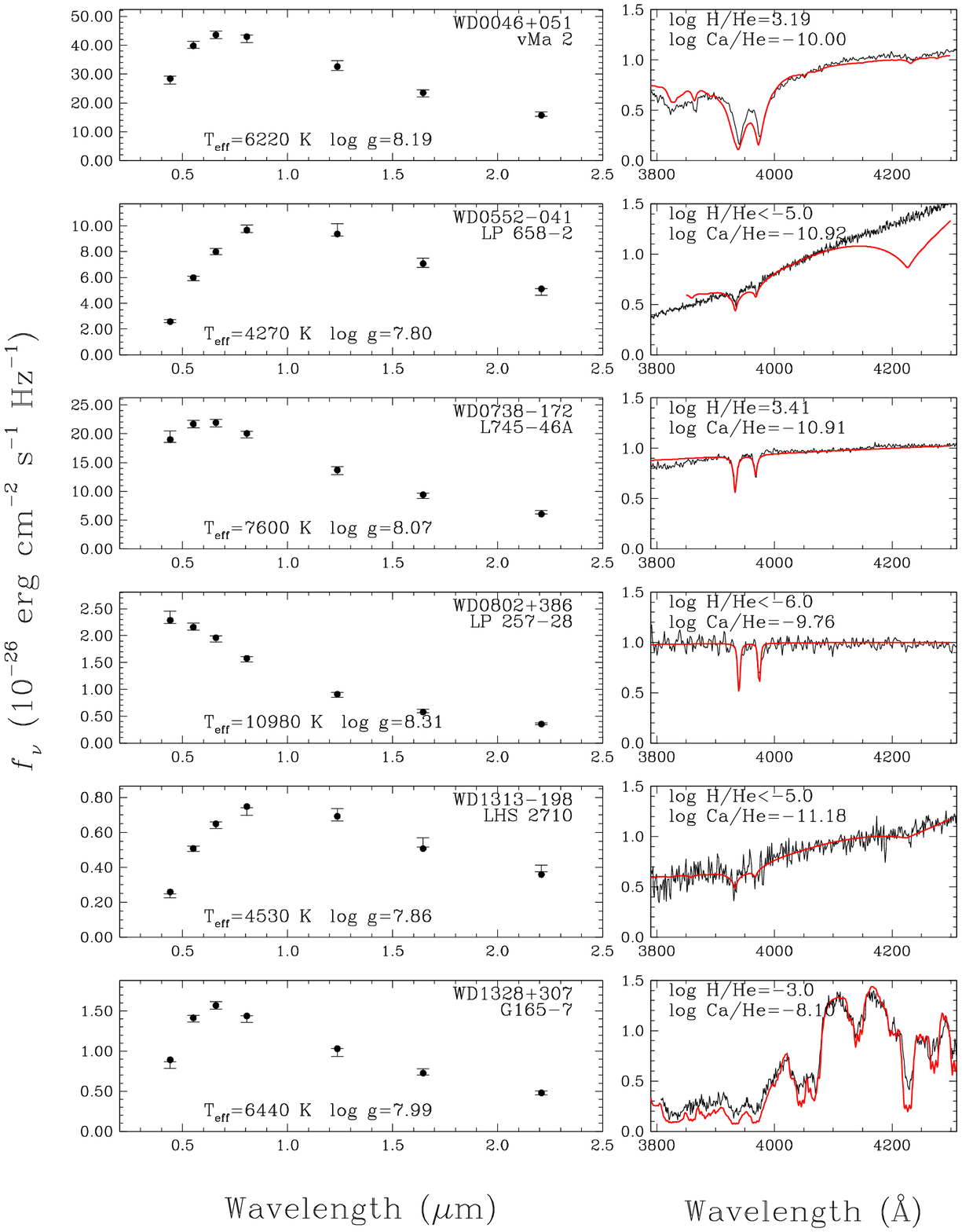] {Fits to the energy distribution ({\it left panels}) 
and calcium lines ({\it right panels}) with helium-rich models
including traces of metals and hydrogen (the abundances are indicated
in each panel). The $BVRI$ and $JHK$ photometric observations are
represented by error bars, while the average model fluxes are shown by
filled circles. [{\it See the electronic version of the Journal for a color version
of this figure.}]\label{fg:f4}}

\figcaption[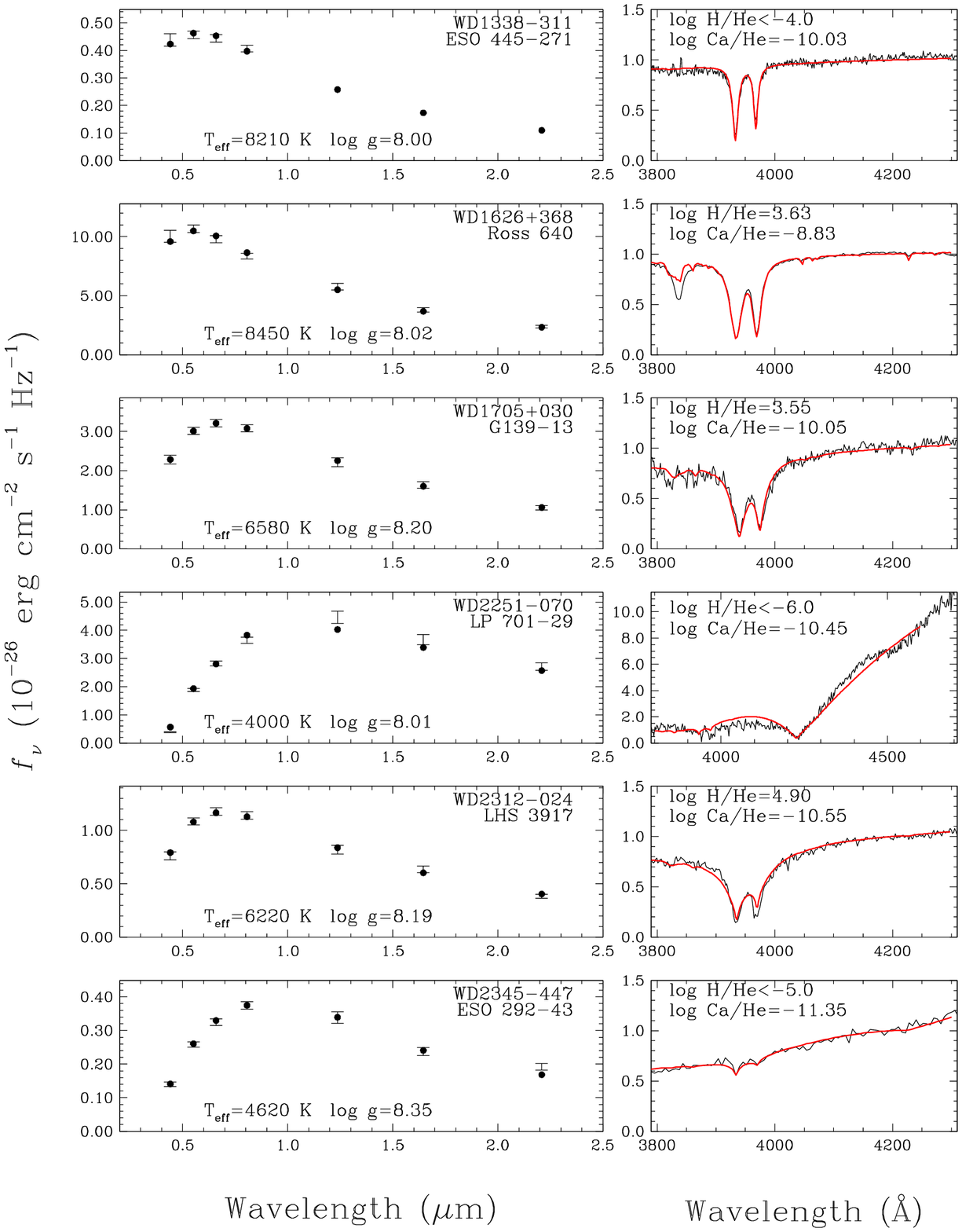] {Same as Fig.~\ref{fg:f4}. \label{fg:f5}}

\figcaption[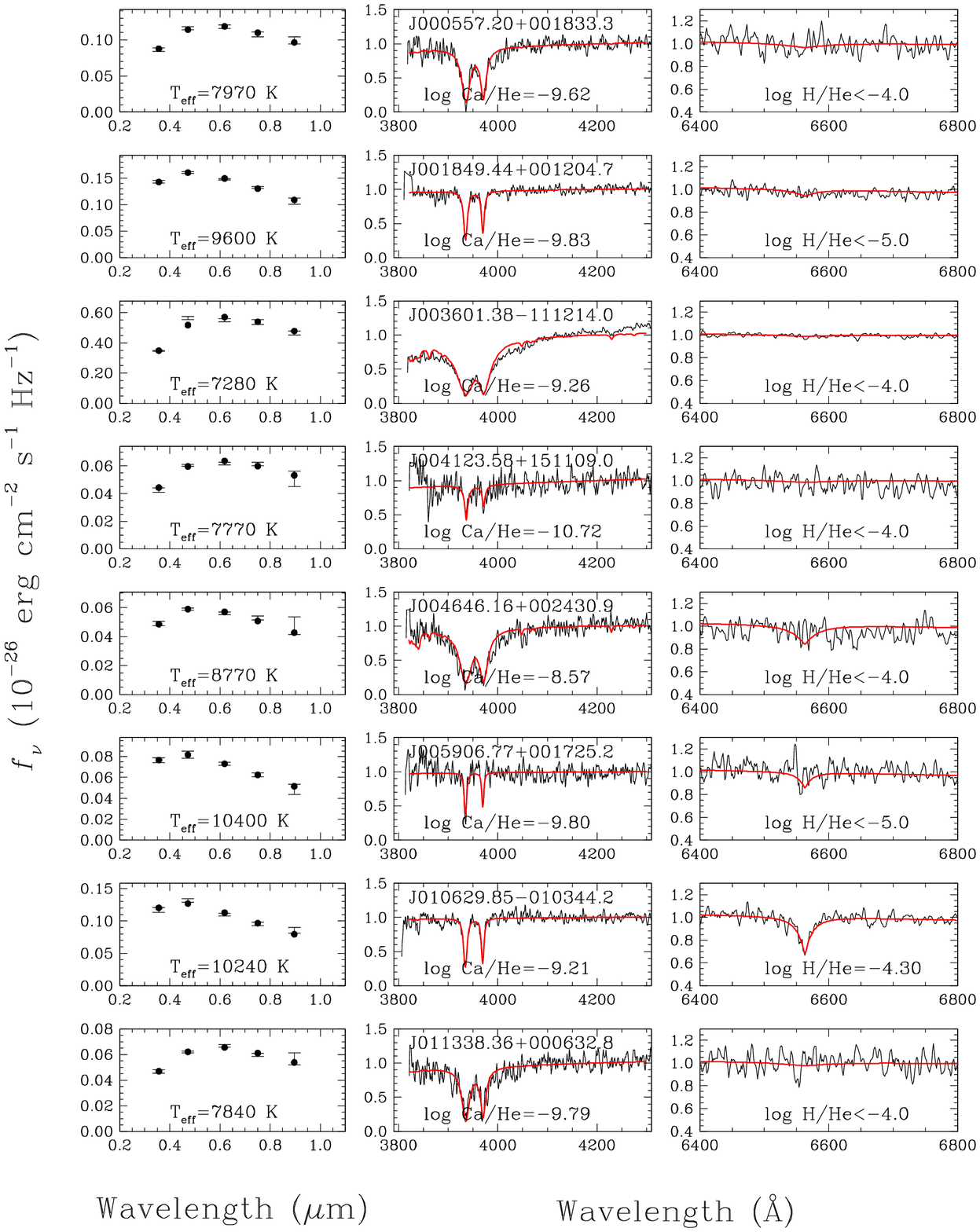] {Fits to the energy distribution, calcium lines, 
and H$\alpha$ profile ({\it from left to right}) for all DZ stars in
our SDSS sample. The $ugriz$ photometric observations are represented
by error bars, while the average model fluxes are shown by filled
circles. The atmospheric parameters are indicated in each panel. Since
some of the spectroscopic observations have low signal-to-noise
ratios, we have applied for clarity a three-point average window to
the data displayed here. [{\it See the electronic version of the
Journal for a color version of this figure.}]\label{fg:f6}}

\figcaption[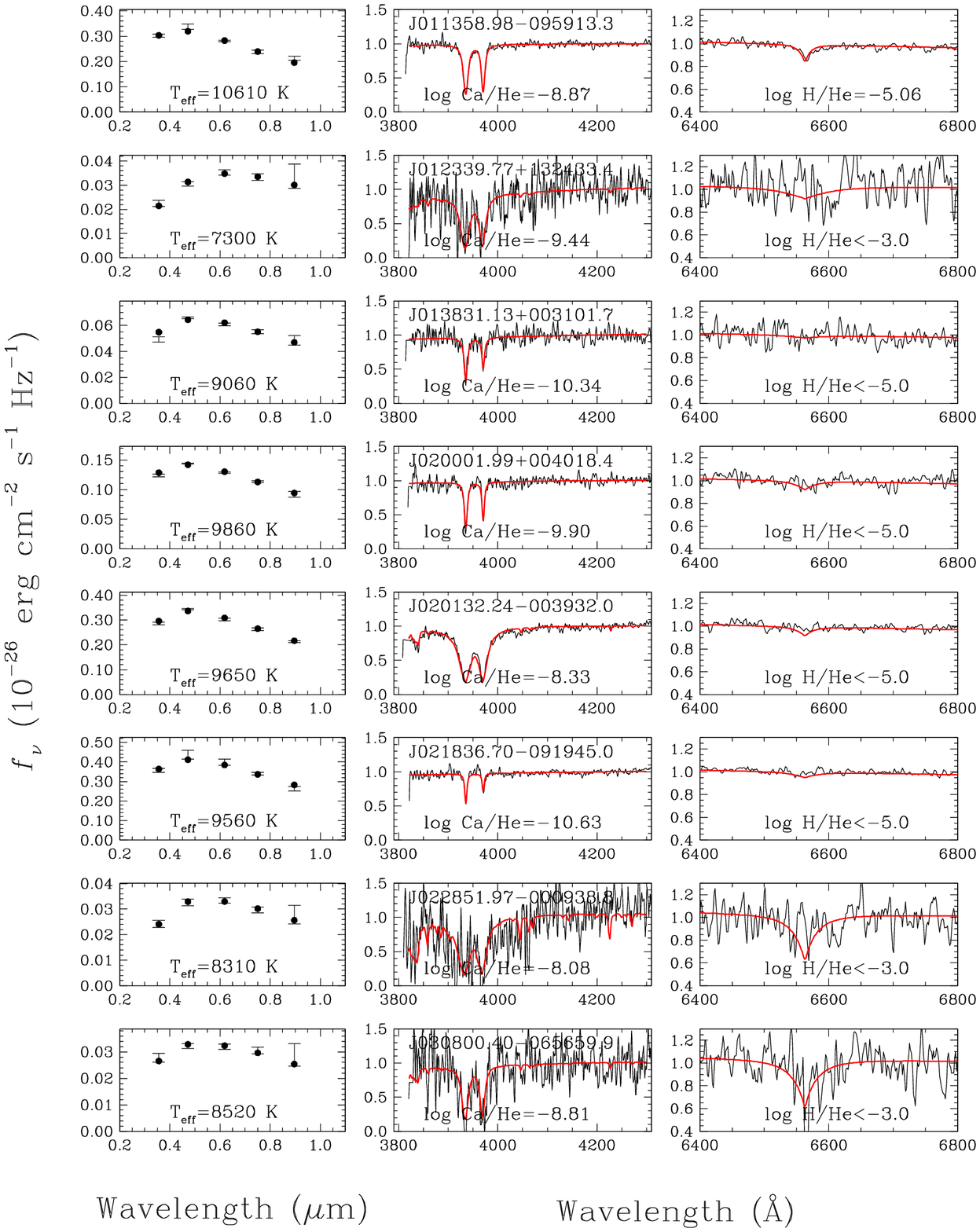] {Same as Fig.~\ref{fg:f6}. \label{fg:f7}}

\figcaption[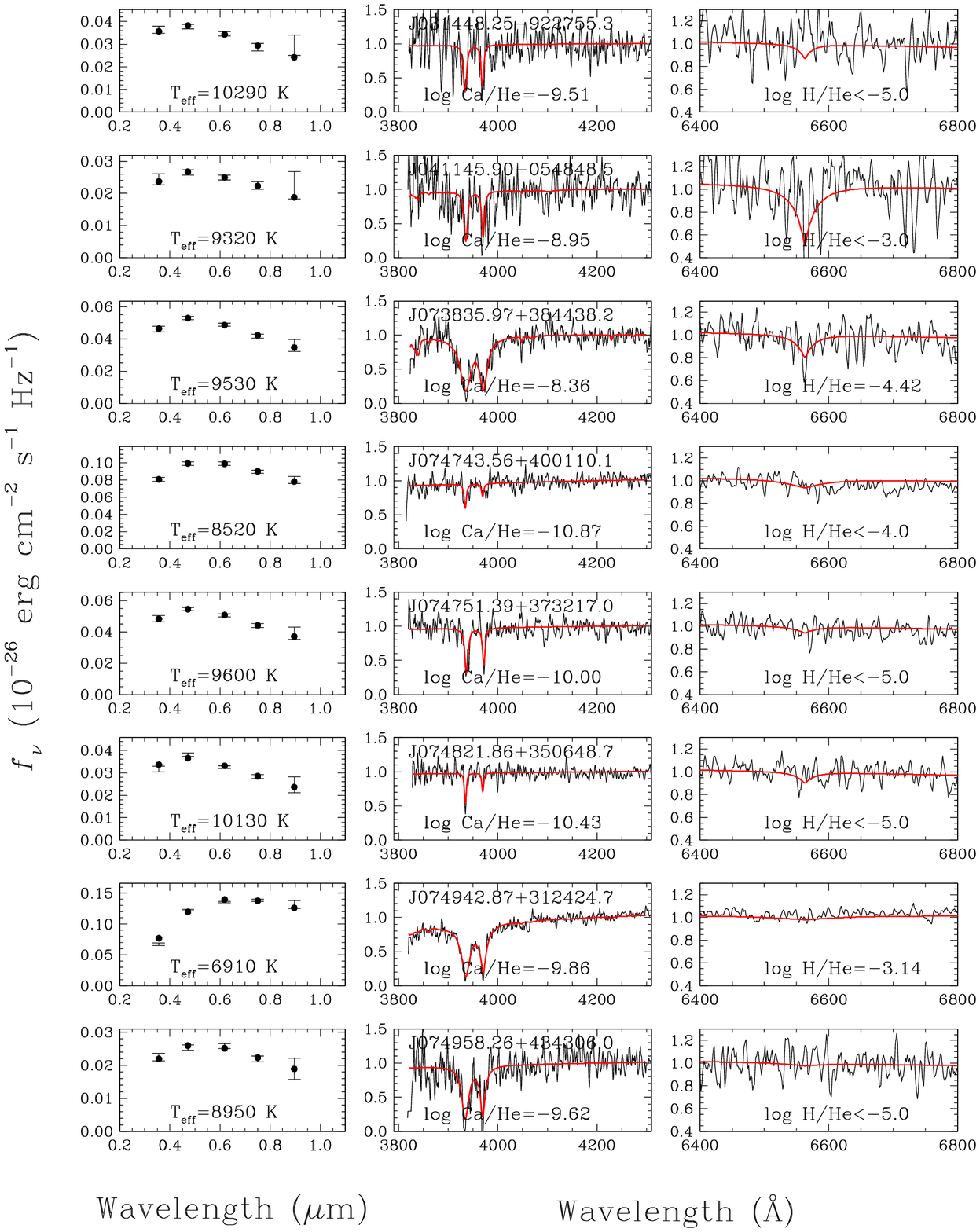] {Same as Fig.~\ref{fg:f6}. \label{fg:f8}}

\figcaption[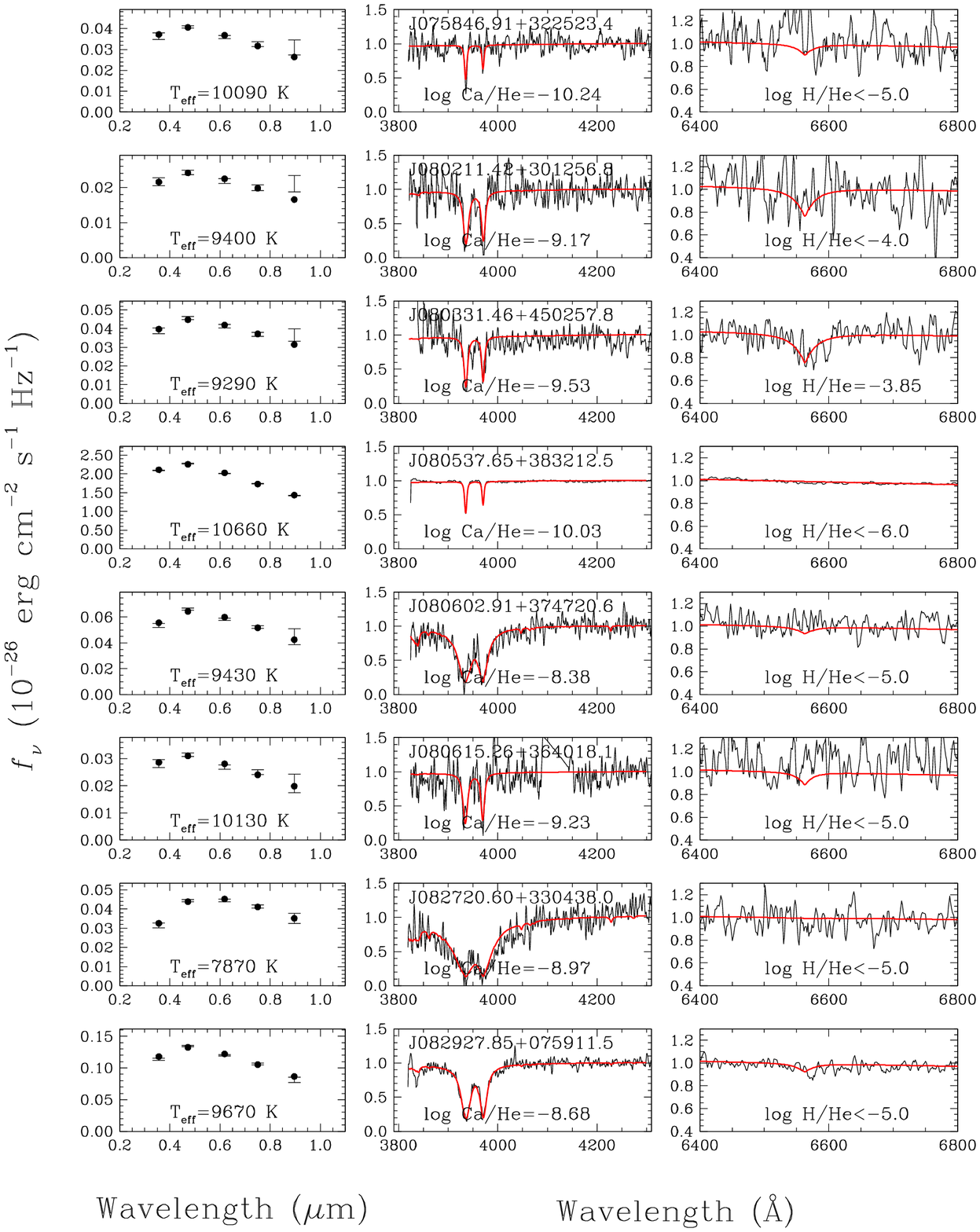] {Same as Fig.~\ref{fg:f6}. \label{fg:f9}}

\figcaption[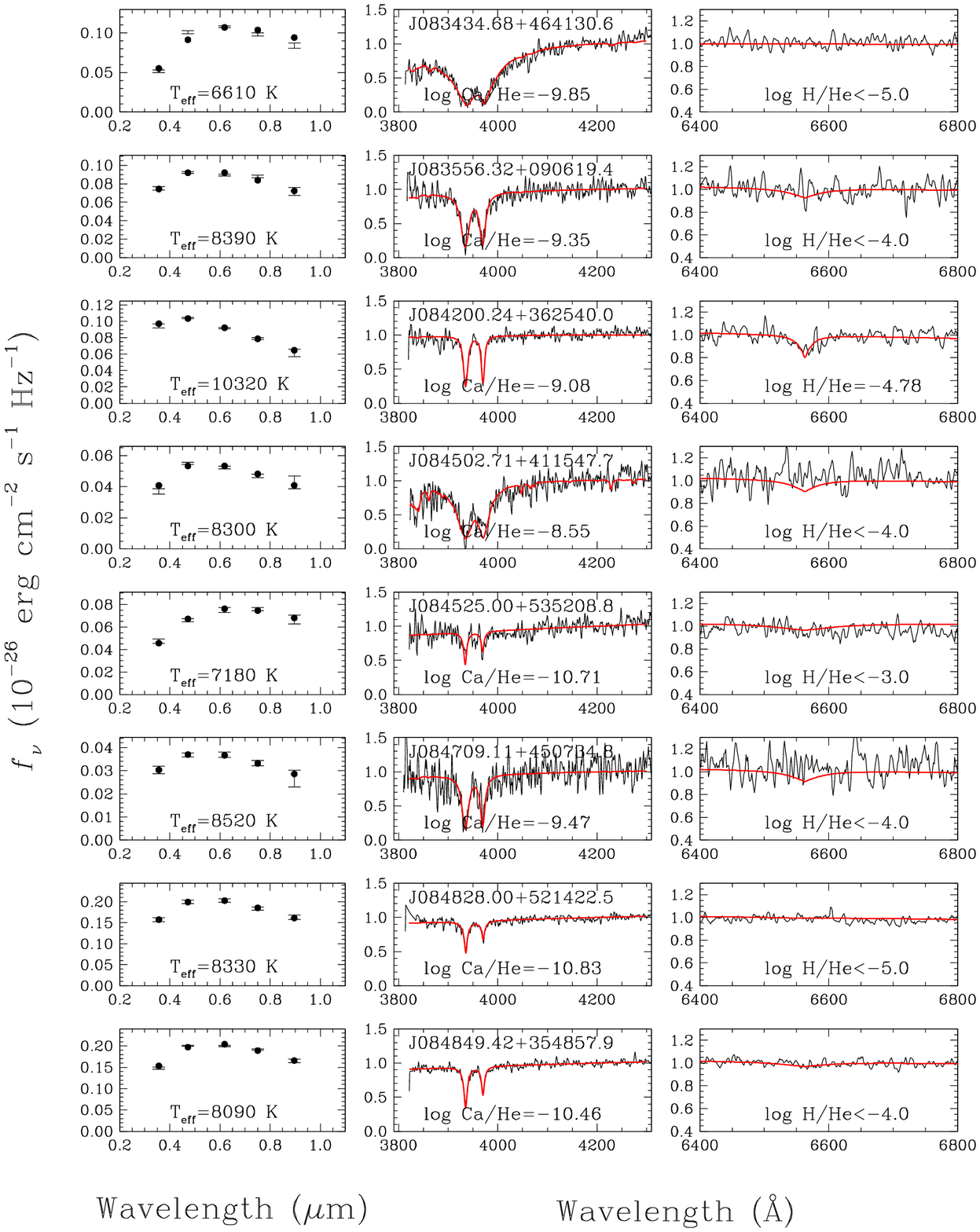] {Same as Fig.~\ref{fg:f6}. \label{fg:f10}}

\figcaption[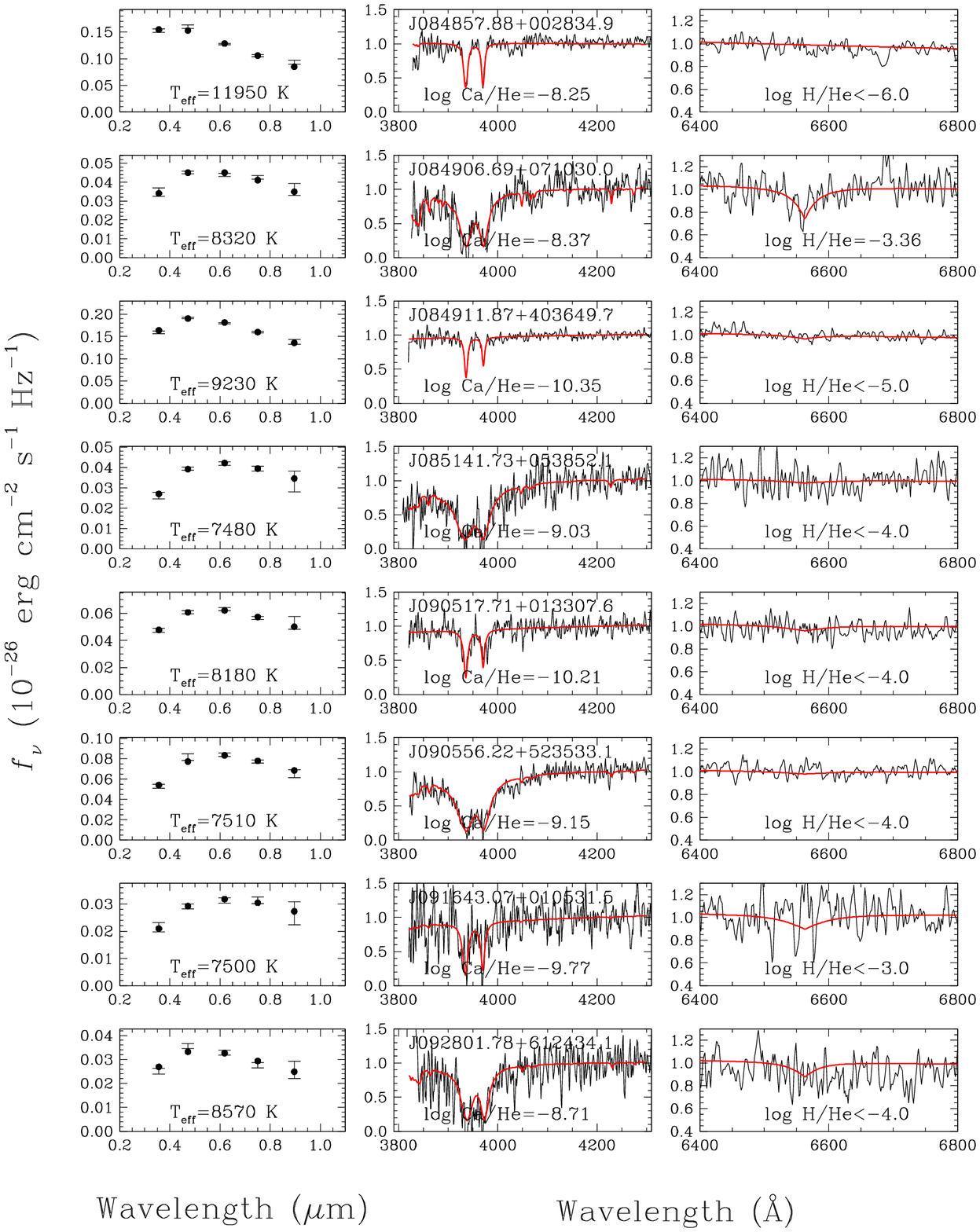] {Same as Fig.~\ref{fg:f6}. \label{fg:f11}}

\figcaption[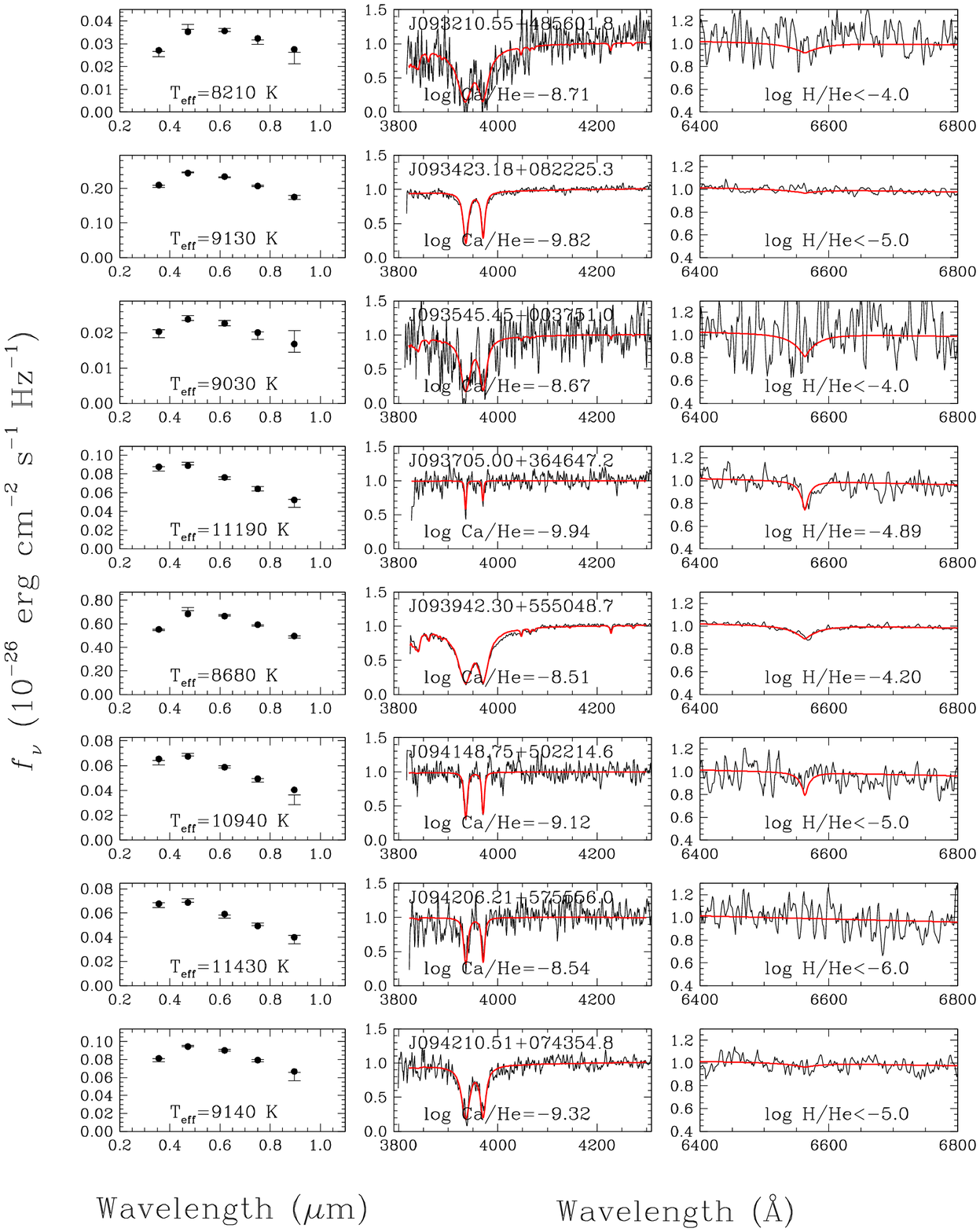] {Same as Fig.~\ref{fg:f6}. \label{fg:f12}}

\figcaption[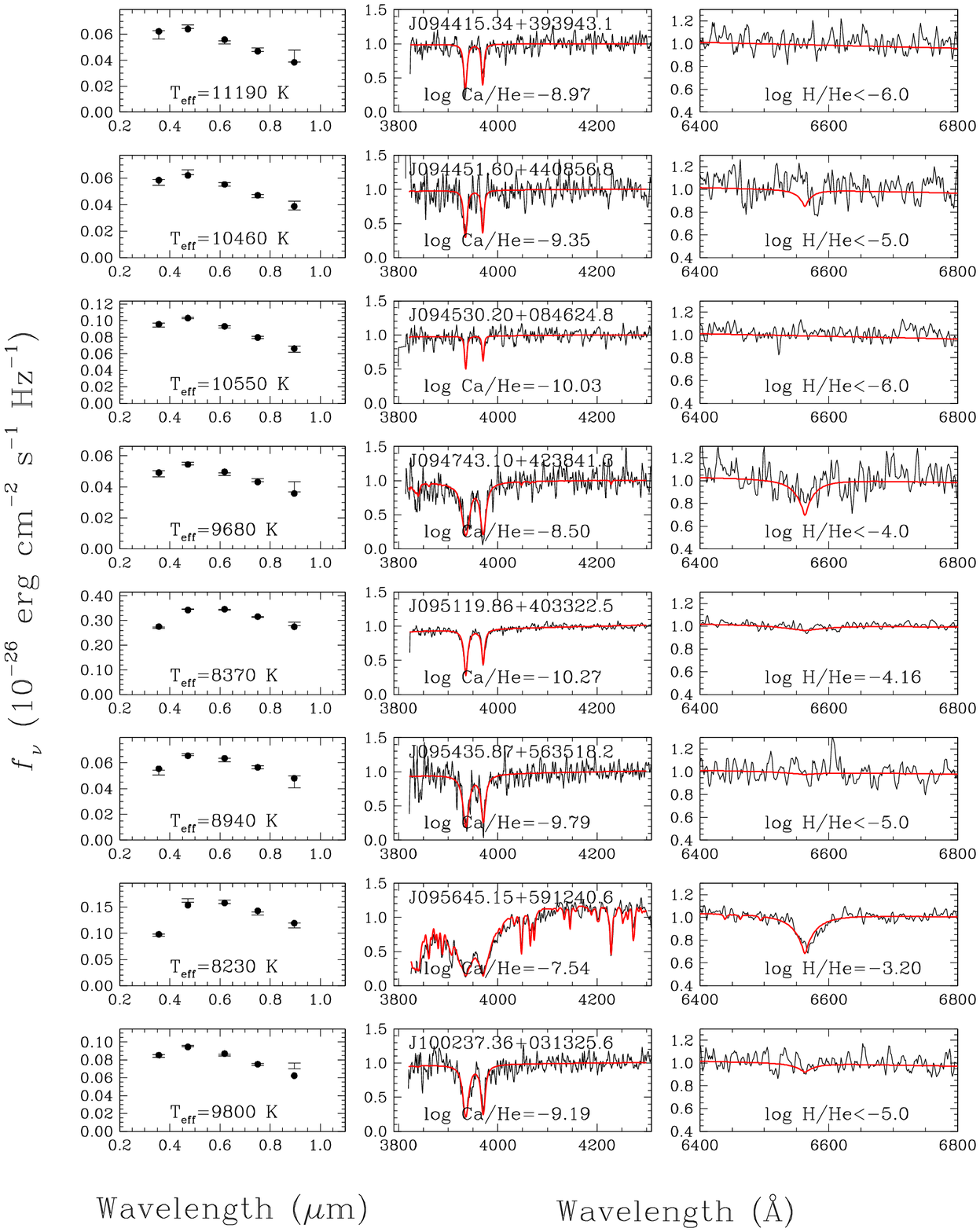] {Same as Fig.~\ref{fg:f6}. \label{fg:f13}}

\figcaption[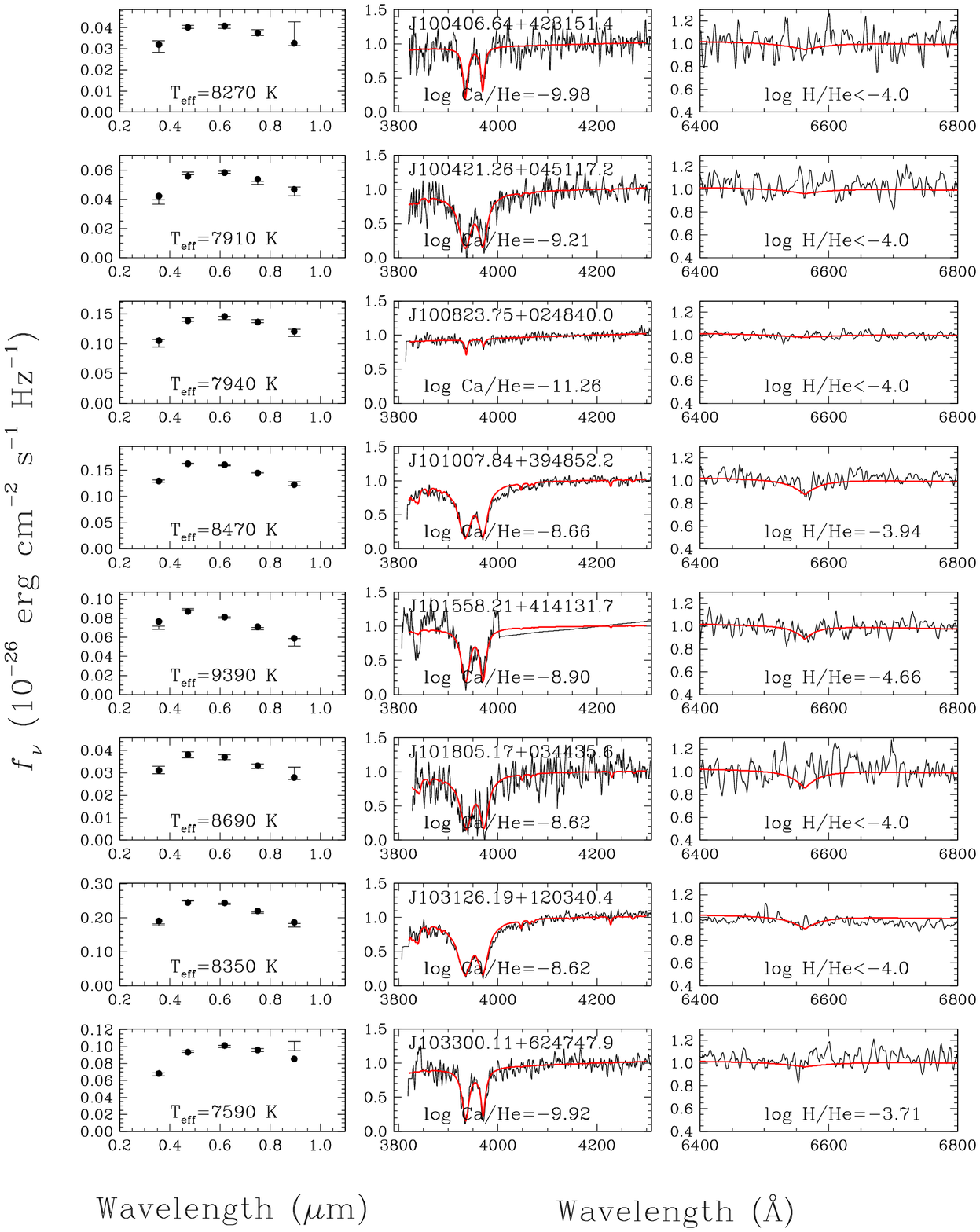] {Same as Fig.~\ref{fg:f6}. \label{fg:f14}}

\figcaption[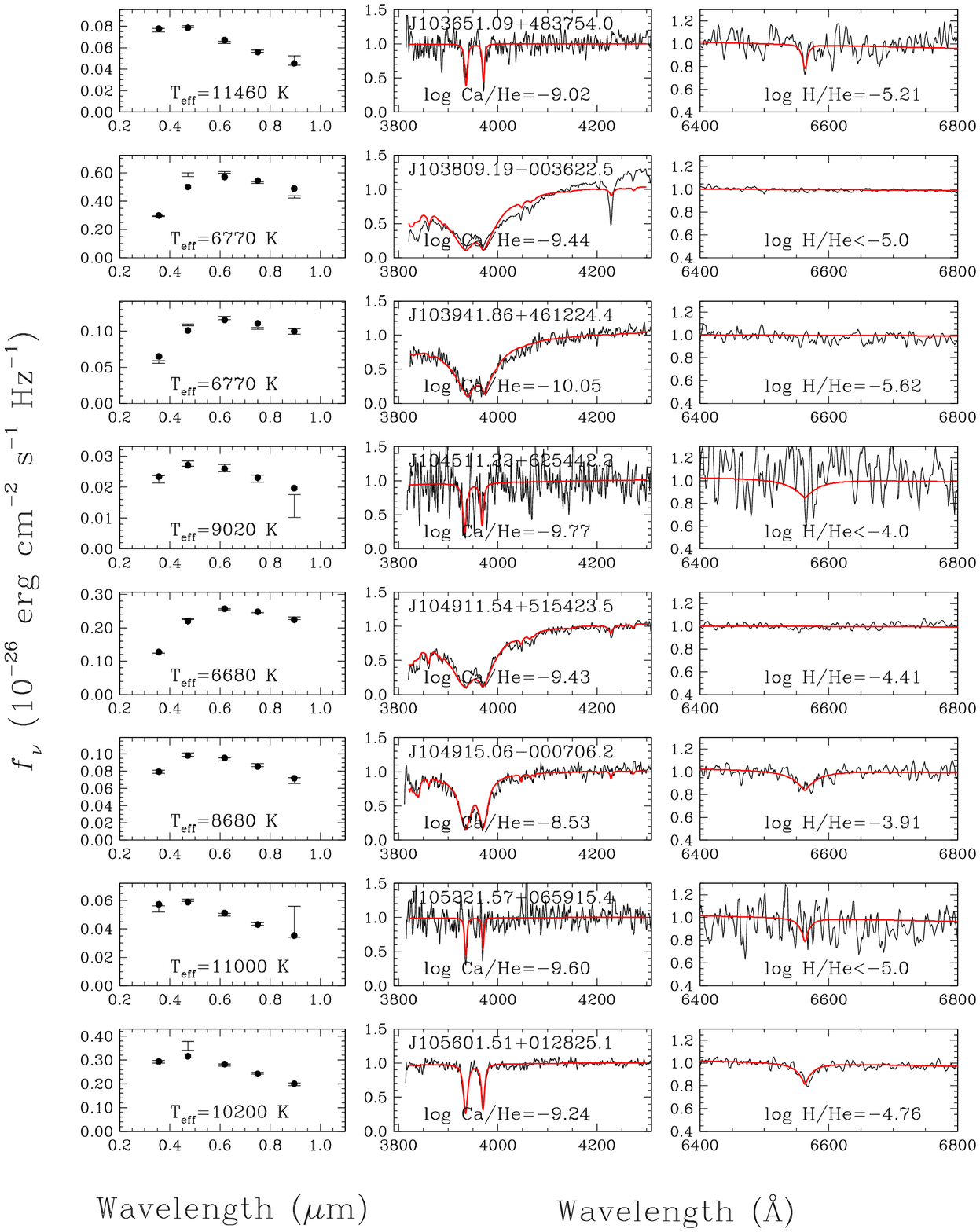] {Same as Fig.~\ref{fg:f6}. \label{fg:f15}}

\figcaption[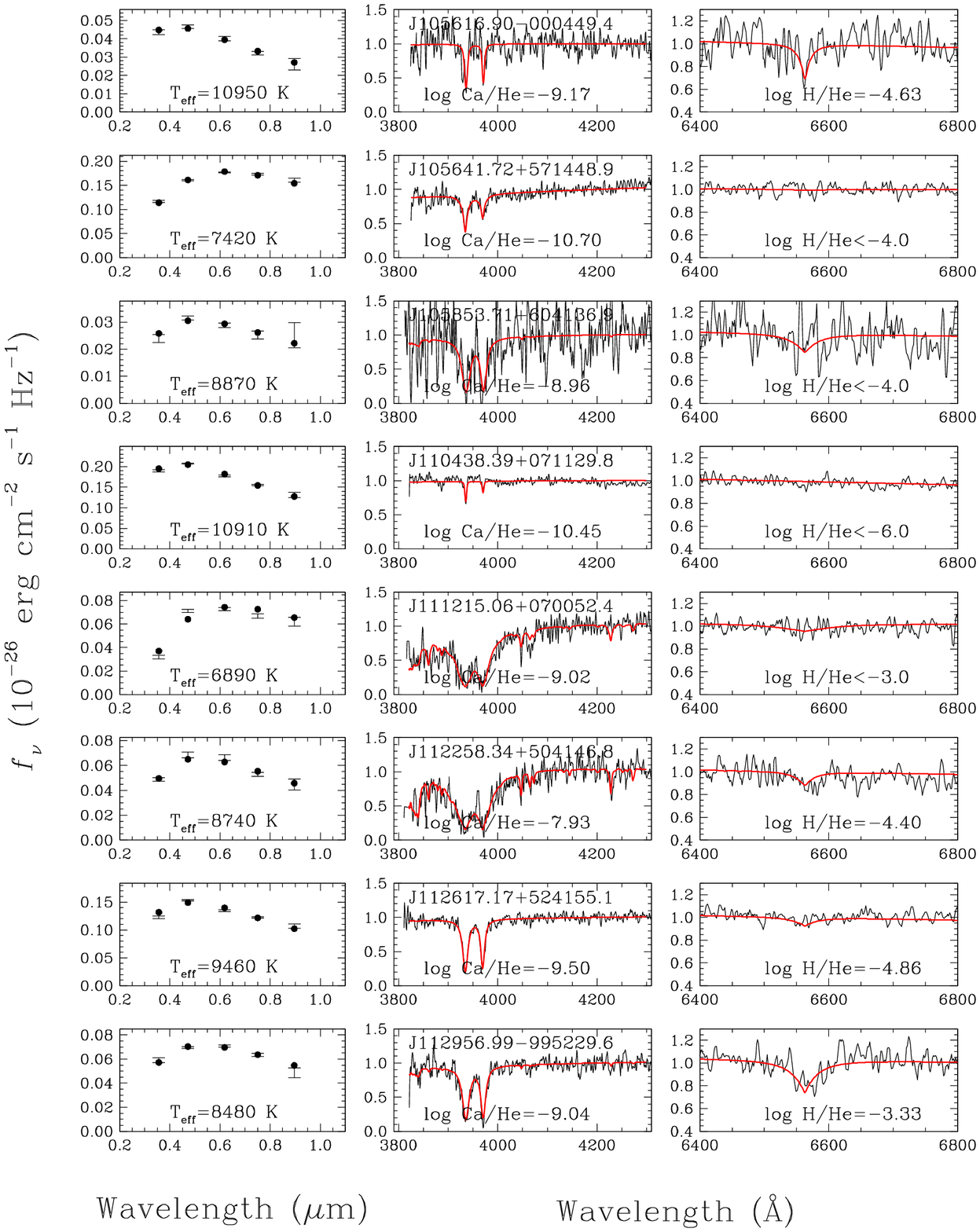] {Same as Fig.~\ref{fg:f6}. \label{fg:f16}}

\figcaption[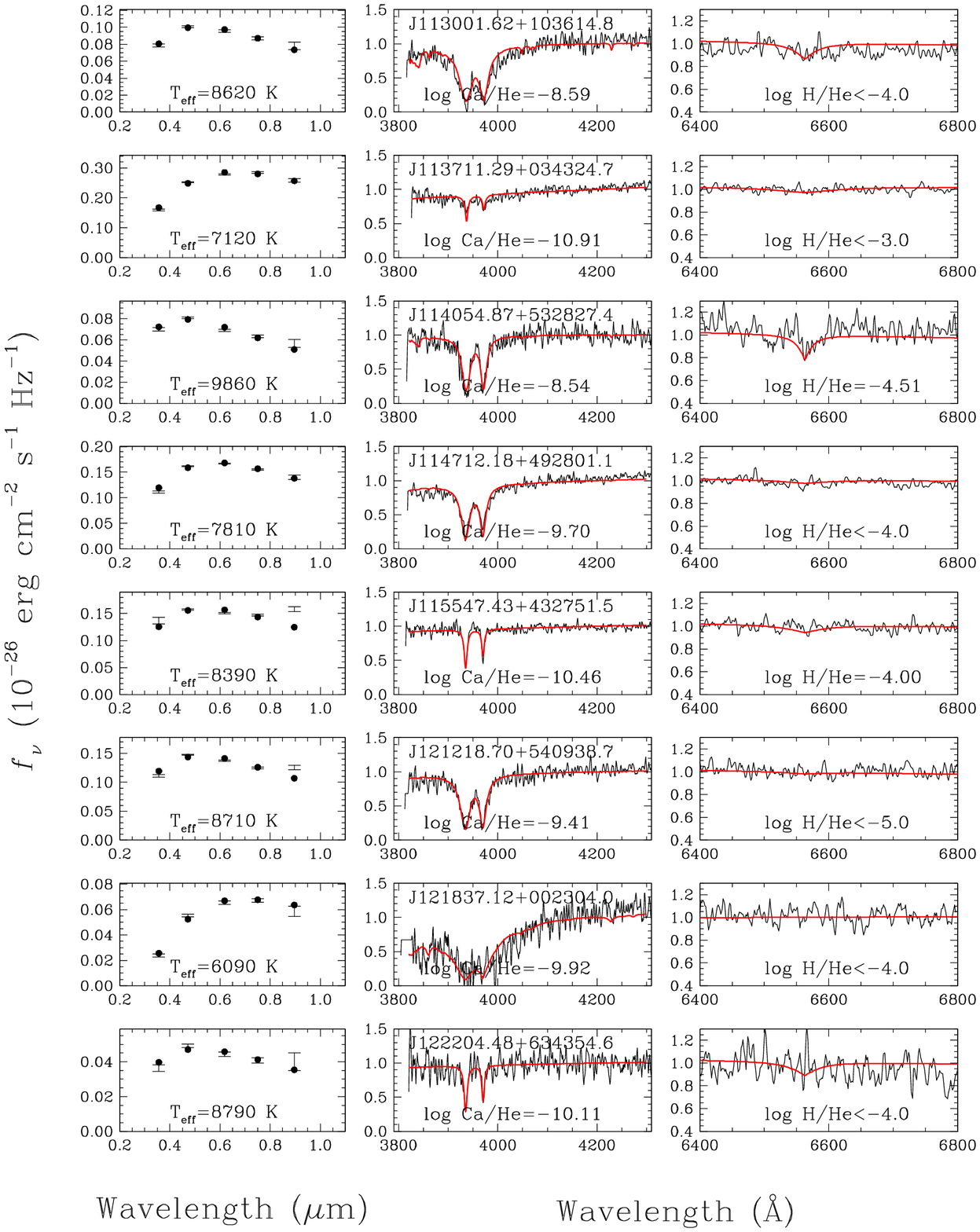] {Same as Fig.~\ref{fg:f6}. \label{fg:f17}}

\figcaption[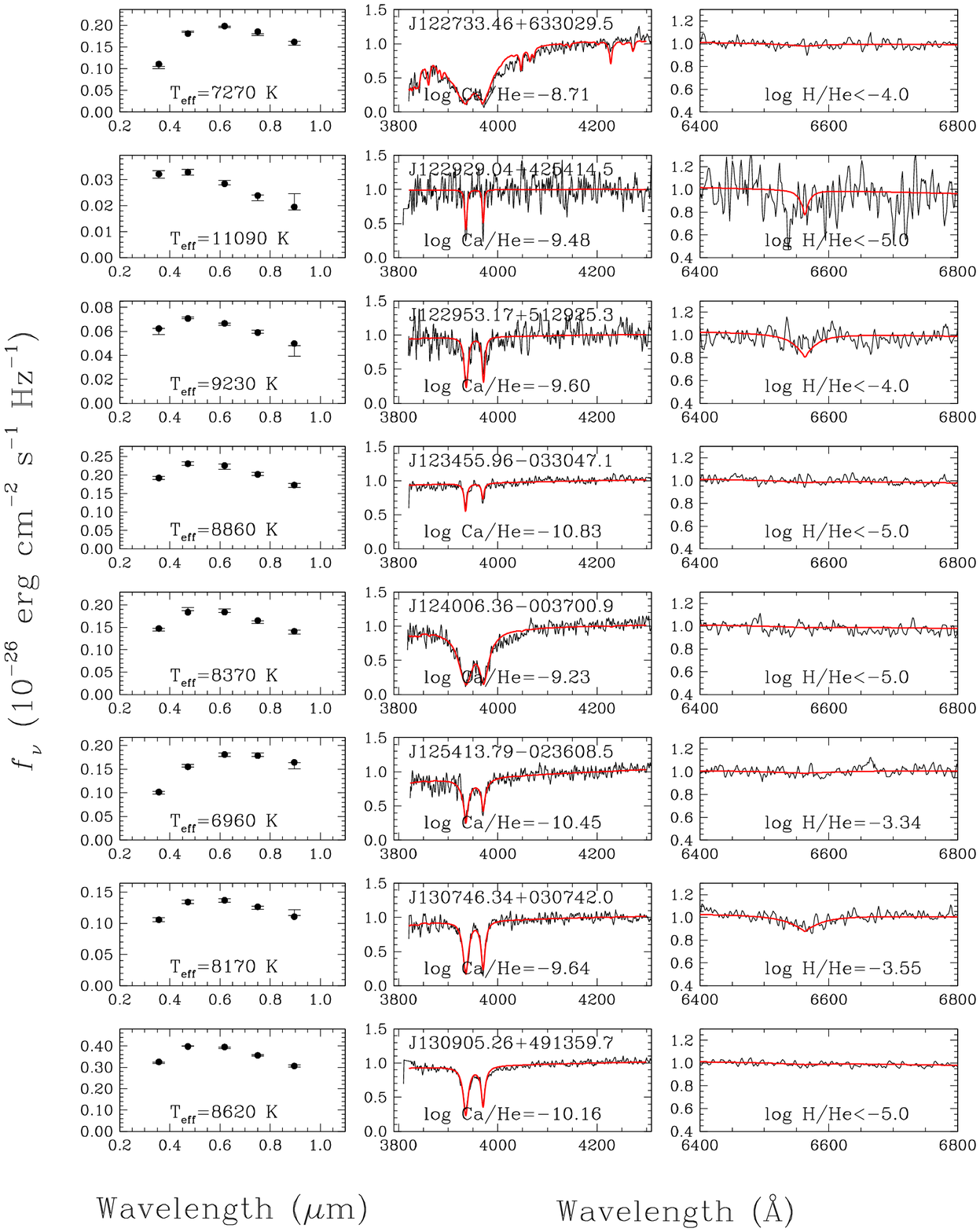] {Same as Fig.~\ref{fg:f6}. \label{fg:f18}}

\figcaption[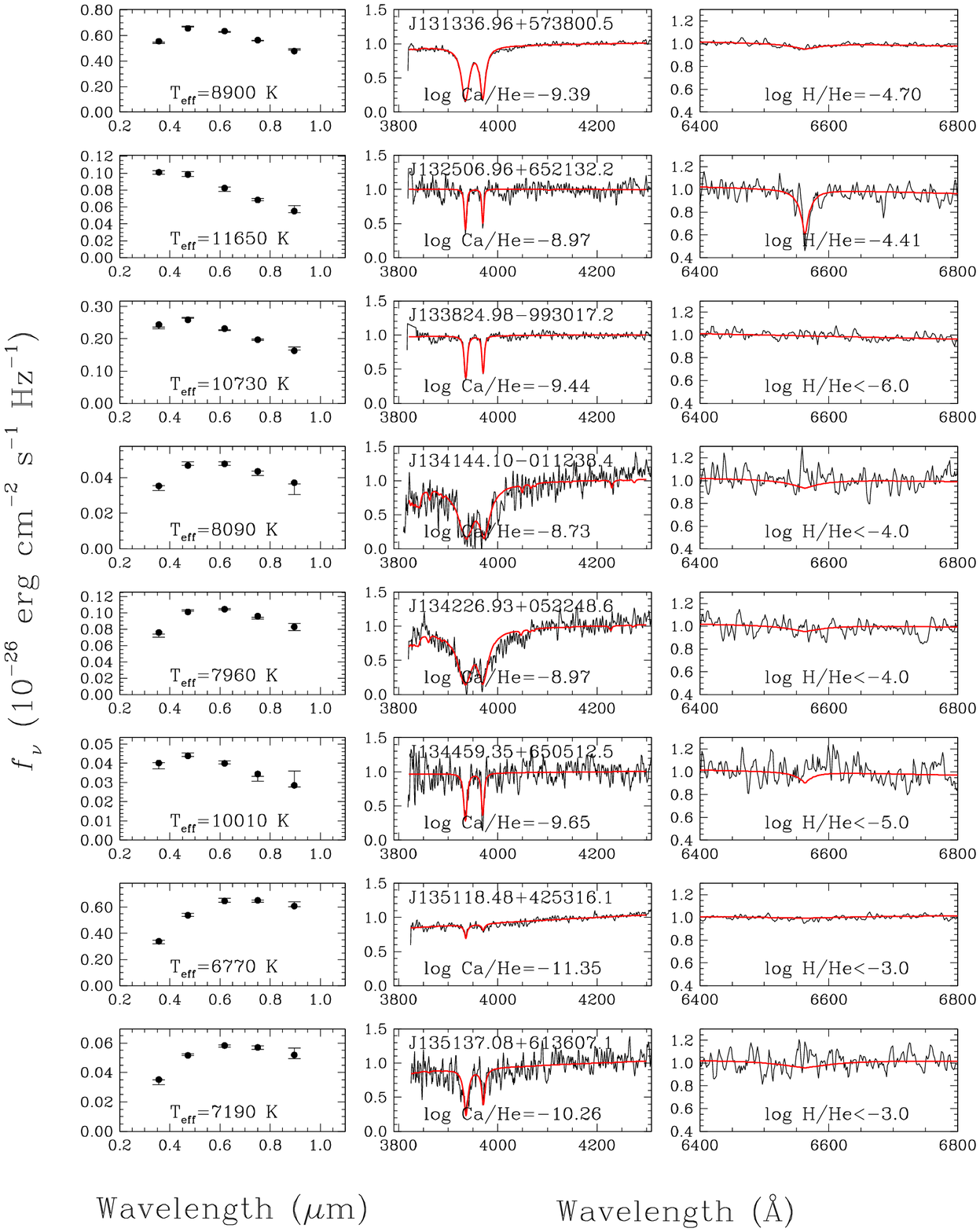] {Same as Fig.~\ref{fg:f6}. \label{fg:f19}}

\figcaption[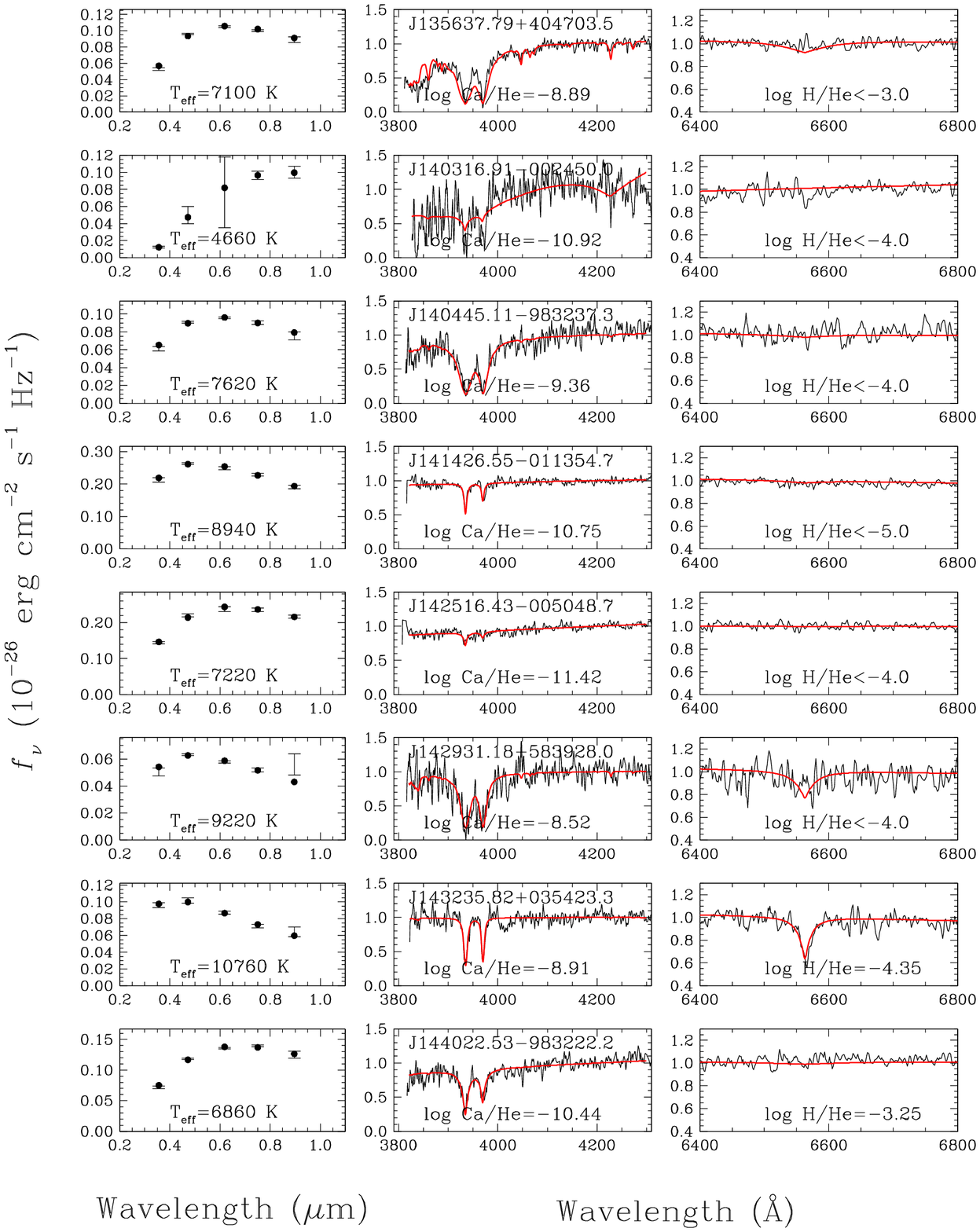] {Same as Fig.~\ref{fg:f6}. \label{fg:f20}}

\figcaption[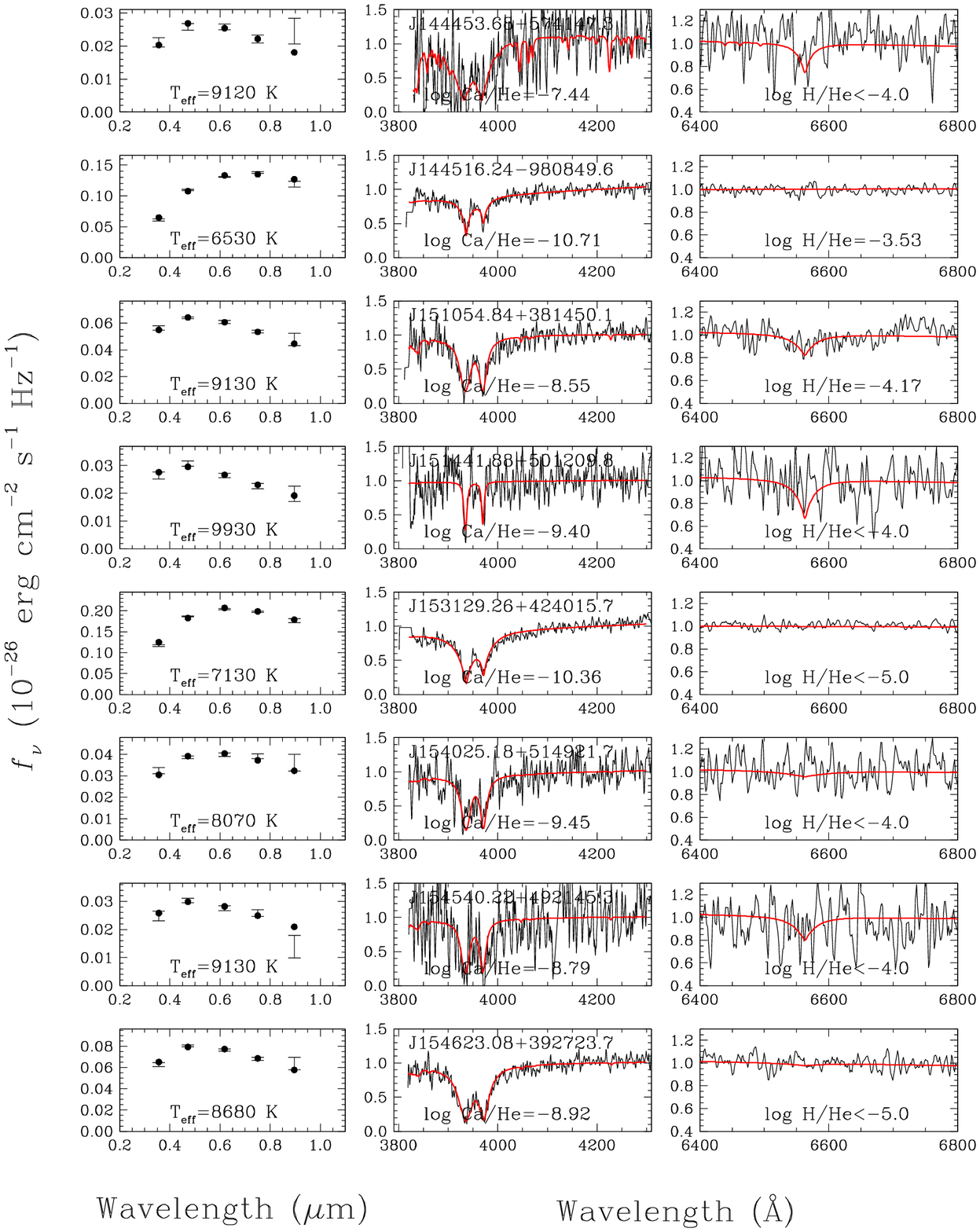] {Same as Fig.~\ref{fg:f6}. \label{fg:f21}}

\figcaption[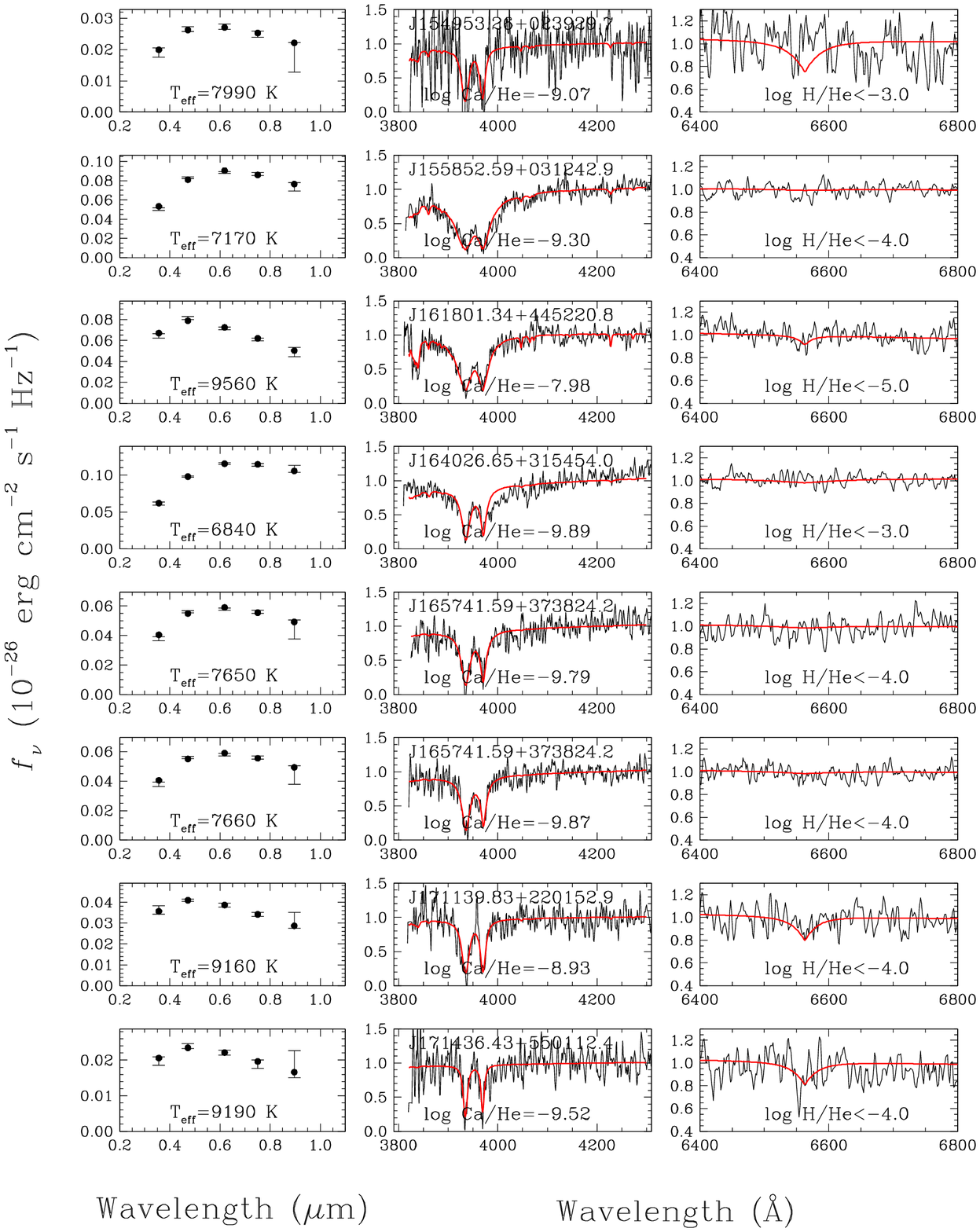] {Same as Fig.~\ref{fg:f6}. \label{fg:f22}}

\figcaption[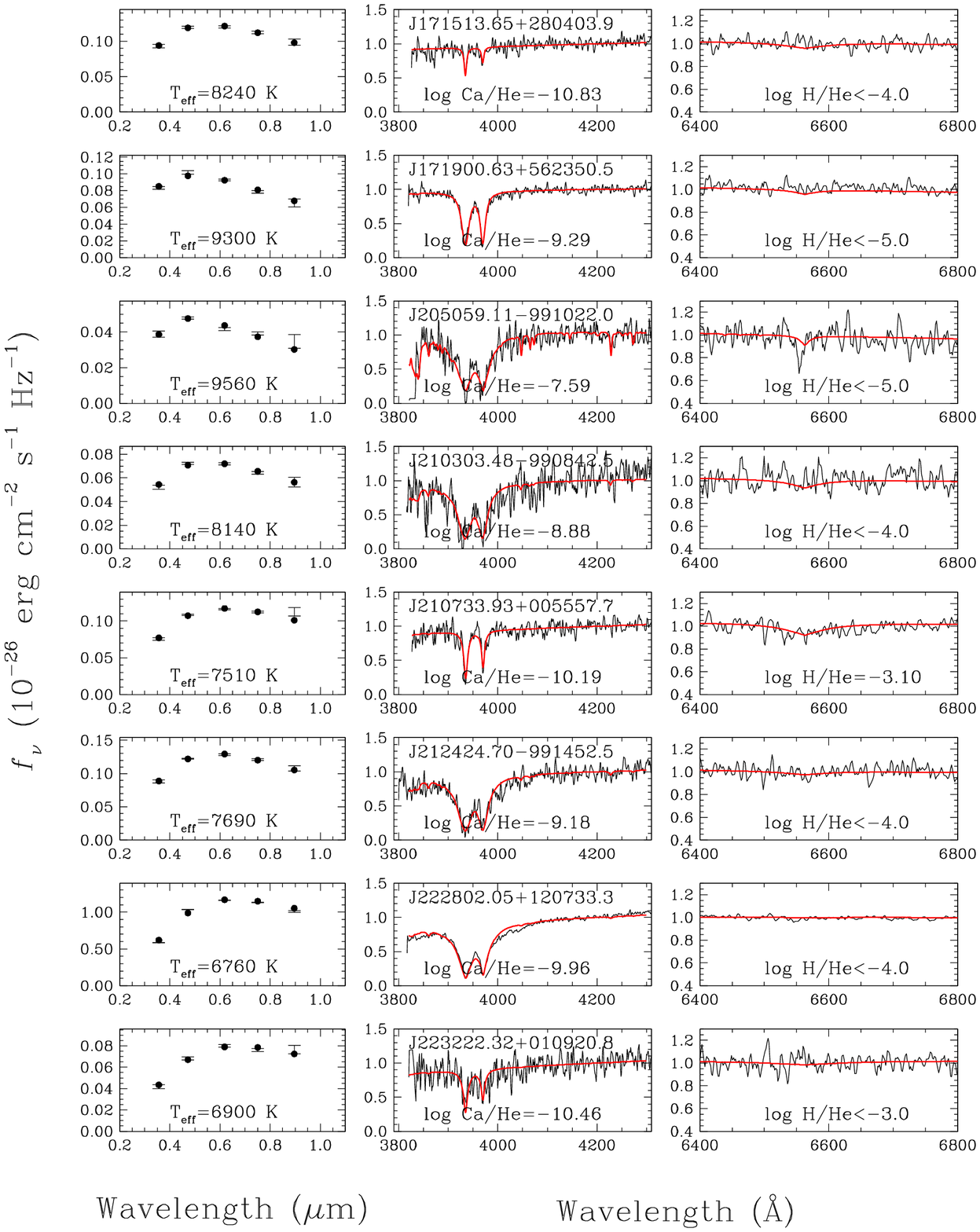] {Same as Fig.~\ref{fg:f6}. \label{fg:f23}}

\figcaption[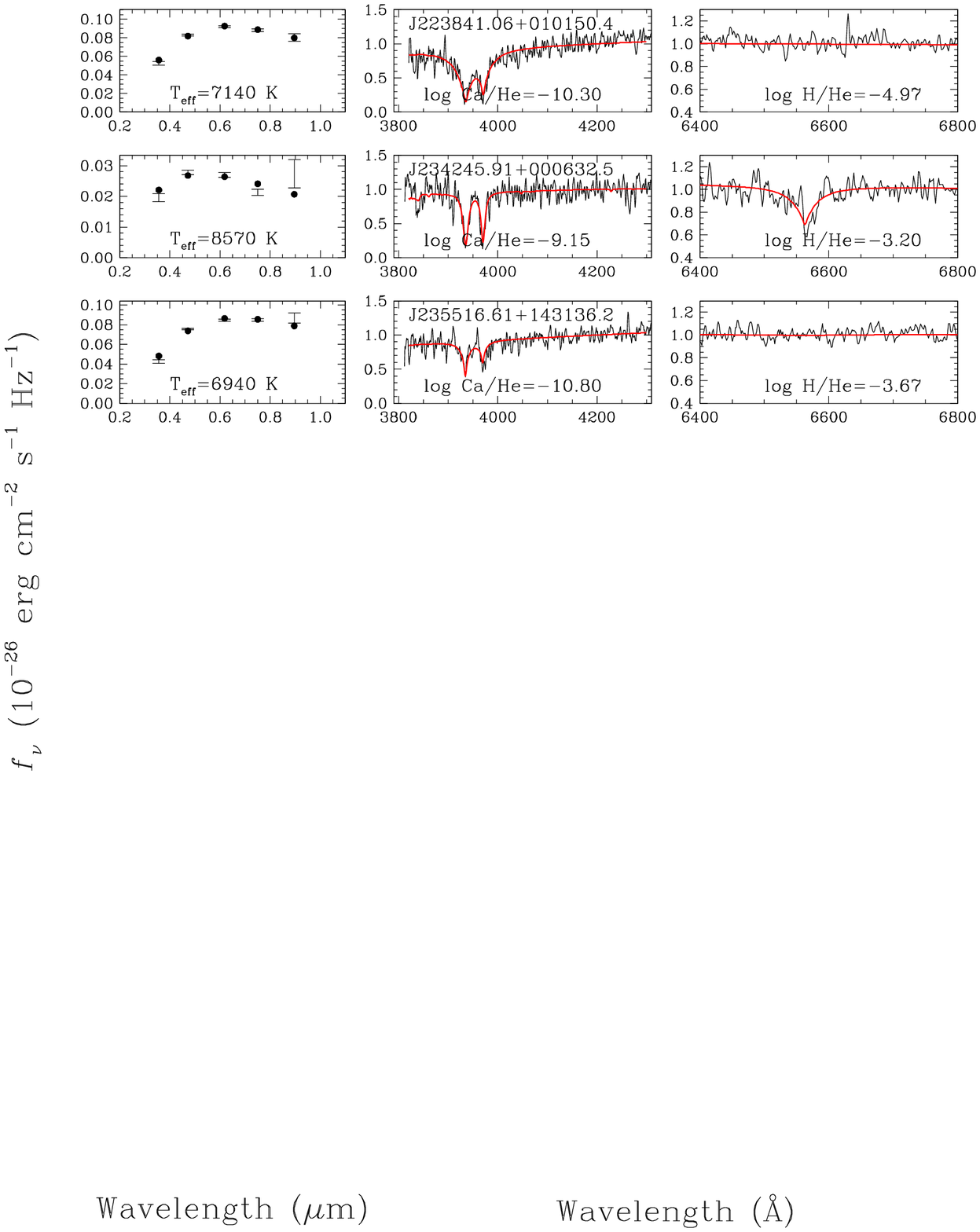] {Same as Fig.~\ref{fg:f6}. \label{fg:f24}}

\figcaption[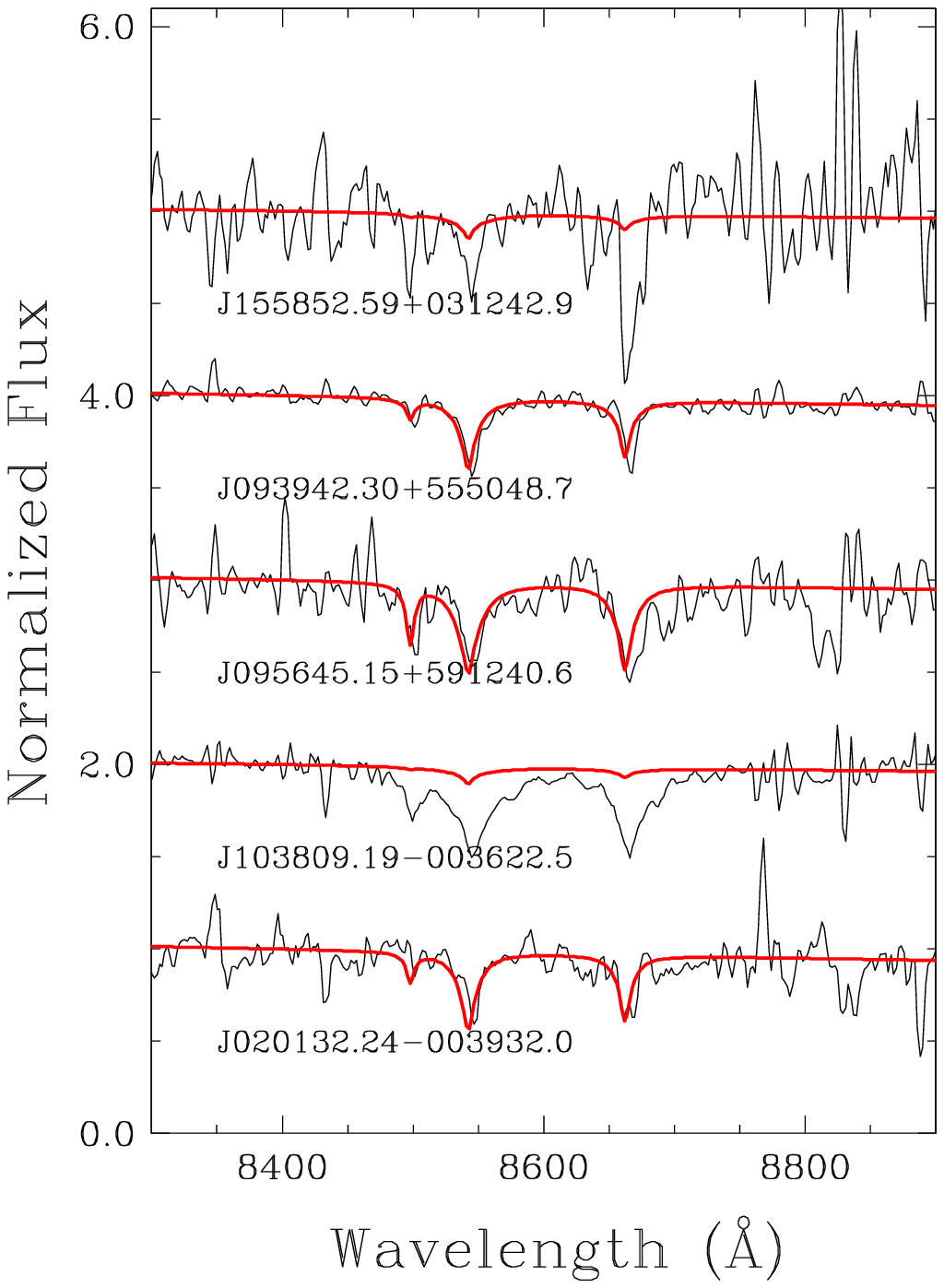] {SDSS spectra with observed Ca~\textsc{ii} triplets 
compared with synthetic spectra interpolated at the atmospheric
parameter solution obtained from the Ca~\textsc{ii} H \& K
lines. [{\it See the electronic version of the Journal for a color version
of this figure.}]\label{fg:f25}}

\figcaption[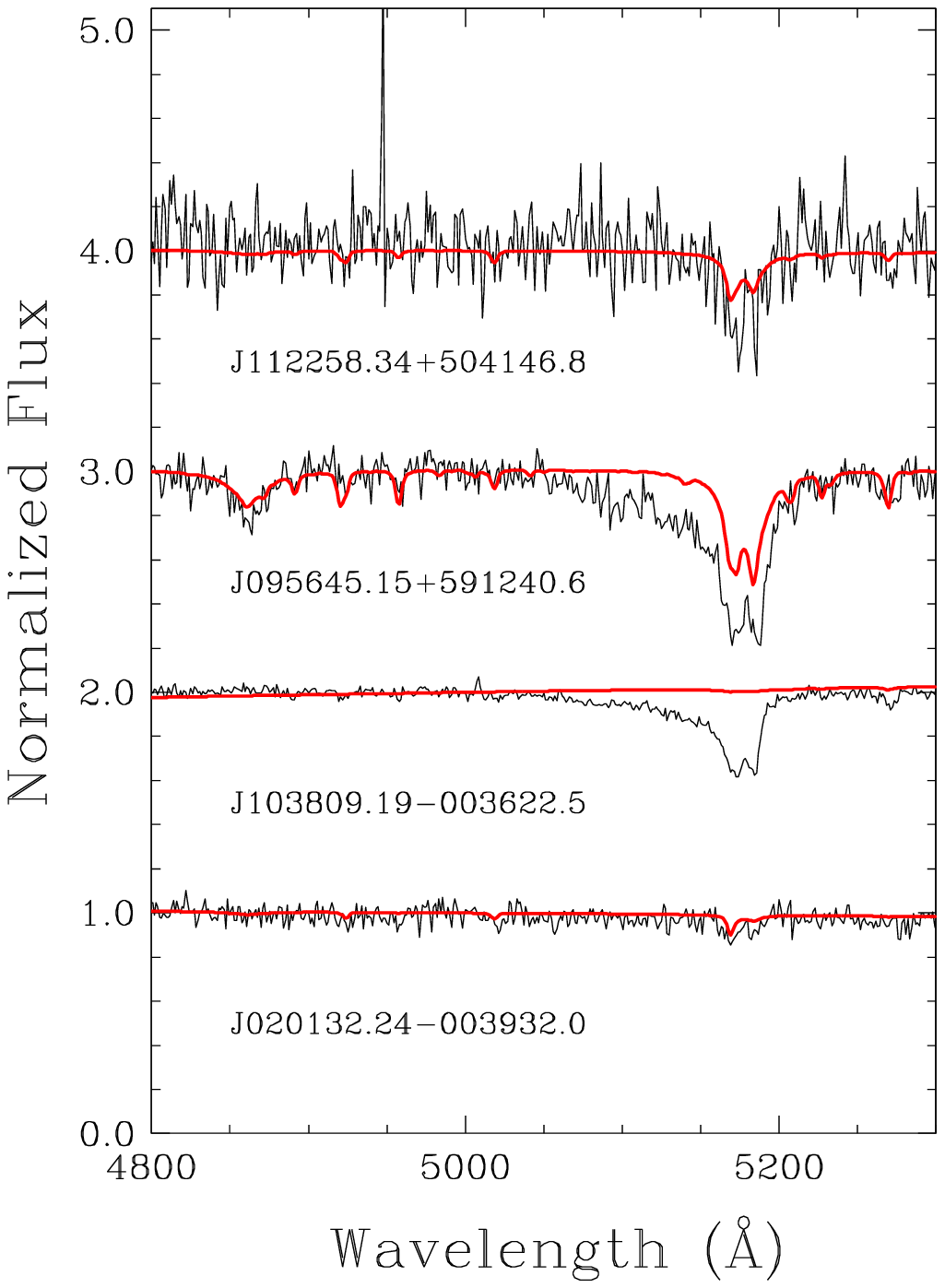] {SDSS spectra of DZ stars showing 
the Mg~\textsc{i} ``b'' blend at $\sim 5175$ \AA\
compared with synthetic spectra interpolated at the atmospheric
parameter solution obtained from the Ca~\textsc{ii} H \& K lines. The
feature at $\sim4860$ \AA\ is H$\beta$.
[{\it See the electronic version of the Journal for a color version
of this figure.}]\label{fg:f26}}

\figcaption[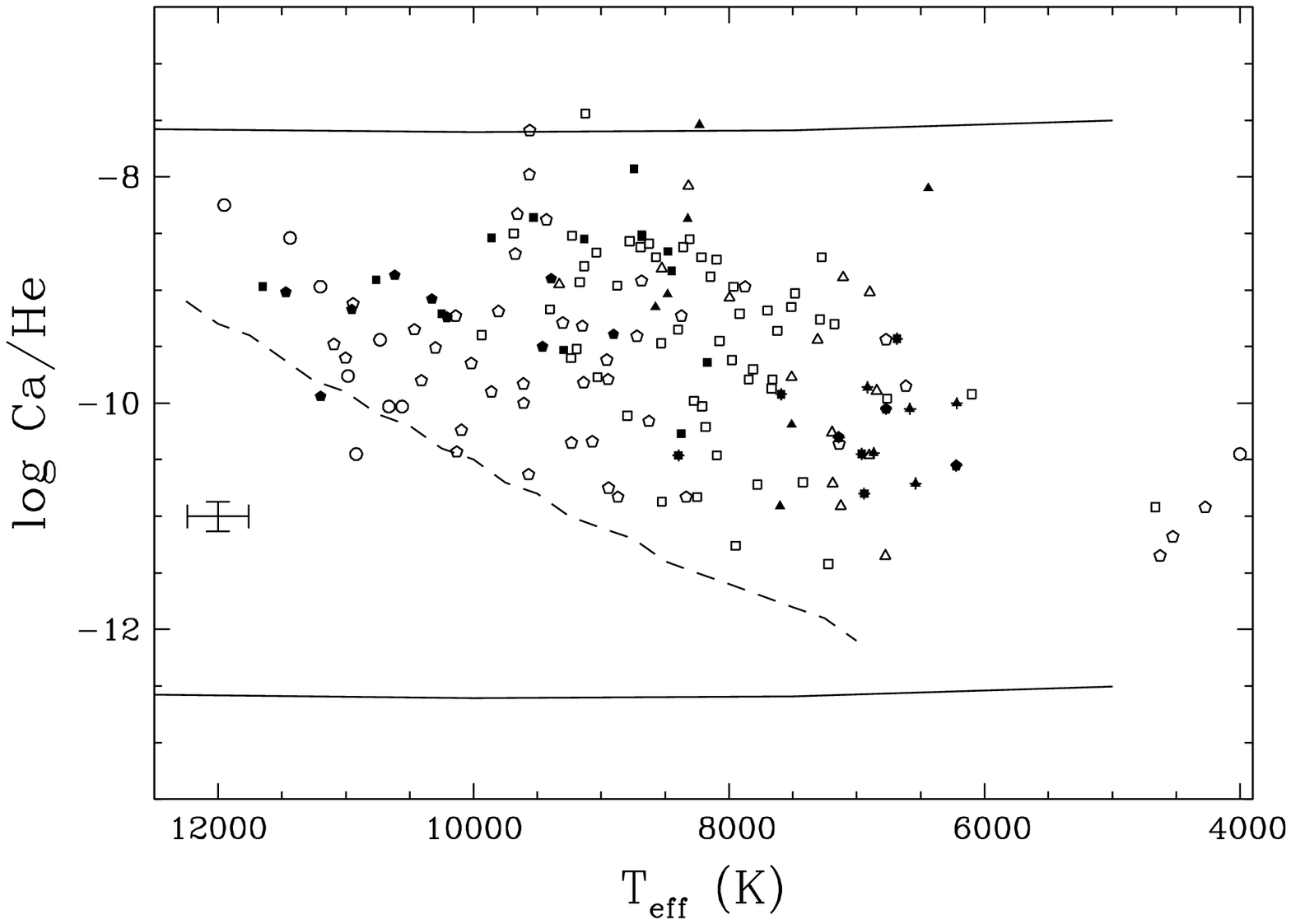] {Relative abundance of calcium with respect to 
helium as a function of effective temperature for the BRL97/BLR01 and
SDSS samples. The two continuous curves define the range of expected
abundances for a low accretion rate in solar proportion of
$5\times10^{-20}\ M_{\odot}$ yr$^{-1}$ ({\it lower curve}) and a high
rate of $5\times10^{-15} M_{\odot}$ yr$^{-1}$ ({\it upper curve}). The
dashed line indicates the detection limit of the Ca~\textsc{ii} H \& K
lines (taken as 5 \AA\ of total equivalent width). The error bars in
the bottom left corner represent the average uncertainties of $\Te$
($\sim 240$~K) and $\log\ ({\rm Ca/He})$ ($\sim 0.15$ dex). Filled
symbols correspond to DZ stars with hydrogen abundances determined
directly from H$\alpha$ or indirectly from the Ca~\textsc{ii} lines
(these are marked with an additional $+$ sign), while open symbols
correspond to objects with only upper limits. The approximate hydrogen
abundance for each object (or upper limits) can be deduced from the
number of sides of the polygon (triangles: $\nh\sim-3$, squares:
$\nh\sim-4$, pentagons: $\nh\sim-5$; circles represent pure helium
solutions).\label{fg:f27}}

\figcaption[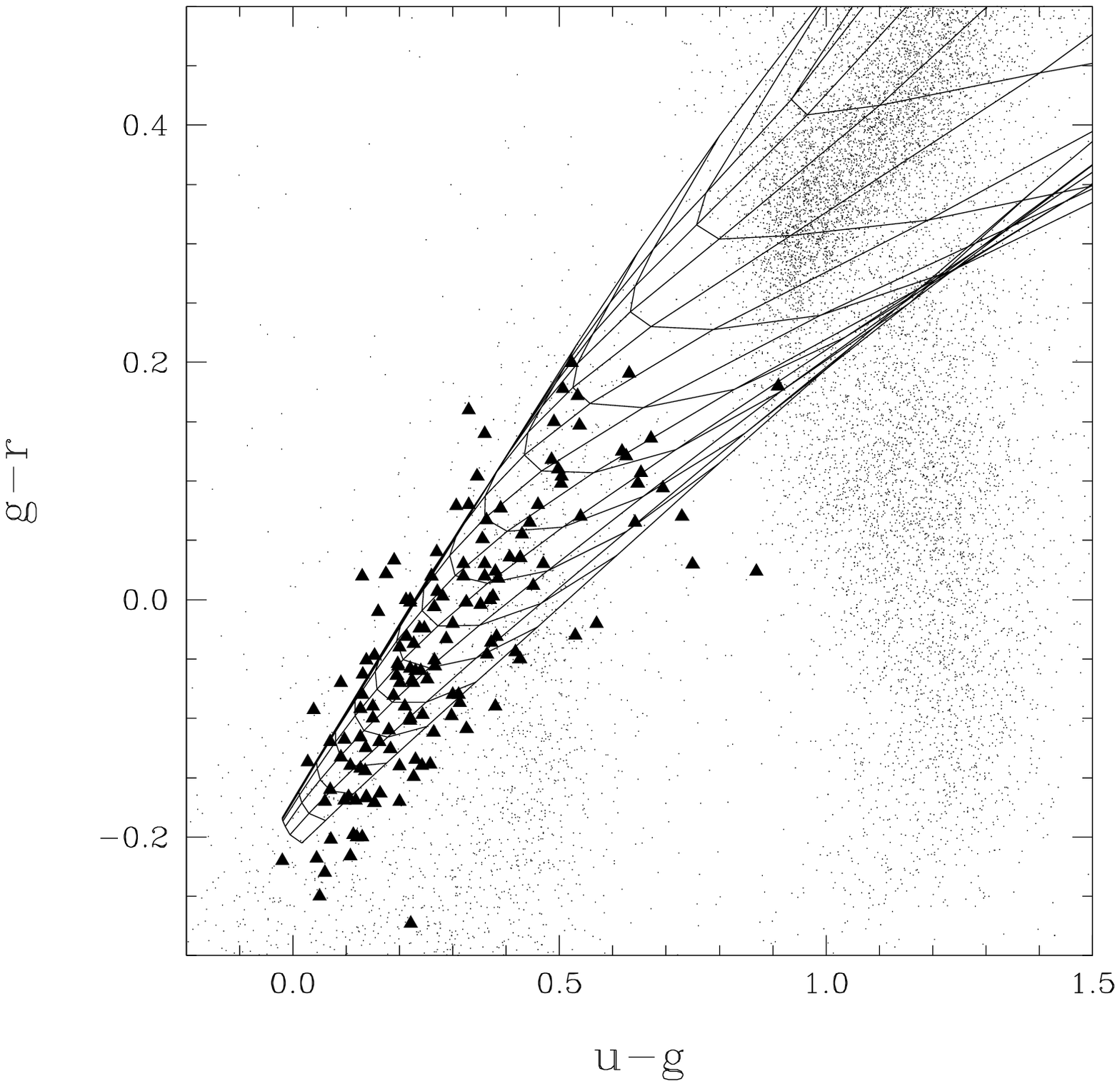] {The ($g-r$, $u-g$) color-color diagram. Small dots
represent objects with stellar images while filled triangles
correspond to the SDSS DZ stars from Table 2. The curves show our DZ
photometric sequences at $\logg=8$, $\nh=-30$, for $\Te=12,000$~K ({\it
bottom left parabola}) down to 5000~K ({\it upper right parabola}) by steps of 500 K, and
for calcium abundances of $\log\ ({\rm Ca/He})=-7$ ({\it bottom sequence}) 
down to $-12$ by steps of 0.5 dex. The uppermost sequence corresponds to the pure
helium models. \label{fg:f28}}

\figcaption[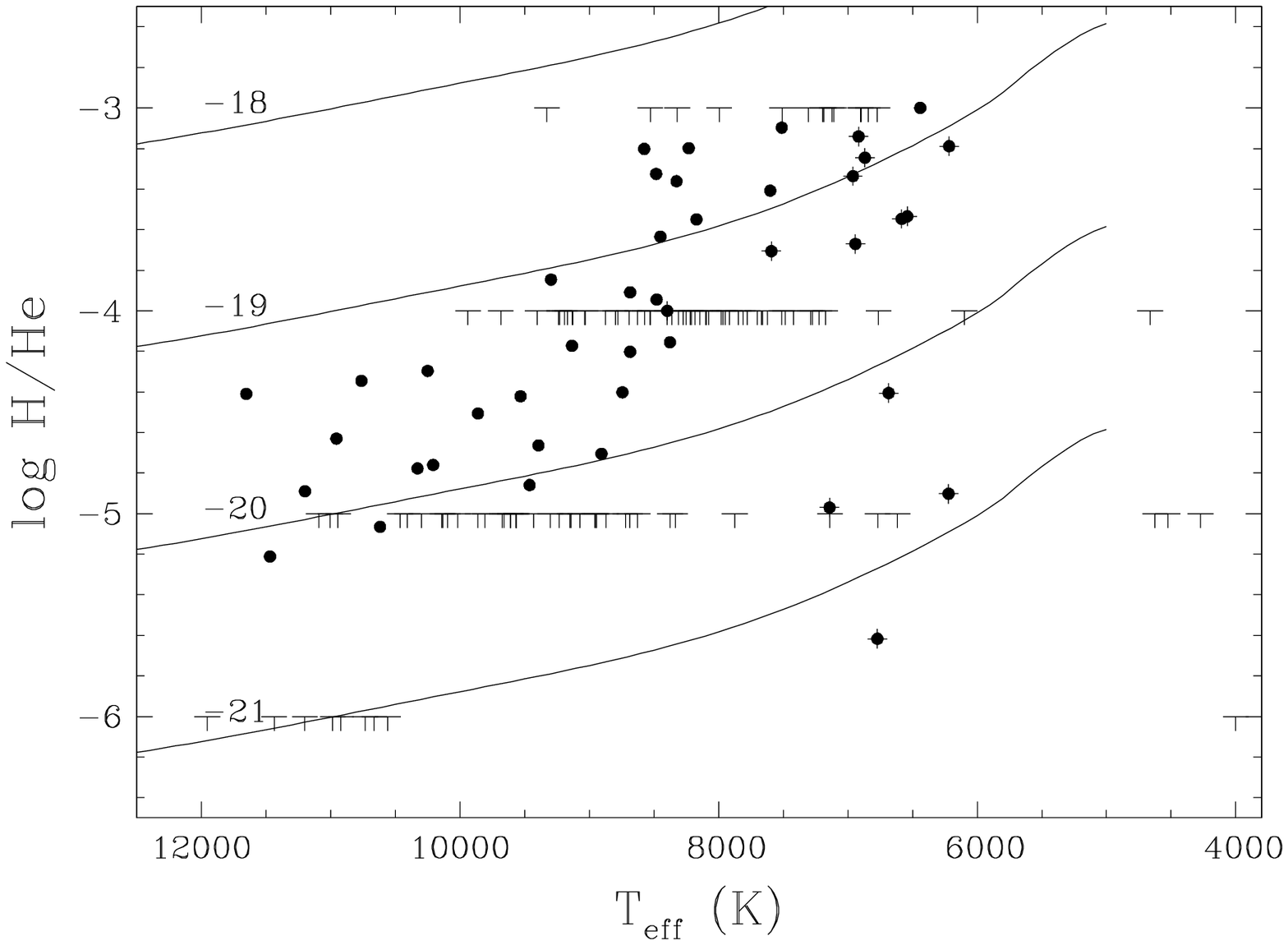] {Hydrogen-to-helium abundance ratios as a function 
of effective temperature for the BRL/BLR and SDSS samples. Filled
circles correspond to objects with hydrogen abundance determinations;
objects with abundances measured indirectly from the calcium lines are
marked with an additional $+$ sign. The crosses indicate white dwarfs
with only upper limit determinations. Solid lines represent the
expected abundances for continuous accretion of material from the ISM
with a solar composition and accretion rates of $10^{-21}$ to
$10^{-18}\ M_{\odot}$ yr$^{-1}$.
\label{fg:f29}}

\figcaption[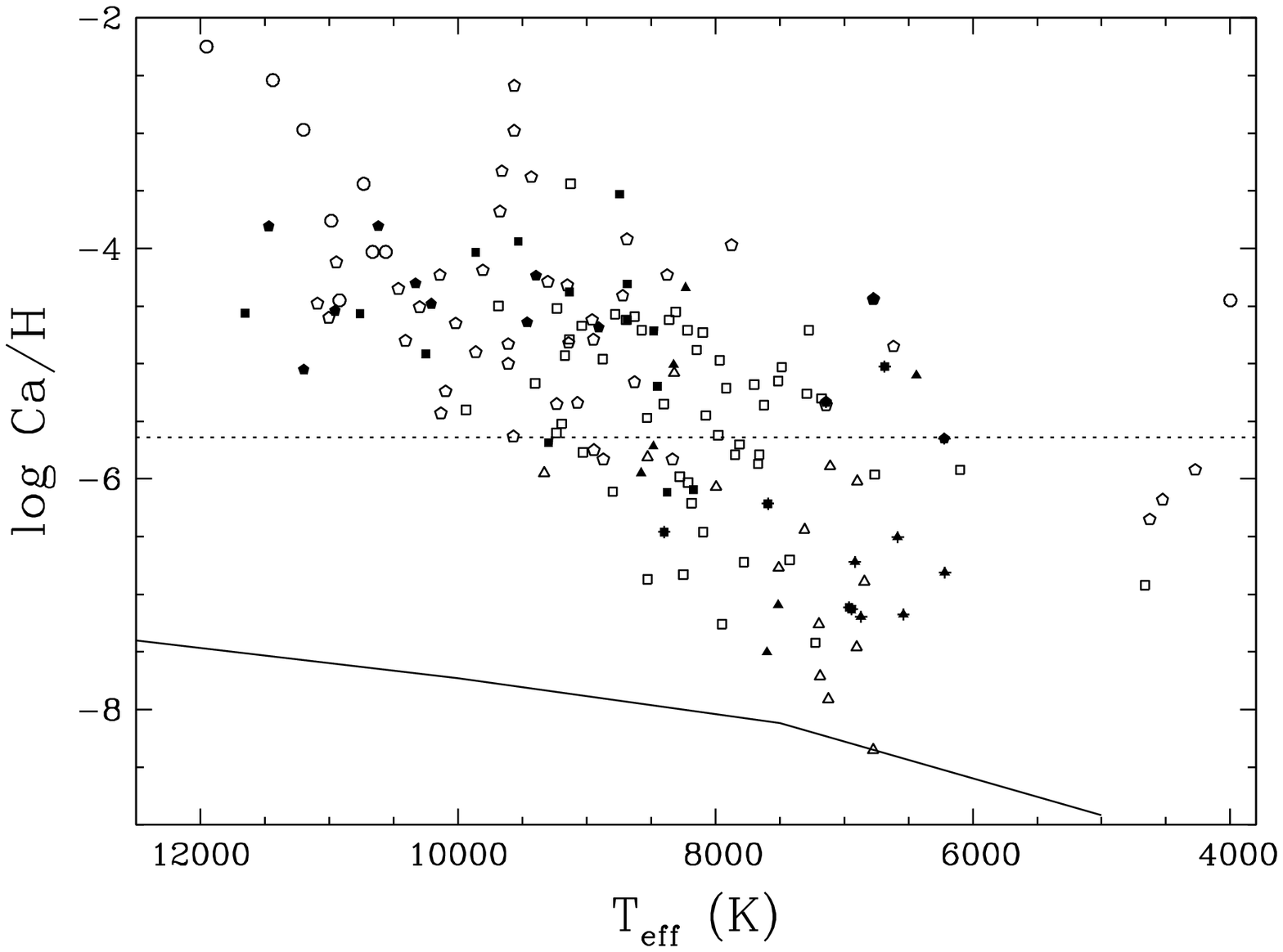] {Calcium-to-hydrogen abundance ratios as a function 
of effective temperature for the BRL/BLR and SDSS samples. The
horizontal dashed line indicates the solar Ca/H abundance ratio, while
the solid line corresponds to the maximum value expected from
accretion of material with solar compositions. The various symbols are
explained in the caption of Figure \ref{fg:f27}.\label{fg:f30}}

\figcaption[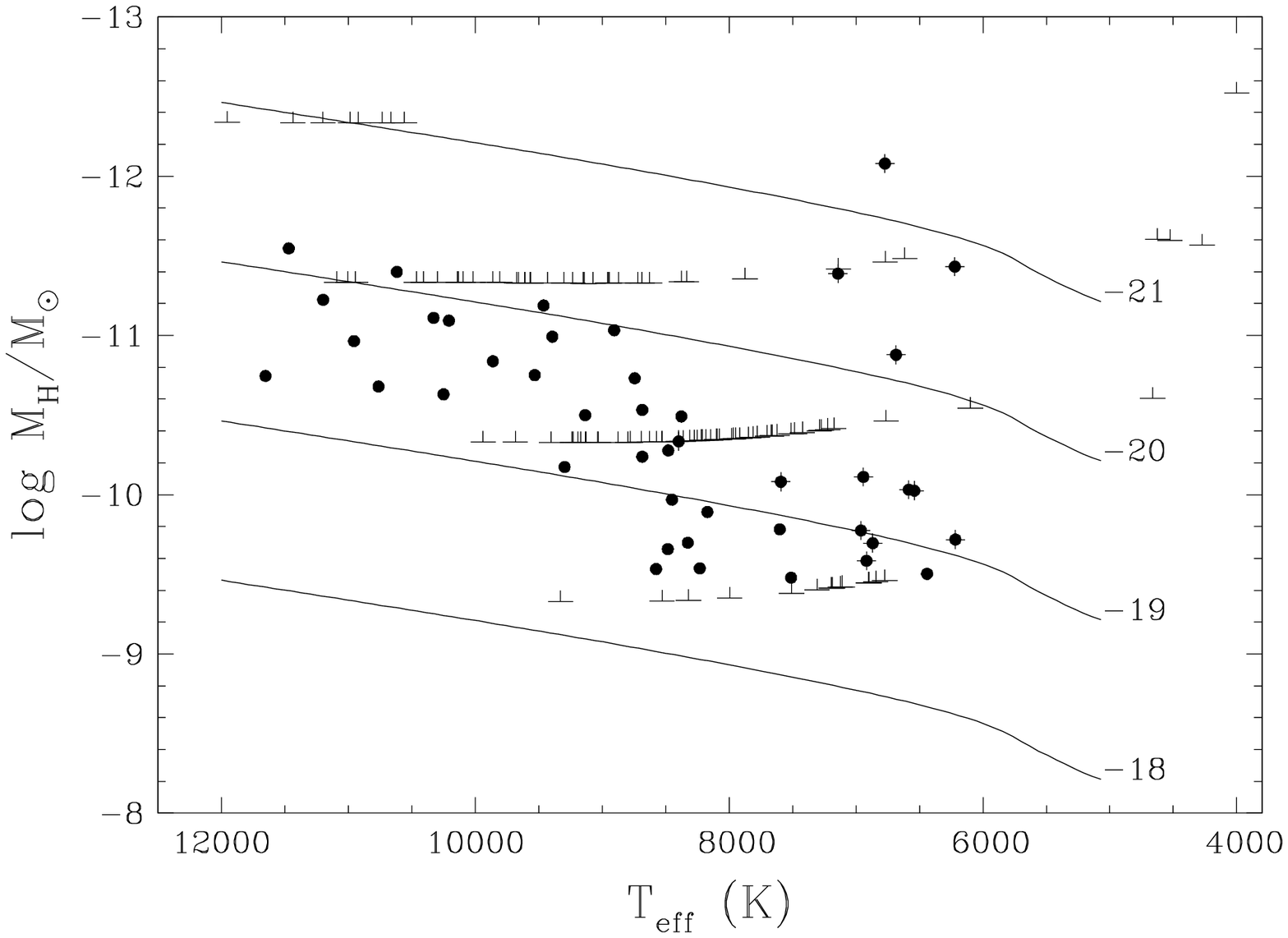] {Total mass of hydrogen present in the helium 
convection zone as a function of effective temperature. The symbols
are the same as in Figure \ref{fg:f29}. The solid curves indicate the
amount of hydrogen expected from continuous accretion of material with
solar composition and accretion rates of $10^{-21}$ to $10^{-18}\
M_{\odot}$ yr$^{-1}$.\label{fg:f31}}

\figcaption[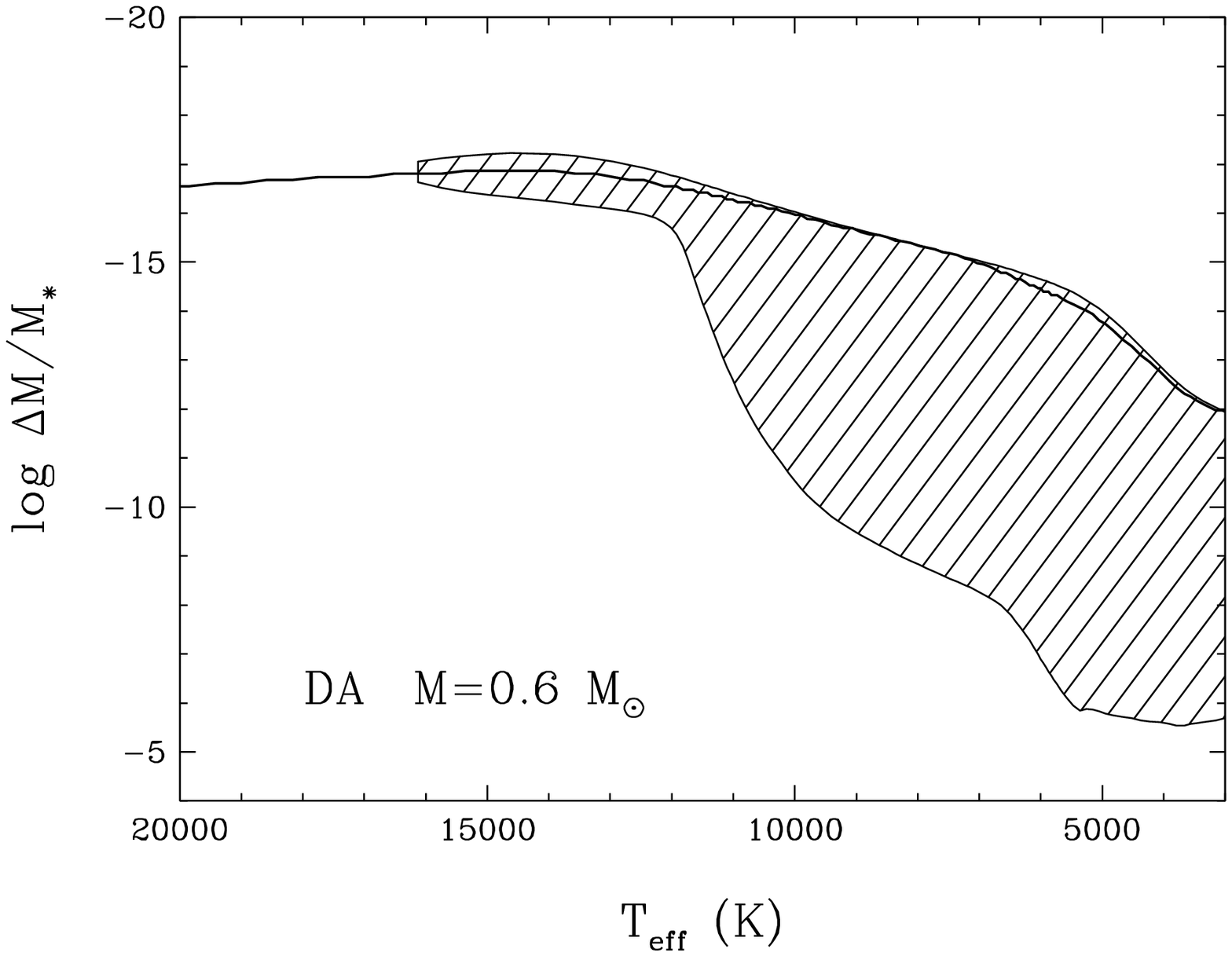] {Location of the hydrogen convection zone ({\it 
hatched area}) as a function of $\Te$ in a 0.6 \msun\ DA white dwarf.
The y-axis expresses on a logarithmic scale the mass fraction $\Delta
M$ above a certain point with respect to the total mass of the star
$M_\star$. The thick solid line indicates the location of the
photosphere.\label{fg:f32}}

\clearpage
\begin{figure}[p]
\plotone{f1}
\begin{flushright}
Figure \ref{fg:f1}
\end{flushright}
\end{figure}

\clearpage
\begin{figure}[p]
\plotone{f2}
\begin{flushright}
Figure \ref{fg:f2}
\end{flushright}
\end{figure}

\clearpage
\begin{figure}[p]
\plotone{f3}
\begin{flushright}
Figure \ref{fg:f3}
\end{flushright}
\end{figure}

\clearpage
\begin{figure}[p]
\plotone{f4}
\begin{flushright}
Figure \ref{fg:f4}
\end{flushright}
\end{figure}

\clearpage
\begin{figure}[p]
\plotone{f5}
\begin{flushright}
Figure \ref{fg:f5}
\end{flushright}
\end{figure}

\clearpage
\begin{figure}[p]
\plotone{f6}
\begin{flushright}
Figure \ref{fg:f6}
\end{flushright}
\end{figure}

\clearpage
\begin{figure}[p]
\plotone{f7}
\begin{flushright}
Figure \ref{fg:f7}
\end{flushright}
\end{figure}

\clearpage
\begin{figure}[p]
\plotone{f8}
\begin{flushright}
Figure \ref{fg:f8}
\end{flushright}
\end{figure}

\clearpage
\begin{figure}[p]
\plotone{f9}
\begin{flushright}
Figure \ref{fg:f9}
\end{flushright}
\end{figure}

\clearpage
\begin{figure}[p]
\plotone{f10}
\begin{flushright}
Figure \ref{fg:f10}
\end{flushright}
\end{figure}

\clearpage
\begin{figure}[p]
\plotone{f11}
\begin{flushright}
Figure \ref{fg:f11}
\end{flushright}
\end{figure}

\clearpage
\begin{figure}[p]
\plotone{f12}
\begin{flushright}
Figure \ref{fg:f12}
\end{flushright}
\end{figure}

\clearpage
\begin{figure}[p]
\plotone{f13}
\begin{flushright}
Figure \ref{fg:f13}
\end{flushright}
\end{figure}

\clearpage
\begin{figure}[p]
\plotone{f14}
\begin{flushright}
Figure \ref{fg:f14}
\end{flushright}
\end{figure}

\clearpage
\begin{figure}[p]
\plotone{f15}
\begin{flushright}
Figure \ref{fg:f15}
\end{flushright}
\end{figure}

\clearpage
\begin{figure}[p]
\plotone{f16}
\begin{flushright}
Figure \ref{fg:f16}
\end{flushright}
\end{figure}

\clearpage
\begin{figure}[p]
\plotone{f17}
\begin{flushright}
Figure \ref{fg:f17}
\end{flushright}
\end{figure}

\clearpage
\begin{figure}[p]
\plotone{f18}
\begin{flushright}
Figure \ref{fg:f18}
\end{flushright}
\end{figure}

\clearpage
\begin{figure}[p]
\plotone{f19}
\begin{flushright}
Figure \ref{fg:f19}
\end{flushright}
\end{figure}

\clearpage
\begin{figure}[p]
\plotone{f20}
\begin{flushright}
Figure \ref{fg:f20}
\end{flushright}
\end{figure}

\clearpage
\begin{figure}[p]
\plotone{f21}
\begin{flushright}
Figure \ref{fg:f21}
\end{flushright}
\end{figure}

\clearpage
\begin{figure}[p]
\plotone{f22}
\begin{flushright}
Figure \ref{fg:f22}
\end{flushright}
\end{figure}

\clearpage
\begin{figure}[p]
\plotone{f23}
\begin{flushright}
Figure \ref{fg:f23}
\end{flushright}
\end{figure}

\clearpage
\begin{figure}[p]
\plotone{f24}
\begin{flushright}
Figure \ref{fg:f24}
\end{flushright}
\end{figure}

\clearpage
\begin{figure}[p]
\plotone{f25}
\begin{flushright}
Figure \ref{fg:f25}
\end{flushright}
\end{figure}

\clearpage
\begin{figure}[p]
\plotone{f26}
\begin{flushright}
Figure \ref{fg:f26}
\end{flushright}
\end{figure}

\clearpage
\begin{figure}[p]
\plotone{f27}
\begin{flushright}
Figure \ref{fg:f27}
\end{flushright}
\end{figure}

\clearpage
\begin{figure}[p]
\plotone{f28}
\begin{flushright}
Figure \ref{fg:f28}
\end{flushright}
\end{figure}

\clearpage
\begin{figure}[p]
\plotone{f29}
\begin{flushright}
Figure \ref{fg:f29}
\end{flushright}
\end{figure}

\clearpage
\begin{figure}[p]
\plotone{f30}
\begin{flushright}
Figure \ref{fg:f30}
\end{flushright}
\end{figure}

\clearpage
\begin{figure}[p]
\plotone{f31}
\begin{flushright}
Figure \ref{fg:f31}
\end{flushright}
\end{figure}

\clearpage
\begin{figure}[p]
\plotone{f32}
\begin{flushright}
Figure \ref{fg:f32}
\end{flushright}
\end{figure}

\end{document}